\documentclass{ws-ijmpa}
\usepackage[super,compress]{cite}
\usepackage{graphicx}
\usepackage{epsfig,simplewick,color}
\usepackage{fancybox,tikz}
\usepackage{amsmath,tcolorbox}
\usepackage{array}
\usepackage[enableskew,vcenter]{youngtab}

\newcommand{\be}{\begin{equation}}
\newcommand{\bea}{\begin{eqnarray}}
\newcommand{\ee}{\end{equation}}
\newcommand{\eea}{\end{eqnarray}}

\begin{document}
\markboth{Robert de Mello Koch, Minkyoo Kim and Augustine Larweh Mahu}{A Pedagogical Introduction to restricted Schur polynomials with applications to Heavy Operators}

%
\catchline{}{}{}{}{}
%

\title{A pedagogical introduction to restricted Schur polynomials with applications to heavy operators
}

\author{Robert de Mello Koch
}

\address{
School of Science, Huzhou University, Huzhou 313000, China \\
\vspace{1mm}
\&
\vspace{1mm}
School of Physics and Mandelstam Institute for Theoretical Physics,\\
University of the Witwatersrand, Wits, 2050,
South Africa\\
robert@zjhu.edu.cn}

\author{Minkyoo Kim}

\address{Center for Quantum Spacetime (CQUeST),
Sogang University Seoul, 121-742, South Korea\\
mimkim80@gmail.com}

\author{Augustine Larweh Mahu}

\address{Department of Mathematics, University of Ghana, LG 62, Legon, Accra, Ghana\\
almahu@ug.edu.gh}

\maketitle


\begin{abstract}
Recent advances in the study of microstates for ${1\over 16}$-BPS black holes have inspired renewed interest in the analysis of heavy operators. For these operators, traditional techniques that work effectively in the planar limit are no longer applicable. Methods that are sensitive to finite $N$ effects are required. In particular, trace relations that connect different multi-trace operators must be carefully considered. A powerful approach to tackling this challenge, which utilizes the representation theory of the symmetric group, is provided by restricted Schur polynomials. In this review, we develop these methods with the goal of providing the background needed for their application to ${1\over 16}$-BPS black holes.

\end{abstract}

\ccode{PACS numbers:11.00.00}


\section{Introduction}

It is reasonable to expect that gauge theory/gravity dualities \cite{Maldacena:1997re,Gubser:1998bc,Witten:1998qj} will enable a detailed understanding of black hole microstates through the language of dual large-N gauge theories. Early progress in this direction successfully reproduced the Bekenstein-Hawking entropy for certain supersymmetric black holes \cite{Strominger:1996sh}, motivating numerous generalizations and refinements. However, until recently, precise insights into black hole entropy remained largely confined to index computations (see \cite{Cabo-Bizet:2018ehj,Choi:2018hmj,Benini:2018ywd} for references relevant to ${1\over 16}$-BPS black holes), which are insensitive to the detailed dynamics of the strongly coupled dual gauge theories -- dynamics expected to play a crucial role.

Recent developments have shifted this paradigm, particularly in the context of ${1\over 16}$-BPS black holes in AdS$_5\times$S$^5$ \cite{Chang:2022mjp,Choi:2022caq,Choi:2023znd,Budzik:2023vtr,Choi:2023vdm,Chang:2024zqi}. The task of counting and constructing black hole microstates can be reformulated in terms of supercharge cohomology, which captures the full spectrum of BPS operators. Advances in this area have driven significant recent progress. Notably, supercharge cohomology provides far more detailed information than indices alone. A key recent insight is the classification of BPS operators into two categories -- `monotone' and `fortuitous' -- as introduced in [\citen{Chang:2024zqi}]. Monotone BPS operators generate infinite sequences of operators as the gauge group rank $N$ increases, whereas fortuitous BPS operators exist only within finite ranges of consecutive ranks and lose their BPS status outside this range. Trace relations and their extensions play a critical role in this classification.

Since 't Hooft's pioneering work \cite{tHooft:1973alw} linking the large-$N$ planar limit to the free limit of a string theory, it has been well understood that Yang-Mills theories admit a systematic $1/N$ expansion. In the double-scaling limit, where $N\to\infty$ while holding $\lambda=g_{YM}^2 N$ fixed, these theories simplify dramatically. A remarkable connection emerges between the genus of ribbon graphs which determines their dependence on the gauge group rank $N$, and the worldsheets of a dual string theory. In particular, the leading (in ${1\over N}$) contribution in the planar limit has been extensively studied in ${\cal N}=4$ super Yang-Mills theory, thanks largely to the discovery of integrability \cite{Beisert:2010jr}. This connection offers a rich interpretation of single-trace operators as states of an integrable spin chain, with the dilatation operator corresponding to the spin chain Hamiltonian. This facilitates a description of the planar spectrum of operator dimensions at large $N$ for arbitrary 't Hooft couplings. However, this planar limit is not suitable for addressing the problem of black hole microstates. It fails to account for finite-$N$ trace relations and many simplifications that arise in the planar limit do not apply to the regimes relevant for black hole microstates. In this non-planar regime, the connection between ribbon graphs and string worldsheets is less important, and non-planar diagram effects become significant.

Addressing black hole microstates requires a novel approach. Finite-$N$ trace relations must be transparently incorporated, and operators with more complex multi-trace structures -- which obscure any straightforward connection to spin chain states -- must be considered. This review argues that approaches using group representation theory, which all build on the foundational insights of [\citen{Corley:2001zk}]  offer the proper framework for tackling this problem. Specifically, we will explore how the basis of gauge theory operators provided by restricted Schur polynomials offers a promising language through which the black hole microstate problem can be approached.

In Section \ref{motivation}, we begin by revisiting the organization of the gauge theory/gravity duality dictionary. Our goal is to highlight that while many intriguing questions can be effectively addressed within the planar limit, others cannot. To set the stage, we provide a brief review of the planar limit in Section \ref{planar}, emphasizing the simplifications that arise in this regime. These simplifications are crucial to understanding why, as we discuss in Section \ref{heavyopdiff}, they no longer apply when dealing with heavy operators.

Having identified the challenges associated with describing heavy operators, we argue in Section \ref{SchurRescue} that approaches based on Schur polynomials offer a way to overcome these difficulties. This framework naturally extends to multi-matrix models through the use of restricted Schur polynomials, which we elaborate on in Section \ref{RestSchurRescue}. While numerous papers have contributed to the theoretical foundations of this area, much of this material falls outside the central narrative of our review. To assist readers interested in deeper exploration, we provide references to relevant literature at the end of each section.

As an application of the restricted Schur polynomial formalism, Section \ref{ExcitedGiant} demonstrates how the states of excited giant graviton branes can be described using this approach. In Section \ref{PakMan}, we explore the counting of restricted Schur operators, and finally, we offer some preliminary remarks on the application of restricted Schur polynomials to black hole physics.

\section{Motivation}\label{motivation}

The AdS/CFT correspondence is a profound insight that deepens our understanding of both quantum gravity in negatively curved spacetimes and the dynamics of strongly coupled field theories. In this duality, operators in the Yang-Mills theory correspond to states in the string theory, and the mapping between them is organized according to the dimensions of the operators.

In free CFT$_4$, scalar fields have a dimension $\Delta = 1$, and the dimension of a composite operator is simply the sum of the dimensions of its individual constituents. For example, an operator constructed from $k$ fields has a dimension $\Delta = k$. Thus, in the free field theory, we can think of the dimension $\Delta$ of a composite operator as a direct reflection of the number of fields used in its construction.

\begin{table}[h]
    \centering
    \begin{tabular}{|>{\centering\arraybackslash}m{7cm}|>{\centering\arraybackslash}m{6cm}|}
        \hline
        \textbf{Dimension of CFT operator ($\Delta$)} & \textbf{String theory state} \\
        \hline
        $\Delta=O(1)$ & KK graviton \\
        \hline
        $\Delta=O(\sqrt{N})$ & BMN string \\
        \hline
        $\Delta=O(N)$ & giant graviton brane \\
        \hline
        $\Delta=O(N^2)$ & new geometry \\
        \hline
    \end{tabular}
    \caption{Mapping between CFT operators and string theory states.}
    \label{tab:cft_string}
\end{table}

To explore KK gravitons and string states, we can focus on the planar limit of the theory. However, when it comes to studying heavy operators such as branes and black holes, as we explain in detail in the next section, the techniques used in the planar limit are insufficient. In these cases, restricted Schur polynomials become an essential tool.

\section{Quick review of relevant aspects of the planar limit}\label{planar}

Consider a theory of a free complex matrix, described by the action
\bea
S&=&\int d^d x {\rm Tr}\left(\partial_\mu Z\partial^\mu Z^\dagger\right)
\eea
We have in mind that $Z$ is one of the scalars of ${\cal N}=4$ super Yang-Mills theory and that we study the free limit of the theory. In this case, these fields transform in the adjoint representation of the $U(N)$ gauge group
\bea
Z&\to& UZU^\dagger\qquad\qquad Z^\dagger \,\,\to\,\, UZ^\dagger U^\dagger\label{gtransform}
\eea
To compute correlators in the free field theory we simply apply Wick's theorem using the following basic contractions
\bea
\langle Z^i_j(x_1)\, (Z^\dagger)^k_l(x_2)\rangle&=&\delta^i_l\delta^k_j G(x_1-x_2)\cr\cr
\langle Z^i_j(x_1)\, Z^k_l(x_2)\rangle&=&0\,\,=\,\,\langle (Z^\dagger)^i_j(x_1)\, (Z^\dagger)^k_l(x_2)\rangle
\eea
In the matrix $Z^i_j$ the row label is $i$ and the column label is $j$. As we see from (\ref{gtransform}) row and column indices transform in a different way under a gauge transformation. Just as in special relativity where we distinguish covariant and contravariant vectors by the position of their indices, here we write the row index as a superscript and the column index as a subscript. In special relativity contracting a contravariant and a covariant index gives a Lorentz invariant. Here contracting an upper and a lower index gives a $U(N)$ invariant. As will become abundantly clear, our interest in what follows is in the colour combinatorics of the problem. Consequently, we drop spacetime dependence and simplify the first relation above to
\bea
\langle Z^i_j\,\, (Z^\dagger)^k_l\rangle&=&\delta^i_l\delta^k_j\label{basicwc}
\eea
Spacetime dependence can always be reinstated using conformal invariance\footnote{For example, the spacetime dependence of two point functions is determined once the dimensions of the operators are specified. Computing these dimensions is a complicated problem in the combinatorics of Feynman diagrams and their color indices. Once this is solved and the dimensions are determined, the two point function with all spacetime dependence is easy to write down.}.

Two natural questions we can ask are
\begin{itemize}
\item What is the complete set of gauge invariant observables constructed from $Z$?
\item What are the correlation functions of these observables?
\end{itemize}
By using only $Z^i_j$ fields (and not $Z^\dagger$s) we are constructing operators that are all ${1\over 2}$-BPS. The advantage of doing this is that the dimensions and correlators of these operators are not corrected \cite{Baggio:2012rr} so that free field results are actually exact. Further, although we have restricted the class of operators we consider, the resulting problem is still rich enough that we encounter all of the complications that arise when going beyond the planar limit and elucidating these complications is our main goal.

A very simple example of a planar operator is
\bea
{\cal O}_n={\rm Tr}(Z^n)
\eea
At large $N$, the two point function of this operator is given by
\bea
\langle {\cal O}_n{\cal O}_m^\dagger\rangle &=&\delta_{nm}n N^n\label{notnorm}
\eea
Correlators of this type are computed by drawing ribbon diagrams. Ribbon diagrams draw lines for the color indices of the matrices. So, for example, the operator
\bea
{\cal O}_3&=&{\rm Tr}(Z^3)=Z^i_jZ^j_kZ^k_i
\eea
is represented as the diagram
\begin{center}
\includegraphics[width=0.5\textwidth]{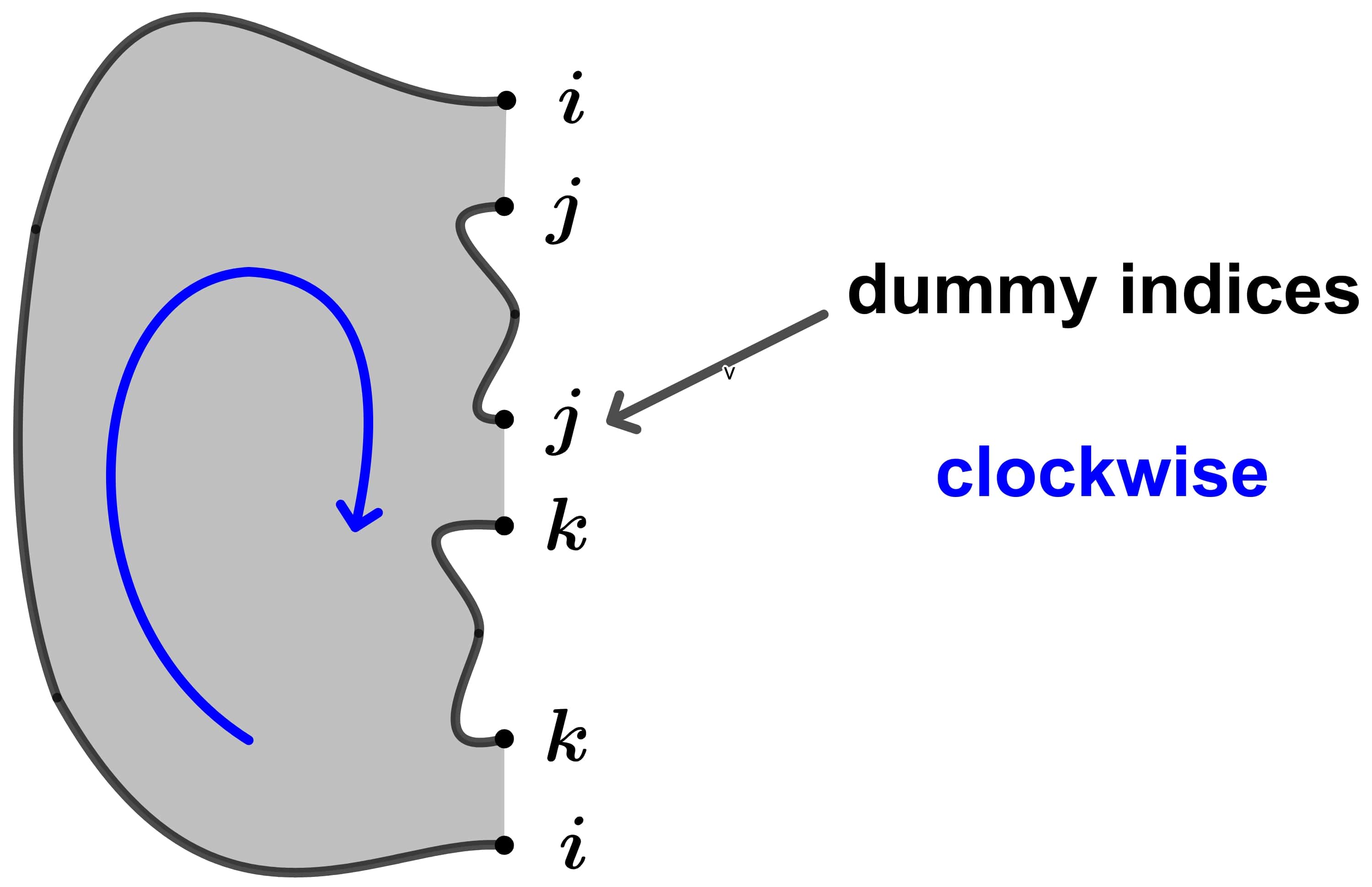}
\end{center}

\noindent
If we traverse the loop clockwise, each gap in the black loop corresponds to a matrix $Z$ and we encounter the row index first and the column index second. The indices $i$, $j$ and $k$ are all summed, i.e. they are dummy indices. Consequently they can freely be relabelled without changing the final answer. For this reason we simply drop dummy indices from the diagram. The Wick contraction also has a diagrammatic interpretation as follows
\begin{center}
\includegraphics[width=0.5\textwidth]{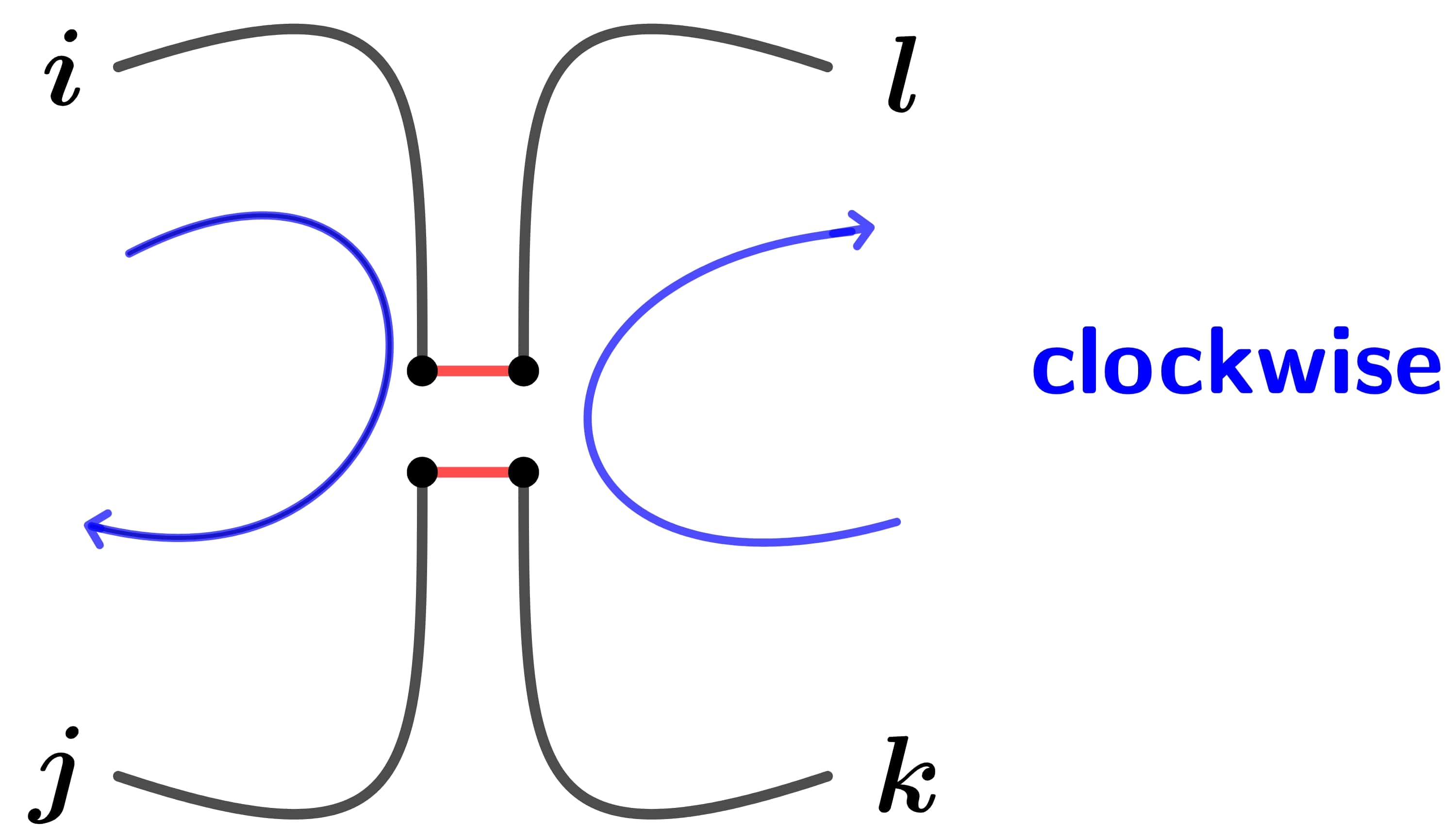}
\end{center}

\noindent
This diagram corresponds to the equation (\ref{basicwc}).  In the end, since we have correlators of gauge invariant operators, every Feynman diagram is composed of a collection of closed loops with alternating red and black lines. These can be translated into sums of Kronecker delta functions and evaluating these sum we easily find that each closed loop supplies a factor of $N$ as illustrated below.
\begin{center}
\includegraphics[width=0.5\textwidth]{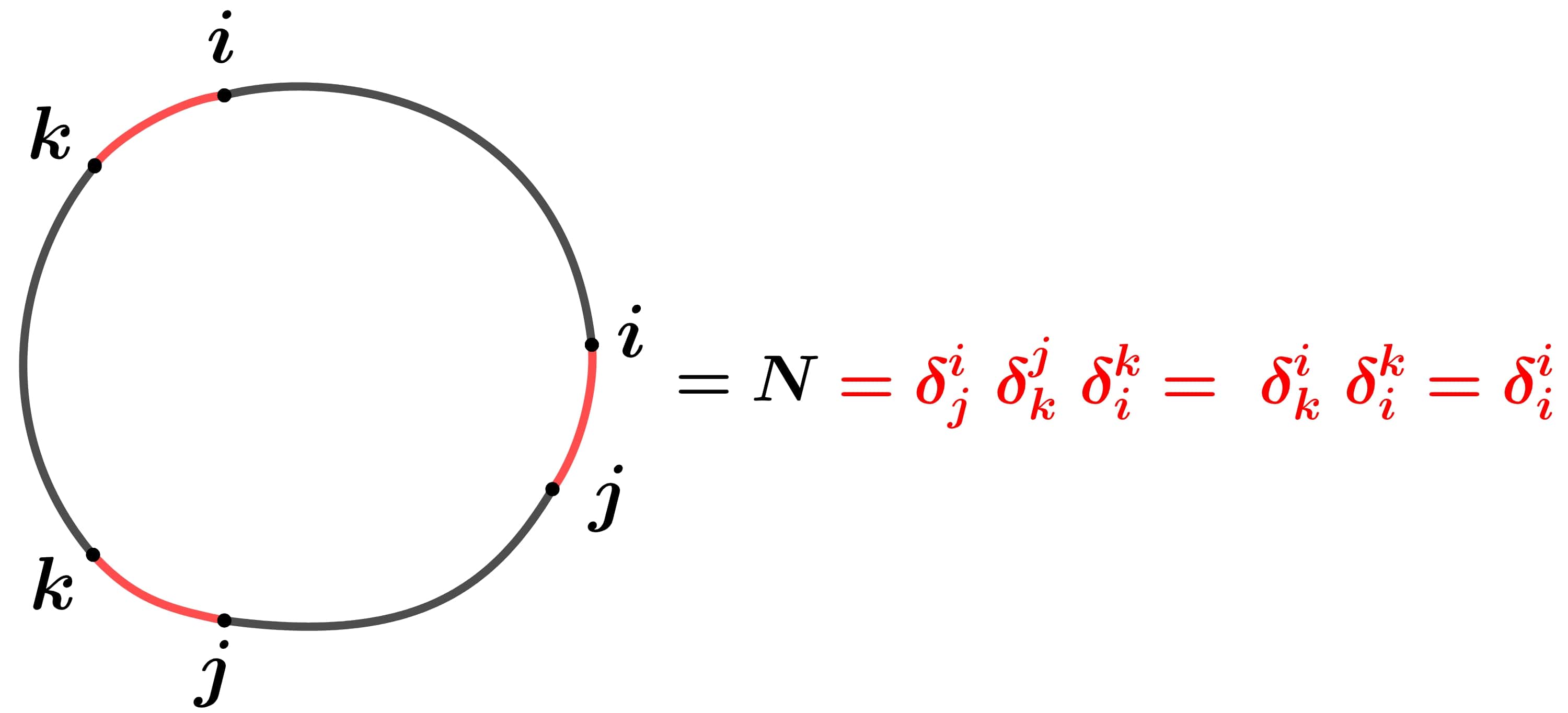}
\end{center}

\noindent
Wick's theorem instructs us to sum over all contractions. Diagrammatically, this means we must draw all possible diagrams. Consequently, the correlator we compute is given by
\begin{center}
\includegraphics[width=0.75\textwidth]{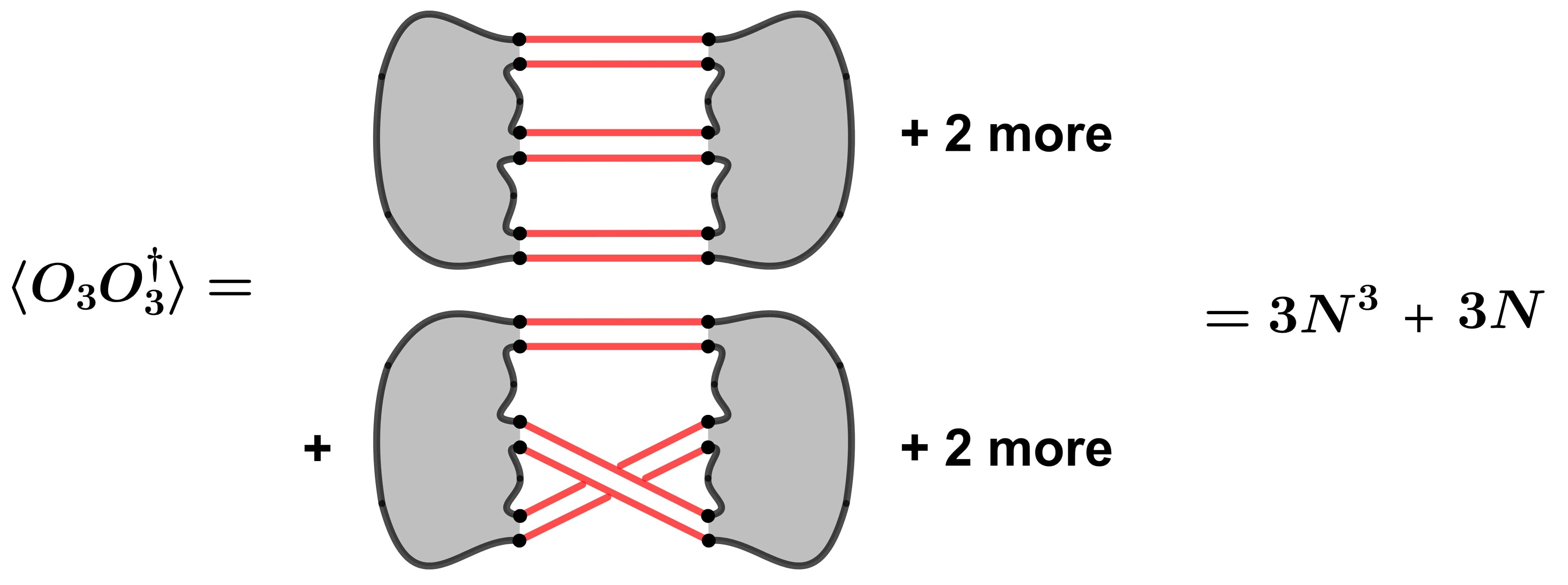}
\end{center}

\noindent
Diagrams with no lines crossing are called planar diagrams. Here is an example:
\begin{center}
\includegraphics[width=0.45\textwidth]{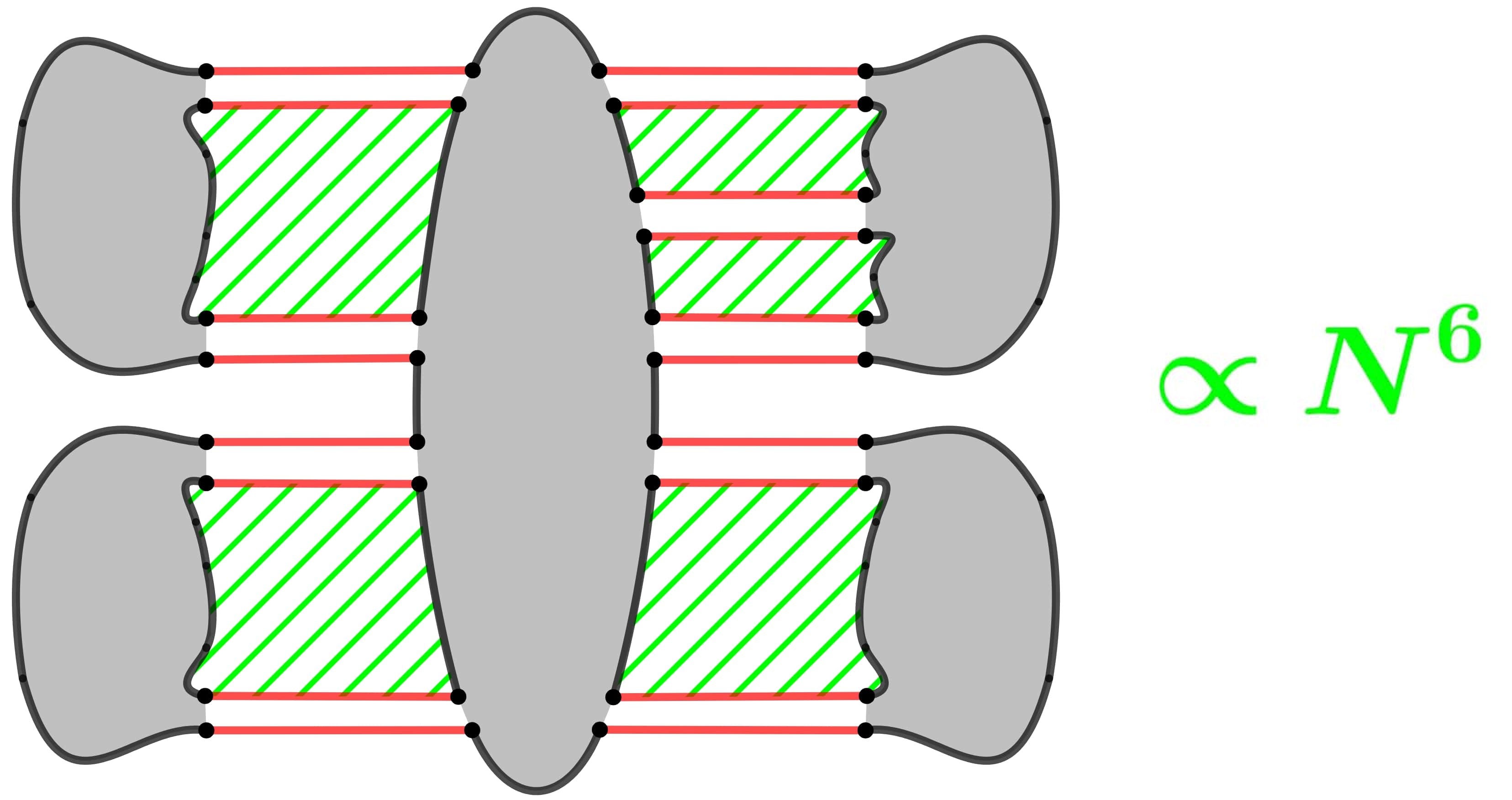}
\end{center}

\noindent
Diagrams with lines crossing are called non-planar diagrams. Here is an example:
\begin{center}
\includegraphics[width=0.45\textwidth]{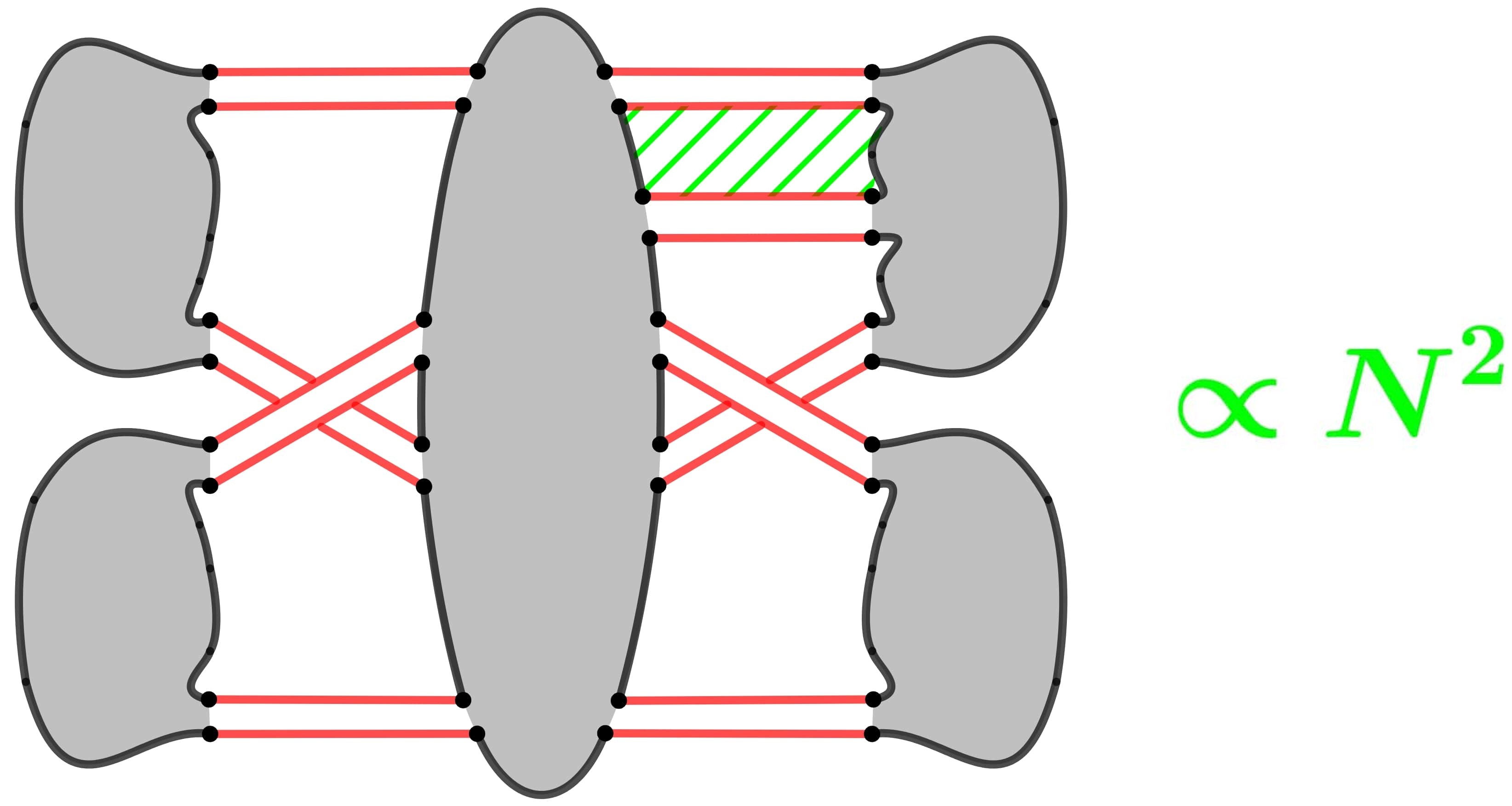}
\end{center}
\begin{tcolorbox}
Planar diagrams always come with more powers of $N$ than non-planar diagrams. Thus, non-planar diagrams can be dropped.
\end{tcolorbox}

{\vskip 1.0cm}

One inconvenient property of the operators we have considered so far is that their two point correlators diverge as we take $N\to\infty$. We can correct this by considering operators that are normalized as follows
\bea
O_n&=&{{\rm Tr}(Z^n)\over\sqrt{n}N^{n\over 2}}
\eea
Indeed, using (\ref{notnorm}) it is simple to verify that
\bea
\langle O_nO_m^\dagger\rangle&=&\delta_{nm}\left(1+O(N^{-2})\right)
\eea
Using these normalized operators we would like to argue that different trace structures do not mix. Towards this end, consider the correlator
\bea
\langle O_2 O_2 O_4^\dagger\rangle &=&{\langle {\rm Tr}(Z^2){\rm Tr}(Z^2){\rm Tr}(Z^{\dagger\, 4})\rangle\over (\sqrt{2}N)^2\sqrt{4}N^2}
\eea
Notice that the operator constructed from $Z$s is a double trace operator while the operator constructed from $Z^\dagger$s is a single trace operator. A typical planar diagram that contributes is given by
\begin{center}
\includegraphics[width=0.5\textwidth]{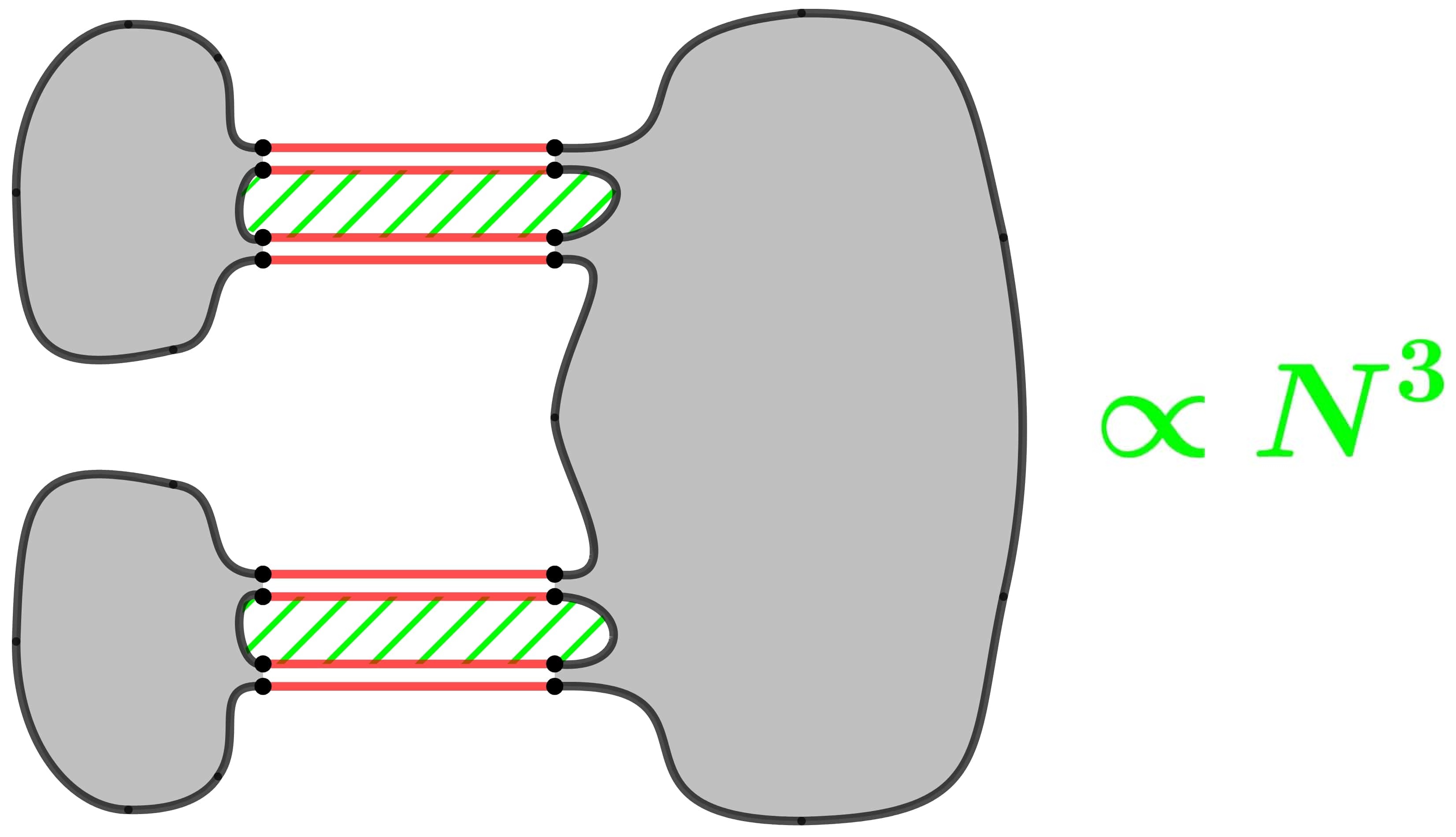}
\end{center}

\noindent
Notice that this diagram is of size $N^3$. Since this is a planar diagram, this is the highest power of $N$ possible for any of the diagram summed. Thus, estimating the size of the correlator using these dominant diagrams we find
\bea
\langle O_2 O_2 O_4^\dagger\rangle &\sim&{N^3 \over N^4}={1\over N}\to 0
\eea
On the other hand 
\bea
\langle O_4 O_4^\dagger\rangle&=&1\qquad\qquad \langle O_2 O_2 O_2^\dagger O_2^\dagger\rangle\,\,=\,\, 2
\eea
This computation illustrates a general result: take two operators, with different trace structures and which are both normalized to have two point functions that are $O(1)$ as $N\to\infty$. The correlator between these two operators vanishes as $N\to\infty$.
\begin{tcolorbox}
At large $N$ we have (for example)
\bea
\langle O_{n_1}O_{n_2}O^\dagger_{n_1+n_2}\rangle =0
\eea
At large $N$ different trace structures do not mix.
\end{tcolorbox}
We see that the trace basis orthogonalizes the large $N$ two points function in the sense that the two point function is diagonal in trace structure. This orthogonality of the different trace structures plays a key role in understanding how the structure of the free supergravity Fock space emerges at large $N$. In this context the number of traces is the number of particles and orthogonality of different trace structures is the orthogonality of states with a different number of particles. 

{\vskip 1.0cm}

\noindent
{\bf Summary of the key ideas:}
{\vskip 0.5cm}
\begin{itemize}
\item[1.] In the $N\to\infty$ limit we need only sum the planar diagrams. This is an enormous simplification: only a small subset of the Feynman diagrams need be summed. To get the large $N$ answer we need not sum everything. 
\item[2.] Different trace structures do not mix. Concretely, correlation functions of normalized operators with different trace structures vanish as we take the $N\to\infty$ limit. So, for example, to compute the anomalous dimensions of operators we can focus on single trace operators. This is a crucial ingredient for the spin chain description of integrability where the single trace operators are identified with states in the closed string Hilbert space.
\end{itemize}

\section{Complications associated with heavy operators}\label{heavyopdiff}

Let us estimate the correlator $\langle O_{J_1}O_{J_2}O^\dagger_{J_1+J_2}\rangle$ by summing only the planar diagrams. The result is
\bea
\langle O_{J_1}O_{J_2}O^\dagger_{J_1+J_2}\rangle&=&J_1J_2(J_1+J_2)\times {1\over\sqrt{J_1}\sqrt{J_2}\sqrt{J_1+J_2}N}
\eea
We have split the answer into two factors. The first factor $J_1J_2(J_1+J_2)$ simply counts the number of planar diagrams. The remaining factor is the product of the $N$ dependence of the planar diagram with the normalization factors of the operators $O_{*}$. Since the growth of the number of Feynman diagrams plays a key role in what follows, we briefly explain this counting. The first contraction contracts any matrix in $O_{J_1}$ with any matrix in $O_{J_1+J_2}^\dagger$. Thus, there are a total of $J_1(J_1+J_2)$ way to perform the first contraction. 
\begin{center}
\includegraphics[width=0.5\textwidth]{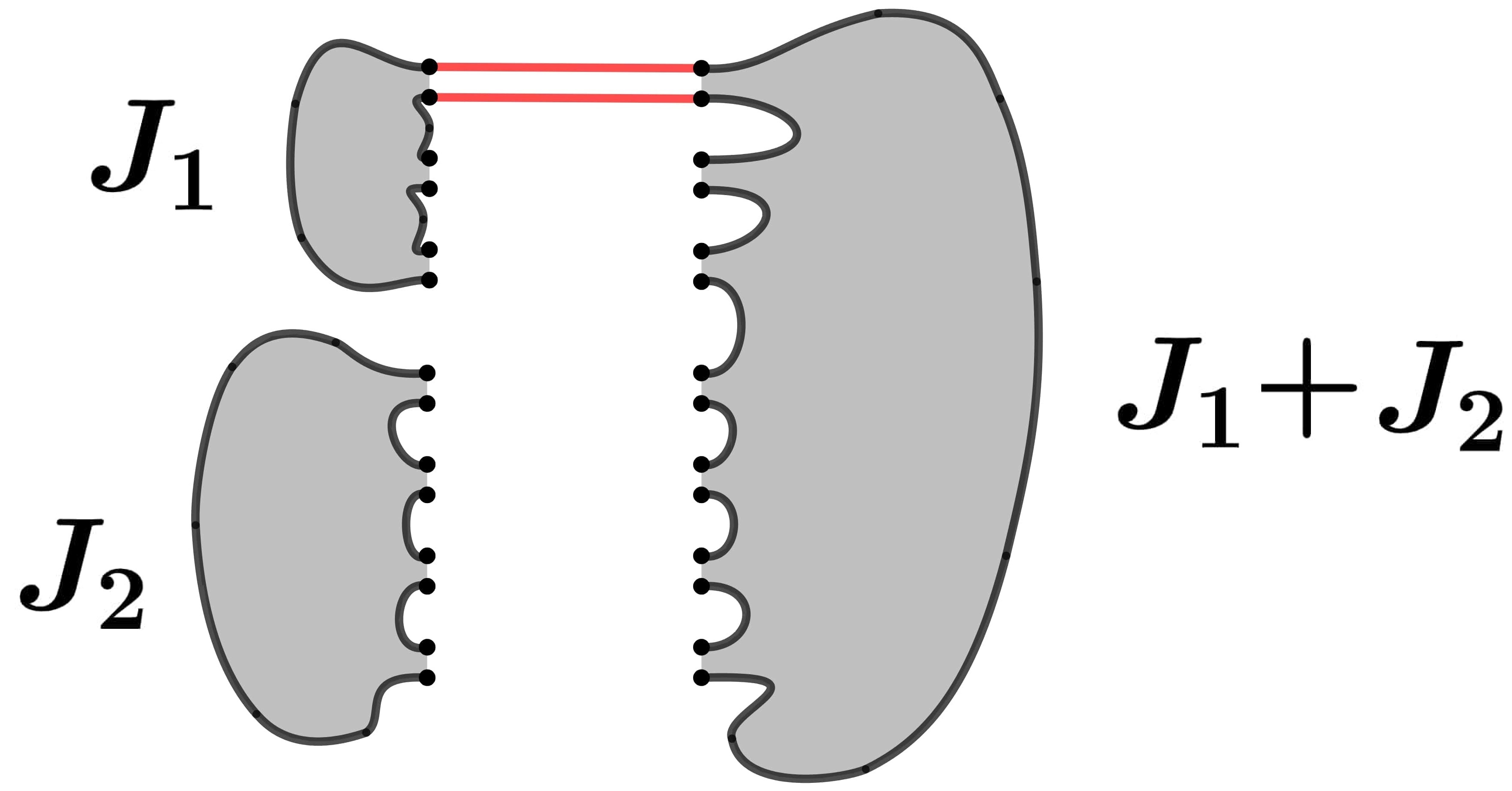}
\end{center}

\noindent
The contractions of the remaining $J_1-1$ fields in $O_{J_1}$ are unique if we are to obtain a planar diagram. 
\begin{center}
\includegraphics[width=0.5\textwidth]{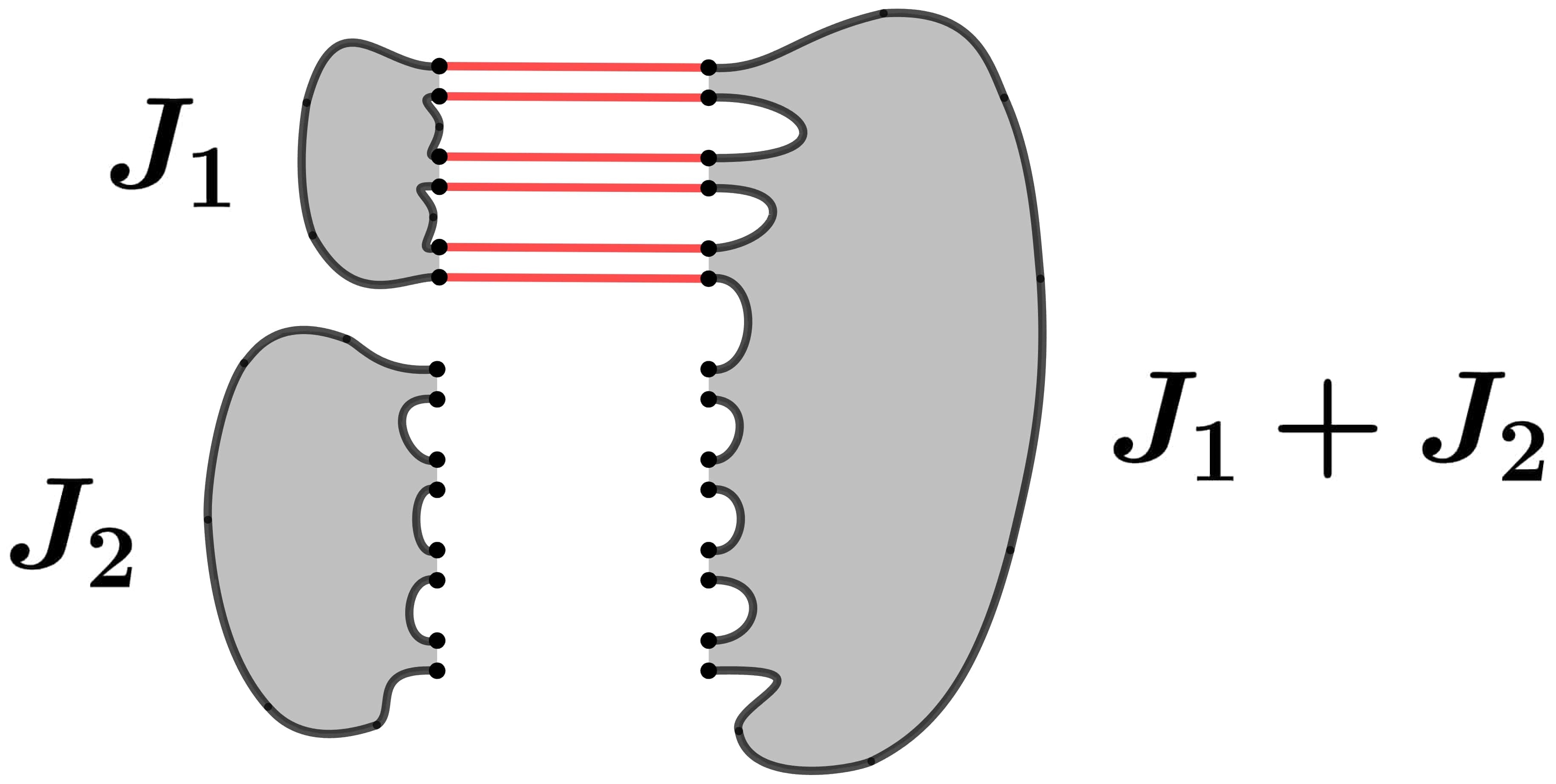}
\end{center}

\noindent
We can contract with any matrix in the $J_2$ trace, so that there are $J_2$ ways to perform the following contraction
\begin{center}
\includegraphics[width=0.5\textwidth]{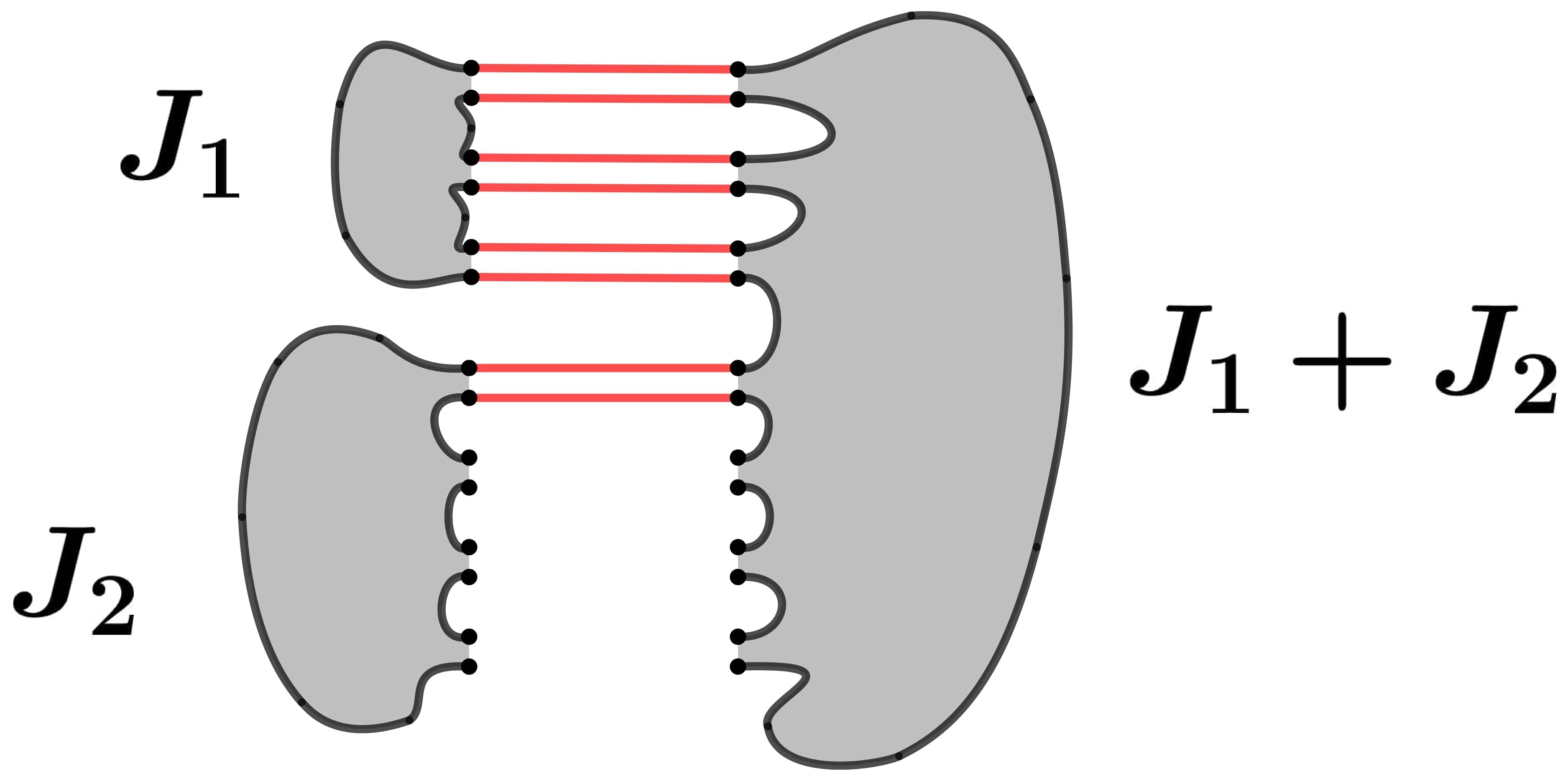}
\end{center}

\noindent
There is now a unique way to complete the diagram. Thus, the total number of diagrams is $J_1 J_2 (J_1+J_2)$ and the result follows. Simplifying a little we have
\bea
\langle O_{J_1}O_{J_2}O^\dagger_{J_1+J_2}\rangle&=&{\sqrt{J_1J_2(J_1+J_2)}\over N}+\cdots
\eea
where $\cdots$ stands for the planar contributions. Clearly, as soon as $J_i\sim N^{2\over 3}$ this correlator no longer vanishes\footnote{This is just a rough estimate. Taking the non-planar diagrams into account, one finds that traces start to mix as soon as $J_i\sim\sqrt{N}$. See, for example, [\citen{Constable:2002hw,Beisert:2002bb}] for nice discussions.}. Clearly for the heavy operators we are interested in where $J_i\sim O(N)$ or $J_i\sim O(N^2)$ mixing between different trace structures is completely unconstrained. Our operators are therefore allowed to have arbitrarily complicated trace structures: in general they will be sums over all possible trace structures which raises a question:
\begin{tcolorbox}
What is a good basis for the description of heavy operators?
\end{tcolorbox}

Before moving on, let's pause to ask why there is mixing between different trace structures? There are two important points to make:
\begin{itemize}
\item The largest diagram which contributes is still of order $N^{-1}$, i.e. this diagram still vanishes as $N\to\infty$. 
\item The number of diagrams $J_1J_2(J_1+J_2)$ diverges to infinity, growing faster than $N$ for operators with $J_i=O(N)$ or $J_i=O(N^2)$.
\end{itemize}
So, individual diagrams are as small as they were for operators in the planar limit. The point is that now the number of diagrams explodes and this overcomes the suppression of individual diagrams.

Consequently, even though the size of each non-planar diagram is still powers of $N$ smaller than a planar diagram, this is not enough to conclude that only planar diagrams contribute. It turns out that the number of non-planar diagrams grows much more rapidly that the number of planar diagrams and because of this the non-planar diagrams can not be neglected.  

There is one more complication associated with heavy operators that we will have to overcome. The matrix $Z^i_j$ is an $N\times N$ matrix so that it has $N$ eigenvalues. Denote these eigenvalues by $\lambda_i$ with $i=1,2,\cdots,N$. The trace of any power of $Z$ can be written as a power sum of the eigenvalues
\bea
{\rm Tr}(Z^n)&=&\sum_{i=1}^N\lambda_i^n
\eea
Given the values of ${\rm Tr}(Z^n)$ for $n=1,2,\cdots,N$ we can solve for the eigenvalues $\lambda_i$ and then compute the values of the traces ${\rm Tr}(Z^n)$ with $n>N$. This leads to identities between the traces. As an example, if $N=2$ there is a relation
\bea
{\rm Tr}(Z)^3-3{\rm Tr}(Z){\rm Tr}(Z^2)+2{\rm Tr}(Z^3)&=&0
\eea
which holds for any 2$\times$2 matrix $Z$. This is an example of a \emph{trace relation}. By solving the complete set of trace relations we could express an ${\rm Tr}(Z^n)$ with $n>N$ as a polynomial in the traces ${\rm Tr}(Z^n)$ with $n\le N$. For a single matrix it is simple to argue that ${\rm Tr}(Z^n)$ with $n\le N$ gives a complete set of invariants in the sense that any gauge invariant operator can be expressed as a polynomial in these traces. For more than a single matrix it is not at all clear how a complete (and not over complete) set could be specified. Heavy operators with $\Delta=O(N)$ or $\Delta=O(N^2)$ most certainly will be sensitive to the trace relations.

{\vskip 0.5cm}

\noindent
{\bf Summary of the key ideas:}
{\vskip 0.5cm}
\begin{itemize}
\item[1.] Since mixing between different trace structures is unconstrained, nothing prevents our operators from having an arbitrarily complicated trace structure. The trace basis that worked well for the planar limit is no longer useful. What is a useful description of the gauge invariant operators? 
\item[2.] Non-planar diagrams can no longer be dropped and there is no obvious subset of diagrams that dominate the $N\to\infty$ limit. It seems we must face the somewhat formidable task of summing all Feynman diagrams.
\item[3.] As a consequence of the trace relations, not all operators that we write down are independent. How do we account for the trace relations?
\end{itemize}

\section{Description of heavy operators using Schur polynomials}\label{SchurRescue}

A fascinating and powerful approach towards the dynamics of heavy operators was pioneered in the remarkable paper [\citen{Corley:2001zk}]. This description makes extensive use of the symmetric group $S_n$ which is the group of permutations of $n$ objects. Why do permutations play a role? There are at least 3 compelling reasons.

{\bf
\begin{itemize}
\item[1.] They provide a useful description for the gauge invariant operators.
\end{itemize}}

To obtain a gauge invariant operator, we need to contract all row and column indices. Here are two simple examples
\bea
{\rm Tr}(Z^2)^2&=&Z^{i_1}{}_{\color{red}i_2}Z^{\color{red}i_2}_{i_1}Z^{\color{blue}i_3}_{\color{green}i_4}Z^{\color{green}i_4}_{\color{blue}i_3}
\eea
\bea
{\rm Tr}(Z^4)&=&Z^{i_1}{}_{\color{red}i_2}Z^{\color{red}i_2}_{\color{blue}i_3}Z^{\color{blue}i_3}_{\color{green}i_4}Z^{\color{green}i_4}{}_{i_1}
\eea
Both operators displayed above have the same collection of upper and lower indices. For both, the lower indices are a permutation of the upper indices. The only difference between them is the permutation that is applied to obtain the lower indices. Thus, different operators correspond to different permutations.

In what follows we use cycle notation to denote the different permutations. Here is an example of the notation
\bea
\sigma=(12)(345)
\eea
This permutation is an element of $S_5$, written $\sigma\in S_5$. This permutation permutes 5 objects. It has two cycles - given in the two round brackets. Each cycle permutes the objects listed, in the order in which they are listed in the cycle. Thus, the above permutation can also be written as a map
\bea
\sigma(1)=2\qquad\sigma(2)=1\qquad\sigma(3)=4\qquad\sigma(4)=5\qquad\sigma(5)=3
\eea

The $N\times N$ matrix $Z$ acts on vectors that belong to an $N$ dimensional vector space, which we denote as $V_N$. To discuss operators constructed using $n$ $Z$ fields we will find it useful to phrase the discussion in the tensor product space $V_N^{\otimes n}$. In this larger vector space (of dimension $N^n$) we find that $Z^{\otimes n}$ acts as a matrix
\bea
(Z^{\otimes n})^I_J &\equiv& Z^{i_1}_{j_1}\cdots Z^{i_n}_{j_n}
\eea
We will also use the following representation of $S_n$ on $V_N^{\otimes n}$
\bea
\sigma^I_J&\equiv&\delta^{i_1}_{j_{\sigma(1)}}\cdots \delta^{i_n}_{j_{\sigma(n)}}
\eea
Using these definitions we have
\bea
{\rm Tr}(\sigma Z^{\otimes n})&=&\sigma^I_J Z^J_I\cr\cr
&=&\delta^{i_1}_{j_{\sigma(1)}}\cdots \delta^{i_n}_{j_{\sigma(n)}}
Z^{j_1}_{i_1}\cdots Z^{j_n}_{i_n}\cr\cr
&=&Z^{j_1}_{j_{\sigma(1)}}\cdots Z^{j_n}_{j_{\sigma(n)}}
\eea
The utility of this language is clearly illustrated in these two examples
\bea
{\rm Tr}(\sigma Z^{\otimes 4})&=&
Z^{i_1}_{i_2}Z^{i_2}_{i_1}Z^{i_3}_{i_4}Z^{i_4}_{i_3}\,\,=\,\,{\rm Tr}(Z^2){\rm Tr}( Z^2)\qquad\qquad\, {\rm for}\qquad \sigma\,\,=\,\,(12)(34)\nonumber
\eea
\bea
{\rm Tr}(\sigma Z^{\otimes 4})&=&
Z^{i_1}_{i_2}Z^{i_2}_{i_3}Z^{i_3}_{i_4}Z^{i_4}_{i_1}\,\,=\,\,{\rm Tr}(Z^4)\qquad\qquad\qquad\quad {\rm for}\qquad \sigma\,\,=\,\,(1234)\nonumber
\eea
What these two examples show is that now different multi-trace structures are all treated on the same footing.

These examples show that we can label different operators with permutations. A natural question now is if there is a unique permutation associated to each operator. The answer is that no there isn't: distinct permutations do not always lead to distinct operators. To see why this is the case, consider the tensor product of three $N\times N$ matrices $A$, $B$ and $C$
\bea
(A\otimes B\otimes C)^I_J&=&A^{i_1}_{j_1}B^{i_2}_{j_2}C^{i_3}_{j_3}
\eea
For $\sigma=(123)$ we have $\sigma^{-1}=(132)$ and
\bea
(\sigma^{-1})^I_J (A\otimes B\otimes C)^J_K (\sigma)^K_L&=&
\delta^{i_1}_{j_3}\delta^{i_2}_{j_1}\delta^{i_3}_{j_2}\,A^{j_1}_{k_1}B^{j_2}_{k_2}C^{j_3}_{k_3}\,\delta^{k_1}_{l_2}\delta^{k_2}_{l_3}\delta^{k_3}_{l_1}\cr\cr
&=&A^{i_2}_{l_2}B^{i_3}_{l_3}C^{i_1}_{l_1}\,\,=\,\,(C\otimes A\otimes B)^I_L
\eea
Thus, the permutation $\sigma$ acting as above simply shuffles the different matrices in the tensor product around. Consequently we have
\bea
(\sigma^{-1})^I_J (Z^{\otimes n})^J_K (\sigma)^K_L&=&(Z^{\otimes n})^I_L
\eea
for any $\sigma\in S_n$. Using this relation it is simple to see that
\bea
{\rm Tr}(\rho\, Z^{\otimes n})&=&{\rm Tr}(\rho\,\sigma^{-1} Z^{\otimes n}\sigma)\,\,=\,\,{\rm Tr}(\sigma\rho\,\sigma^{-1} Z^{\otimes n})
\eea
Thus, two permutations $\rho_1$ and $\rho_2$ related by
\bea
\rho_1=\sigma\rho_2\sigma^{-1}
\eea
lead to the same gauge invariant operator. Thus, our operators correspond to conjugacy classes and there is a unique gauge invariant operator associated to each conjugacy class.

\begin{table}[h]
    \centering
    \begin{tabular}{|>{\centering\arraybackslash}m{5cm}||>{\centering\arraybackslash}m{3cm}||>{\centering\arraybackslash}m{6cm}|}
      \hline
      \textbf{$[\sigma]$} & Structure &\textbf{Operator} \\
      \hline
      $(1)(2)(3)$  & $(\cdot)(\cdot)(\cdot)$& ${\rm Tr}(Z)^3$ \\
      \hline
      $(12)(3)$, $(13)(2)$, $(23)(1)$ &$(\cdot\cdot)(\cdot)$ & ${\rm Tr}(Z){\rm Tr}(Z^2)$ \\
      \hline
      $(123)$, $(132)$ & $(\cdot\cdot\cdot)$& ${\rm Tr}(Z^3)$ \\
      \hline
    \end{tabular}
    \caption{Conjugacy classes of $S_3$ and the associated gauge invariant operator.}
    \label{tab:cft_string}
\end{table}

Notice that in the above table all possible trace structures appear for operators constructed using three matrices and note also that the form of the operator has a transparent relation to the structure of the conjugacy class.

To summarize, we have obtained a useful description of the gauge invariant operators of the theory: for every permutation $\sigma\in S_n$ we can construct a gauge invariant operator as follows
\bea
{\rm Tr}(\sigma Z^{\otimes n})\label{giotrb}
\eea
There is a distinct gauge invariant operator for each conjugacy class of $S_n$.

{\bf
\begin{itemize}
\item[2.] The permutation language allows us to sum all possible Feynman diagrams, in free field theory.
\end{itemize}}

We can use Wick's theorem to compute any correlation functions of the form
\bea
\langle (Z^{\otimes n})^I_J(Z^{\dagger\,\otimes n})^K_L\rangle
\eea
We simply sum over all possible ways of pairing $Z$ and $Z^\dagger$ fields, replacing each pair with the basic Wick contraction given in (\ref{basicwc}). Notice that there are $n!$ ways to perform this pairing so that we have to sum $n!$ different terms, each of which is some Feynman diagram.

To motivate the formula that follows, consider computing the correlator for $n=3$
\bea
\langle Z^{i_1}_{j_1}Z^{i_2}_{j_2}Z^{i_3}_{j_3}(Z^\dagger)^{k_1}_{l_1}(Z^\dagger)^{k_2}_{l_2}(Z^\dagger)^{k_3}_{l_3}\rangle
\eea
The Wick contraction that pairs the first $Z$ with the second $Z^\dagger$, the second $Z$ with the third $Z^\dagger$ and the third $Z$ with the first $Z^\dagger$ is given by
\bea
\delta^{i_1}_{l_2}\delta^{i_2}_{l_3}\delta^{i_3}_{l_1}\delta^{k_1}_{j_3}\delta^{k_2}_{j_1}\delta^{k_3}_{j_2}&=&\big((123)\big)^I_L\,\big((132)\big)^K_J
\eea
This is an illustration of a general rule: every possible Wick contraction corresponds to a permutation and the complete correlator can be written as
\bea
\langle (Z^{\otimes n})^I_J(Z^{\dagger\, n})^K_L\rangle&=&\sum_{\sigma\in S_n}(\sigma)^I_L (\sigma^{-1})^K_J
\eea
Consequently, when using the permutation language the sum over all Wick contractions is replaced by a sum over the symmetric group. Below we will show that for certain observables it is possible to evaluate this sum very explicitly. For these observables we are able to sum the complete set of all Feynman diagrams.

{\bf
\begin{itemize}
\item[3.] Using the permutation language we can take all of the trace relations into account.
\end{itemize}}

To make this point we need to review some background. We will need some aspects of Schur-Weyl duality, which is a fundamental result in the representation theory of the symmetric group $S_n$ and the unitary group $U(N)$. This duality establishes a deep relationship between the representations of these two groups when they act on a tensor product space. Recall that
\bea
(Z^{\otimes n})^I_J &\equiv& Z^{i_1}_{j_1}\cdots Z^{i_n}_{j_n}
\eea
acts as a matrix in the tensor product space $V_N^{\otimes n}$.  We can define an action of $U(N)$ on the same space by declaring that for every element $U\in U(N)$ we have the operator
\bea
(U^{\otimes n})^I_J &\equiv& U^{i_1}_{j_1}\cdots U^{i_n}_{j_n}
\eea
Thus, $U(N)$ acts in exactly the same way on each factor of $V_N$ in the tensor product. The action of $S_n$ is as before, i.e. for $\sigma\in S_n$ we have
\bea
(\sigma)^I_J&=&\delta^{i_1}_{j_{\sigma(1)}}\cdots\delta^{i_n}_{j_{\sigma(n)}}
\eea
so that $S_n$ permutes the order of factors in the tensor product, but otherwise leaves each unchanged. Clearly then, these actions of $S_n$ and $U(N)$ on $V_N^{\otimes n}$ commute. As usual, when two operators commute they can be simultaneously diagonalized. Following this to its conclusion, mathematicians have proved that\footnote{The notation $s\vdash n$ says that $s$ is a partition of $n$. Reading each part of the partition as a row length, we can also read $s\vdash n$ as ``$s$ is a Young diagram with $n$ boxes''. In (\ref{SWDuality}) $V^s_{U(N)}$ is a vector space carrying the representation $s$ of $U(N)$ and $V^s_{S_n}$ is a vector space carrying the representation $s$ of $S_n$}
\bea
V_N^{\otimes n}=\bigoplus_{s\vdash n} V^s_{U(N)}\otimes V^s_{S_n}\label{SWDuality}
\eea
The sum here runs over all Young diagrams with $n$ boxes. There are a few fascinating things that this formula makes clear:
\begin{itemize}
\item[1.] Representation of both $U(N)$ and $S_n$ are labelled by Young diagrams with $n$ boxes. For representations of $U(N)$ the Young diagram can have any number of boxes, but it must have no more than $N$ rows. For $S_n$ the Young diagram must have exactly $n$ boxes.
\item[2.] When we take the tensor product of $n$ $Z$ fields, we find many representations $s$ of $U(N)$ appear. To distinguish these different identical copies we need a multiplicity label. The above formula makes it clear that the multiplicities of the $U(N)$ representations $s$ are organized by an $S_n$ representation with the same Young diagram $s$.
\item[3.] When we take the tensor product of $n$ $Z$ fields, we also find many representations $s$ of $S_n$ appear. A second way to read the above formula is that the multiplicities of the $S_n$ representations $s$ are organized by a $U(N)$ representation with the same Young diagram $s$.
\end{itemize}

A simple sanity check of (\ref{SWDuality}), is to compute dimensions on both sides and verify that they match. For this we need to compute dimensions of irreducible representations of both $U(N)$ and $S_n$.

To compute the dimensions of representations of $S_n$, we associate to each box of the Young diagram a hook length. The hook is a line drawn through the Young diagram. It enters the diagram from the bottom and leaves the diagram to the right. The elbow of the hook lies in a specific box. The hook length of that box is the number of boxes the elbow passes through. In the figure below two hooks and their lengths are shown
\begin{center}
\includegraphics[width=0.2\textwidth]{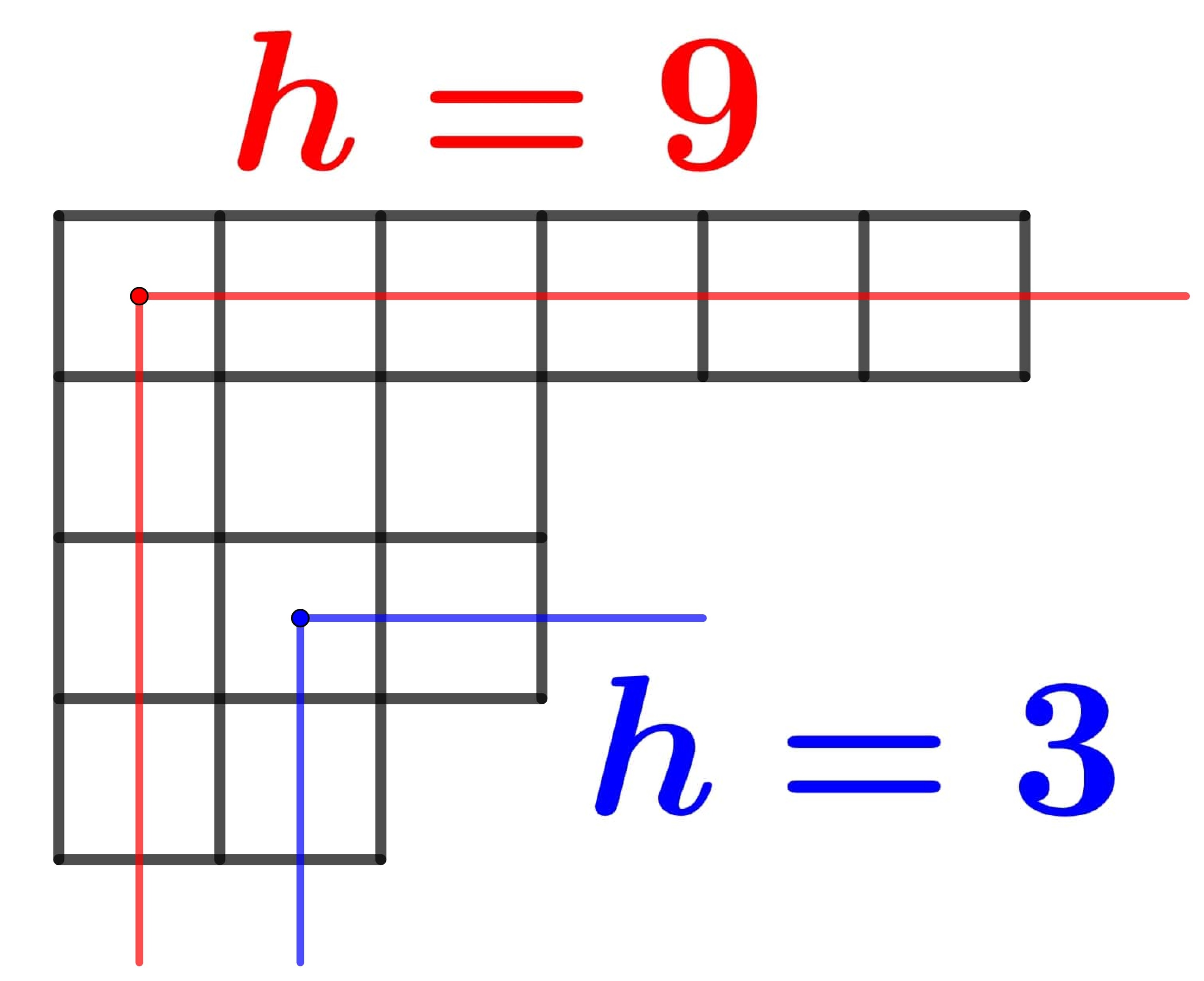}
\end{center}

\noindent
The Young diagram with all hook lengths filled in is given by
\bea
\young(986321,542,431,21)
\eea
The dimension of an $S_n$ representation labelled by a Young diagram $s\vdash n$ is given by $n!$ divided by the product of the hook lengths. For the Young diagram shown above the dimension is
\bea
{\rm dim}_{\tiny\yng(6,3,3,2)}&=&{14!\over 9\cdot 8\cdot 6\cdot 3\cdot 2\cdot 1\cdot 5\cdot 4\cdot 2\cdot 4\cdot 3\cdot 1\cdot 2\cdot 1}\,\,=\,\,35035
\eea
Thus, each of the 14! matrices of this representation would be 35035$\times$35035 dimensional matrices. 

To compute the dimensions of a $U(N)$ representation labelled by a Young diagram $s$, we associate a second number to each box of the Young diagram known as the factor of the box. A box in row $i$ and column $j$ has factor $N-i+j$. A Young diagram with both the hook lengths and the factors indicated is shown below.
\begin{center}
\includegraphics[width=0.5\textwidth]{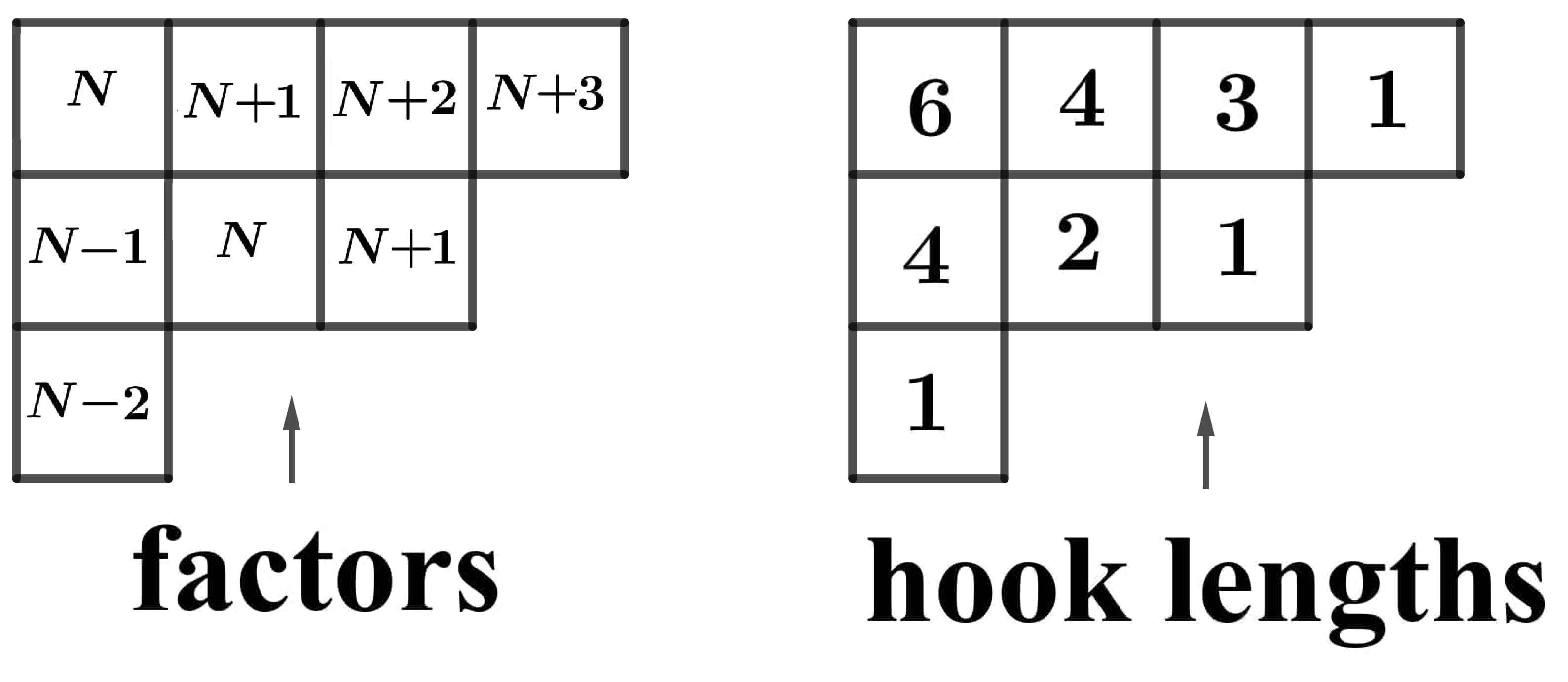}
\end{center}

\noindent
The dimension of the $U(N)$ representation is given by the product of the factors divided by the product of the hooks. For the Young diagram shown we have
\bea
{\rm Dim}_{\tiny\yng(4,3,1)}={N(N+1)(N+2)(N+3)(N-1)N(N+1)(N-2)\over 6\cdot 4\cdot 3\cdot 1\cdot 4\cdot 2\cdot 1\cdot 1}
\eea

With these dimension formulas in hand we can now return to the Schur-Weyl decomposition of $V_N^{\otimes n}$ and check that dimensions work out correctly. Consider the case that $n=3$. There are three Young diagrams
\bea
s=\left\{{\tiny\yng(3)}\,,\,{\tiny \yng(2,1)}\, ,\,{\tiny \yng(1,1,1)}\right\}
\eea
The dimensions of these as $U(N)$ and $S_3$ representations are given in the table below.

\begin{table}[h]
    \centering
    \begin{tabular}{|>{\centering\arraybackslash}m{5cm}||>{\centering\arraybackslash}m{3cm}||>{\centering\arraybackslash}m{6cm}|}
      \hline
      \textbf{Young\,\,diagram} &\textbf{${\rm dim}_s$} &\textbf{${\rm Dim}_s$}\\
      \hline
      {\tiny $\yng(3)$}  & 1 & ${N(N+1)(N+2)\over 6}$ \\
      \hline
      {\tiny $\yng(2,1)$} & 2 & ${N(N^2-1)\over 3}$ \\
      \hline
      {\tiny $\yng(1,1,1)$}& 1 & ${N(N-1)(N-2)\over 6}$ \\
      \hline
    \end{tabular}
    \caption{Dimensions of representation for $n=3$.}
    \label{tab:cft_string}
\end{table}
The Schur-Weyl decomposition of $(V_N)^{\otimes 3}$ is
\bea
V_N^{\otimes 3}&=&\left(V^{U(N)}_{\tiny\yng(3)}\otimes V^{S_3}_{\tiny\yng(3)}\right)\oplus\left(V^{U(N)}_{\tiny\yng(2,1)}\otimes V^{S_3}_{\tiny\yng(2,1)}\right)\oplus\left(V^{U(N)}_{\tiny\yng(1,1,1)}\otimes V^{S_3}_{\tiny\yng(1,1,1)}\right)
\eea
The dimension of $V_N^{\otimes 3}$ is $N^3$. This can be compared to
\bea
&&{\rm dim}_{\tiny\yng(3)}\,{\rm Dim}_{\tiny\yng(3)}+{\rm dim}_{\tiny\yng(2,1)}\,{\rm Dim}_{\tiny\yng(2,1)}+{\rm dim}_{\tiny\yng(1,1,1)}\,{\rm Dim}_{\tiny\yng(1,1,1)}\cr\cr
&=&1\cdot{N(N+1)(N+2)\over 6}+2\cdot{N(N^2-1)\over 3}+1\cdot{N(N-1)(N-2)\over 6}\cr\cr
&=& N^3
\eea

Another consequence of Schur-Weyl duality is a formula for the characters of $U(N)$ (denoted by $\chi_R(U)$) in terms of those of $S_n$ (denoted $\chi_R(\sigma)$)
\bea
\chi_R(U)&=&{1\over n!}\sum_{\sigma\in S_n}\chi_R(\sigma){\rm Tr}(\sigma U^{\otimes n})
\eea
where $R$ is any Young diagram with $n$ boxes, written as $R\vdash n$. This formula is most easily understood by noting that
\bea
(P_R)^I_J&=&{1\over n!}\sum_{\sigma\in S_n}\chi_R(\sigma)\sigma^I_J
\eea
is a projection operator acting in $V_N^{\otimes n}$. For any projection operator we expect
\bea
P_R\cdot P_S&\propto&\delta_{RS}P_R
\eea
For a properly normalized projection operator we would have an equality above, the left and right hand sides would not simply be proportional. $P_R$ projects onto the $V^R_{U(N)}\otimes V^R_{S_n}$ subspace of $V_N^{\otimes n}$. The product $P_R\cdot P_S$ is easily evaluated with the help of Schur's orthogonality relation which reads
\bea
\sum_{g\in{\cal G}} \Gamma_R(g)_{ab}\Gamma_S(g^{-1})_{cd}&=&\delta_{RS}\delta_{ad}\delta_{bc}\, {|{\cal G}|\over d_R}\label{FOR}
\eea
where ${\cal G}$ is any finite group, $|{\cal G}|$ is the order of ${\cal G}$, $R$ and $S$ label irreducible representations and $d_R$ is the dimension of irreducible representation $R$. $\Gamma_R(g)_{ab}$ is the $d_R\times d_R$ matrix representing $g$ in the irreducible representation $R$. When we say we have a representation of the group, we mean we have found a set of matrices that solve the equation
\bea
\Gamma(g_1)\Gamma(g_2)&=&\Gamma(g_1g_2)
\eea
On the left hand side of this equation there is the usual matrix product between two matrices, while on the right hand side we have a product between two elements of the group, specified by the multiplication table of the group. If we specialize (\ref{FOR}) to $S_n$, then $R$ and $S$ are Young diagrams with $n$ boxes, $|{\cal G}|=|S_n|=n!$ and $d_R$ is what we have been calling ${\rm dim}_R$. Schur's orthogonality relation is an extremely important result, well worth memorizing.

Using Schur's orthogonality relation is it easy to prove that
\bea
P_R P_S &=&{1\over{\rm dim}_R}P_R\delta_{RS}
\eea
Thus
\bea
\hat{P}_R&=&{\rm dim}_R P_R
\eea
is a properly normalized projection operator and we have
\bea
\hat{P}_R\hat{P}_S&=&\delta_{RS}\hat{P}_R
\eea
Thus, the trace of $\hat{P}_R$ is the dimension of the space it projects to so that
\bea
{\rm Tr}(\hat{P}_R)&=&{\rm dim}_R{\rm Dim}_R
\eea
and hence
\bea
{\rm Tr}(P_R)&=&{\rm Dim}_R
\eea
We will make good use of this formula below.

We need one more property of the projectors $P_R$: they commute with all elements $\rho\in S_n$. To prove this, start from
\bea
P_R  \rho&=&{1\over n!}\sum_{\sigma\in S_n}\chi_R(\sigma)\sigma\rho
\eea
We want to pull $\rho$ out on the left to prove that $\rho$ commutes with $P_R$. Thus, it seems we should trade $\sigma$ for $\tilde{\sigma}$ where
\bea
\sigma\rho&=&\rho\tilde{\sigma}
\eea
We will want to trade the sum over $\sigma$ for a sum over $\tilde{\sigma}$. To make this trade, note that if $\sigma_1\ne\sigma_2$, then multiplying on the right by $\rho$ implies that $\sigma_1\rho\ne\sigma_2\rho$, and now multiplying on the left with $\rho^{-1}$ implies $\rho^{-1}\sigma_1\rho\ne\rho^{-1}\sigma_2\rho$, i.e. $\tilde{\sigma}_1\ne \tilde{\sigma}_2$. Thus, a sum over all values of $\sigma$ can be traded for a sum over all values of $\tilde{\sigma}$. Finally, it is a simple exercise to show that
\bea
\chi_R(\sigma)&=&\chi_R(\tilde{\sigma})
\eea
Thus, we now have
\bea
P_R\rho&=&{1\over n!}\sum_{\sigma\in S_n}\chi_R(\sigma)\sigma\rho\cr\cr
&=&{1\over n!}\sum_{\tilde{\sigma}\in S_n}\chi_R(\tilde{\sigma})\rho\tilde{\sigma}\cr\cr
&=&\rho P_R
\eea
which proves that $P_R$ commutes with any $\rho\in S_n$.

You probably already know how $P_R$ acts, but the language being used in this discussion might not be familiar. $P_R$ is known as a Young projector. Consider the simplest possible case of $n=2$ which is enough to illustrate the point. For $n=2$ there are two Young diagrams
\bea
{\tiny\yng(2)}\qquad\qquad{\tiny\yng(1,1)}
\eea
S$_2$ has two elements: $S=\{(1)(2),(12)\}$. It is simple to check that both of these representations are one dimensional. The characters for these two representations (which are the group elements since they are $1\times 1$ matrices) are given by
\bea
\chi_{\tiny\yng(2)}((1)(2))&=&1\,\,=\,\,\chi_{\tiny\yng(2)}((12))
\eea
and
\bea
\chi_{\tiny\yng(1,1)}((1)(2))&=&1\qquad \chi_{\tiny\yng(1,1)}((12))\,\,=\,\,-1
\eea
Thus, the projectors $P_R$ associated to these two representations are given by
\bea
(P_{\tiny\yng(2)})^I_J&=& {1\over 2}\left( \big((1)(2)\big)^I_J+\big((12)\big)^I_J\right)\,\,=\,\, {1\over 2}\left(\delta^{i_1}_{j_1}\delta^{i_2}_{j_2}+\delta^{i_1}_{j_2}\delta^{i_2}_{j_1}\right)
\eea
\bea
(P_{\tiny\yng(1,1)})^I_J&=& {1\over 2}\left( \big((1)(2)\big)^I_J-\big((12)\big)^I_J\right)\,\,=\,\, {1\over 2}\left(\delta^{i_1}_{j_1}\delta^{i_2}_{j_2}-\delta^{i_1}_{j_2}\delta^{i_2}_{j_1}\right)
\eea
These projectors act on a tensor with two indices in the obvious way
\bea
P^I_J T^J &=&P^{i_1i_2}_{j_1j_2}T^{j_1j_2}
\eea
For the projectors defined above this action is
\bea
(P_{\tiny\yng(2)}T)^I&=&{T^{i_1i_2}+T^{i_2i_1}\over 2}
\eea
\bea
(P_{\tiny\yng(1,1)}T)^I&=&{T^{i_1i_2}-T^{i_2i_1}\over 2}
\eea
which is projecting onto the symmetric and anti-symmetric components of $T^{i_1i_2}$ respectively. In general, a projector $P_R$ with a more complicated Young diagram $R$ will symmetrize indices corresponding to boxes in the same row and it will anti-symmetrize indices corresponding to boxes in the same column.

With the background already developed, it is possible to make contact with the trace relations. The foundation for this discussion is from {\it invariant theory}. Hilbert produced many of the key results in this field. See [\citen{sneddon}] for a readable introduction.

An {\it invariant} of a set of tensors in $d$ dimensions is a polynomial function of the components of those tensors that is invariant under some group of transformations of the tensors. In our discussion the tensors are $Z^i_j$ and the group of transformations is $U(N)$ acting as
\bea
Z&\to&UZU^\dagger
\eea
Invariant theory considers two problems:
\begin{itemize}
\item the nature of the invariants,
\item the nature of the relationship between them.
\end{itemize}

There are two central theorems:

{\vskip 0.5cm}

\noindent
{\bf The First Fundamental Theorem:} Any polynomial invariant of a set of tensors can be expressed as a linear combination of complete contractions of products of these tensors.

{\vskip 0.5cm}

For us, it means that our invariants are general multi-trace operators.

{\vskip 0.5cm}

\noindent
{\bf The Second Fundamental Theorem:} Any identity between the invariants of a set of tensors in $d$-dimensions can be obtained as a consequence of the fact that antisymmetrizing over $d+1$ indices will annihilate the tensor.

{\vskip 0.5cm}

For our problem, this implies that the trace relations can {\it all} be obtained as statements of the form
\bea
{\rm Tr}(P_R Z^{\otimes n})&=&0\label{TRasPE}
\eea
where $R$ is a Young diagram with $n>N$ boxes ($R\vdash n$) and $R$ has more than $N$ rows.

One more idea is needed - related to the Fourier transform. The usual Fourier transform can be written as
\bea
\tilde{f}(k)&=&\int dx \, e^{ikx} f(x)\qquad\qquad f(x)\,\,=\,\,\int{dk\over 2\pi}e^{-ikx}\tilde{f}(k)
\eea
The key idea behind the Fourier transform is the fact that the plane waves are a complete and orthogonal set of functions, which is embodied in the formulas
\bea
\int_{-\infty}^\infty \, dx\, e^{ikx}e^{-ik'x}&=&2\pi\delta(k-k')\cr\cr
\int_{-\infty}^\infty \, {dk\over 2\pi}\, e^{ikx}e^{-ikx'}&=&\delta(x-x')
\eea
Notice that $e^{ikx}$ has both a momentum ($k$) and a position ($x$) label.

The statement (\ref{TRasPE}) makes it clear that the trace relations are closely related to Young diagrams. The gauge invariant operators defined by (\ref{giotrb}) are related to conjugacy classes. At this point it is useful to introduce a Fourier transform between functions defined on the conjugacy classes (like our gauge invariant operators) and functions defined on the irreducible representations (i.e. the Young diagrams).

If $\sigma$ is a permutation, denote the conjugacy class it belongs to by $[\sigma]$. The Fourier transform is
\bea
f([\sigma])&=&\sum_{R\vdash n}\chi_R(\sigma) \tilde{f}_R\qquad \sigma\in S_n\cr\cr
\tilde{f}_R&=&{1\over n!}\sum_{\sigma\in S_n} \chi_R(\sigma)f([\sigma])
\eea
This Fourier transform works\footnote{We have stated this Fourier transform for the symmetric group, but it is valid for any finite group.} because the group characters are a complete and orthogonal set of functions, which is embodied in the formulas
\bea
\sum_{\sigma\in S_n}\chi_R(\sigma)\chi_S(\sigma)&=&n!\delta_{RS}\cr\cr
\sum_R\chi_R(\sigma)\chi_R(\rho)&=&{n!\over |[\sigma]|}\delta_{[\sigma]\,[\rho]}
\eea
where $\delta_{[\sigma]\,[\rho]}=1$ if $\rho$ and $\sigma$ belong to the same conjugacy class and it is zero otherwise, and $|[\sigma]|$ denotes the number of elements in conjugacy class $[\sigma]$. Notice that $\chi_R(\sigma)$ has both a Young diagram ($R$) and a conjugacy class ($[\sigma]$) label.

We now use this Fourier transform to move from observables that take values on the conjugacy classes $[\sigma]$ to observables that take values on the irreducible representations $R$ as follows
\bea
\chi_R(Z)&=&{1\over n!}\sum_{\sigma\in S_n}\chi_R(\sigma){\rm Tr}(\sigma Z^{\otimes n})
\eea
The $\chi_R(Z)$ are known as Schur polynomials. The $\chi_R(Z)$ with $R$ having no more than $N$ rows are linearly independent. In particular there are {\it no} trace relations between them. In this basis the trace relations are the statement that
\bea
\chi_R(Z)&=&0
\eea
when $R$ has more than $N$ rows, i.e. the Schur polynomial itself vanishes. In the trace basis the trace relations was a complicated polynomials in the traces. By moving to the Schur basis we have solved all of the trace relations.

Notice that we can write
\bea
\chi_R(Z)&=&{1\over n!}\sum_{\sigma\in S_n}\chi_R(\sigma){\rm Tr}(\sigma Z^{\otimes n})\,\,=\,\,{\rm Tr}(P_R Z^{\otimes n})
\eea
since the projection operator is
\bea
P_R&=&{1\over n!}\sum_{\sigma\in S_n}\chi_R(\sigma)\sigma 
\eea

Now let's compute the two point function of Schur polynomials to illustrate how easy it is to sum all of the Feynman diagrams (and not just the planar diagrams) for certain observables.
\bea
\langle \chi_R(Z) \chi_S(Z^\dagger)\rangle &=&
\langle {\rm Tr}(P_R\,Z^{\otimes n}) {\rm Tr}(P_S\, Z^{\dagger\, \otimes n})\rangle
\,\,=\,\, (P_R)^I_J (P_S)^K_L\langle (Z^{\otimes n})^J_I(Z^{\dagger\,\otimes n})^L_K\rangle\cr\cr
&=&(P_R)^I_J (P_S)^K_L\sum_{\sigma\in S_n}(\sigma)^J_K (\sigma^{-1})^L_I
\,\,=\,\,\sum_{\sigma\in S_n}{\rm Tr}(\sigma P_S\sigma^{-1}P_R)\cr\cr
&=&n!{\rm Tr}(P_SP_R)\,\,=\,\,\delta_{RS}{n!\over {\rm dim}_R}{\rm Tr}(P_R)\cr\cr
&=&\delta_{RS}{n!{\rm Dim}_R\over {\rm dim}_R}\,\,=\,\,\delta_{RS}f_R
\eea
In the final equality $f_R$ is the product of the factors (one for each box) of Young diagram $R$.

Here is a simple example of the above formula
\bea
\langle\chi_{\tiny\yng(4,2,2)}(Z)\chi_{\tiny\yng(4,2,2)}(Z^\dagger)\rangle&=&N(N+1)(N+2)(N+3)(N-1)N(N-2)(N-1)\nonumber
\eea
To obtain this result, $8!=40320$ diagrams were summed.

{\vskip 0.5cm}

\noindent
{\bf Summary of the key ideas:}
{\vskip 0.5cm}
\begin{itemize}
\item[1.] The Schur polynomials $\chi_R(Z)$ labeled by a Young diagram with $n$ boxes ($R\vdash n$) provide the complete set of holomorphic observables constructed using $n$ fields. 
\item[2.] These observables are a complicated linear combination of all possible trace structures. The fact that different trace structures mix is not a problem.
\item[3.] By only using Schur polynomials labeled by Young diagrams $R$ with no more than $N$ rows, we do not need to worry about the trace relations.
\item[4.] The Schur polynomials diagonalize the free field two point function
\bea
\langle\chi_R(Z)\chi_S(Z^\dagger)\rangle&=&\delta_{RS}f_R
\eea
All diagrams (not just the planar ones!) have been summed to obtain this result. 
\end{itemize}

{\vskip 0.5cm}

These results are nice, but they are all for a single matrix. One way to generalize these results to more than one matrix is to consider restricted Schur polynomials. This is the topic of the next Section.

{\vskip 0.5cm}

\noindent
{\bf Guide to further reading:}

The group theory approach to the study of local operators in Yang-Mills theory has its origin in the remarkable pioneering paper [\citen{Corley:2001zk}]. Quickly after this initial paper the work [\citen{Corley:2002mj}], which stressed the role of projection operators in the group theory approach, appeared. This has been an important insight that has inspired progress in many different directions. The original paper [\citen{Corley:2001zk}] deals with gauge group $U(N)$. The extension to gauge group $SU(N)$ was first explored in [\citen{deMelloKoch:2004crq}]. An important follow up was given in [\citen{Brown:2007bb}]. Schur polynomials bases for $SO(N)$ and $Sp(N)$ gauge groups were given in [\citen{Caputa:2013hr,Caputa:2013vla}]. The construction of Schur polynomials for fermionic matrix fields was given in [\citen{Berenstein:2019esh}]. Even more recently, extensions to permutation invariant matrix models were pursued in [\citen{Ramgoolam:2018xty,Ramgoolam:2019ldg}].

As far as applications to physics goes, the first hint that Schur polynomials were important came from the observation that the string exclusion principle \cite{McGreevy:2000cw} naturally suggests that sphere giant gravitons are dual to sub-determinant operators \cite{Balasubramanian:2001nh,Aharony:2002nd,Berenstein:2003ah}. The original paper [\citen{Corley:2001zk}] then quickly suggested a correspondence between any system of giant gravitons \cite{Grisaru:2000zn,Hashimoto:2000zp} and specific Schur polynomials. For an insightful discussion of the physics of Schur polynomials see also [\citen{Berenstein:2004kk}]. There are interesting ways in which the gravitational physics of heavy operators has been recovered from Schur polynomials. First, a Schur polynomial describing an operator with a bare dimension of order $N^2$ can be put into correspondence with one of the ${1\over 2}$-BPS backgrounds of type IIB supergravity constructed in [\citen{Lin:2004nb}]; see [\citen{deMelloKoch:2009pnn}] for a review. Second, the emission/absoption of a point graviton by a giant graviton can be described as a semi-classical process, using a saddle point of the gravity description [\citen{Bissi:2011dc,Caputa:2012yj,Lin:2012ey}]. These papers establish that Schur polynomials are a correct description of giant graviton branes in the dual CFT.

As we have mentioned, when the bare dimension of operators increase, the planar limit becomes a less and less useful description of the physics. Using the failure of factorization as a diagnostic, the paper [\citen{Garner:2014kna}] gave a clear quantitative description of the failure of the planar approximation. 

Finally for a lucid and readable description of Schur-Weyl duality, see [\citen{Ramgoolam:2008yr}].

\section{Restricted Schur Polynomials}\label{RestSchurRescue}

The restricted Schur polynomials were introduced in [\citen{Bhattacharyya:2008rb}]. Fortunately, given the discussion of the previous section, restricted Schur polynomials are a natural generalization of the Schur polynomials and we will get to the punchline quite directly. The starting point is to imagine constructing observables using $n$ $Z$ fields and $m$ $Y$ fields. As a consequence of the fact that all indices must be contracted to produce a gauge invariant operator, the group $S_{n+m}$ naturally appears
\bea
Z^{i_1}_{i_{\sigma(1)}}\cdots Z^{i_n}_{i_{\sigma(n)}}Y^{i_{n+1}}_{i_{\sigma(n+1)}}\cdots Y^{i_{n+m}}_{i_{\sigma(n+m)}}&\equiv&{\rm Tr}(\sigma Z^{\otimes n}Y^{\otimes m})\qquad \sigma\in S_{n+m}
\eea
Again, not every distinct permutation leads to a distinct operators. This is because there is a symmetry which acts by swapping $Z$ fields with each other and/or by swapping $Y$ fields with each other. The subgroup swapping $Z$ fields is $S_n$ and the subgroup swapping $Y$ fields is $S_m$. Thus, the group that swaps both $Z$ fields with themselves and $Y$ fields with themselves is $S_n\times S_m$. Consequently
\bea
\sigma Z^{\otimes n}Y^{\otimes m}\sigma^{-1}=Z^{\otimes n}Y^{\otimes m}\qquad \sigma\in S_n\times S_m
\eea
so that now the group $S_n\times S_m$ also plays a role. This group is a subgroup of $S_{n+m}$. As a consequence of this symmetry we have ($\rho\in S_{n+m}$ and $\sigma\in S_n\times S_m$)
\bea
{\rm Tr}(\rho Z^{\otimes n}Y^{\otimes m})&=&{\rm Tr}(\rho\sigma Z^{\otimes n}Y^{\otimes m}\sigma^{-1})\cr\cr
&=&{\rm Tr}(\sigma^{-1}\rho\sigma Z^{\otimes n}Y^{\otimes m})
\eea
Thus, if two permutations $\rho_1$ and $\rho_2$ are related as
\bea
\rho_1 &=&\sigma^{-1}\rho_2\sigma\qquad\qquad\rho_1,\rho_2\in S_{n+m}\quad \sigma\in S_n\times S_m
\eea
then they define the same gauge invariant operator. We say that $\rho_1$ and $\rho_2$ belong to the same restricted conjugacy class.

Just as characters form a complete set of functions on conjugacy classes, restricted characters provide a complete set of functions on restricted conjugacy classes. To advance this theory further, we need to introduce one more concept: an irreducible representation of a group typically becomes a reducible representation when restricted to a subgroup. To illustrate the point, consider a vector
\bea
\vec{x}&=&(x_1,x_2,x_3,x_4)
\eea
which transforms in a four dimensional irreducible representation of SO(4). Consider restricting SO(4) to the SO(2) subgroup that rotates only the first two components of this vector into each other. The vector $\vec{x}$ can now be decomposed into three components that don't mix under the action of this subgroup as follows
\bea
\vec{x}&=&\vec{v}_1+\vec{v}_2+\vec{v}_3
\eea
where
\bea
\vec{v}_1&=&(x_1,x_2,0,0)\cr\cr
\vec{v}_2&=&(0,0,x_3,0)\cr\cr
\vec{v}_3&=&(0,0,0,x_4)
\eea
We see that the 4 dimensional representation of SO(4) decomposes into the direct sum of a two dimensional representation and two one dimensional representations
\bea
{\bf 4}&\to& {\bf 2}\oplus{\bf 1}\oplus{\bf 1}
\eea 
We must introduce a {\it multiplicity index} to {\it distinguish repeated representations}. Elements of the group that belong to the subgroup take the following block diagonal form (non-zero matrix elements are denoted $\times$ and vanishing matrix elements by $0$)
\begin{center}
\includegraphics[width=0.75\textwidth]{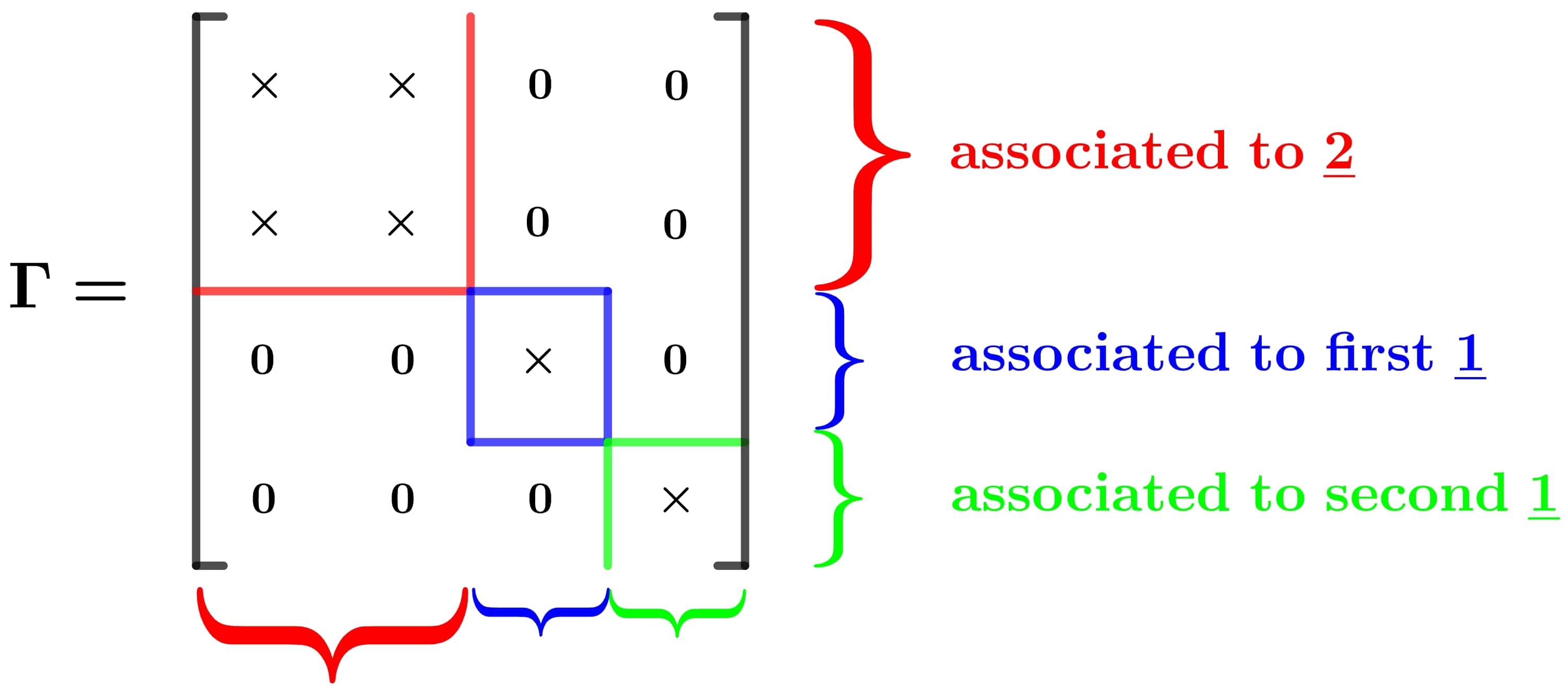}
\end{center}

Elements of the group that don't belong to the subgroup are not block diagonal, but we can still associate row and column indices to representations of the subgroup.
\begin{center}
\includegraphics[width=0.75\textwidth]{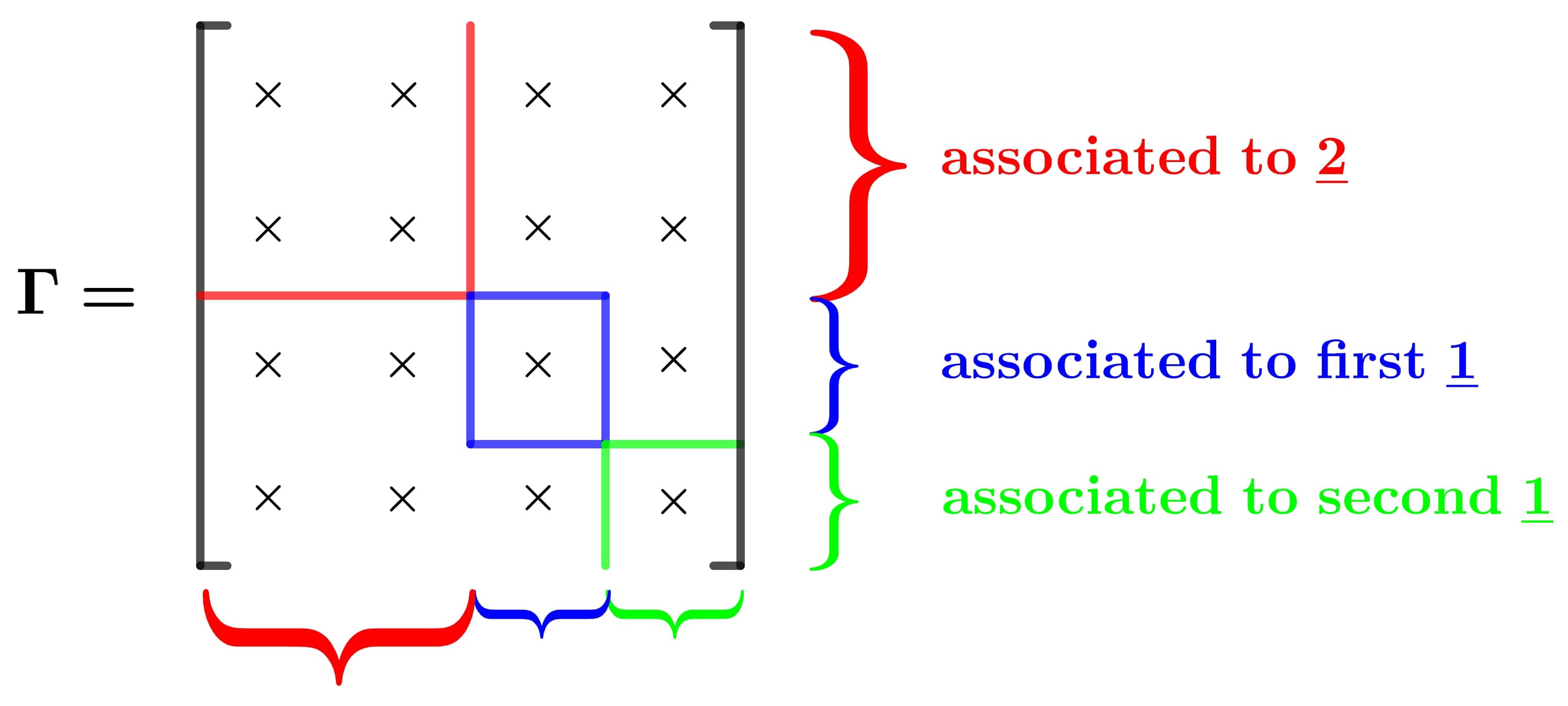}
\end{center}

We are now ready to introduce the idea of a restricted character. To define the restricted character we will use, start with an irreducible representation of $S_{n+m}$ labelled by Young diagram $R\vdash n+m$. Now restrict to the $S_n\times S_m$ subgroup. The irreducible representations of this subgroup are labelled by a pair of Young diagrams $(r,s)$ with $r$ an irreducible representation of $S_n$ ($r\vdash n$) and $s$ an irreducible representation of $S_m$ ($s\vdash m$). By definition, the character is equal to a trace over the group element
\bea
\chi_R(\sigma)&=&{\rm Tr}(\Gamma_R(\sigma))\qquad R\vdash n+m\qquad \sigma\in S_{n+m}
\eea
The restricted character is obtained by restricting the trace to some representation of the subgroup. More than that, we are free to use which copy of the subgroup we use for the row and column indices independently. 

Here is an example to illustrate the idea: consider the representation
\bea
R&=&{\tiny \yng(3,2,1)}\qquad {\rm dim}_R=16
\eea
of $S_6$ and consider the $S_3\times S_3$ subgroup. $R$ contains the following representations of the subgroup
\bea
({\tiny \yng(2,1)},{\tiny \yng(3)}),\qquad({\tiny\yng(2,1)},{\tiny\yng(1,1,1)}),\qquad
({\tiny\yng(3)},{\tiny\yng(2,1)}),\qquad({\tiny\yng(1,1,1)},{\tiny\yng(2,1)})
\eea
as well as two copies of
\bea
({\tiny\yng(2,1)},{\tiny\yng(2,1)})
\eea
Choose the basis of $R$ so that the first two blocks of $\Gamma_R(\sigma)$ are these two $({\tiny\yng(2,1)},{\tiny\yng(2,1)})$ representations. The rows and columns of the matrix $\Gamma_R(\sigma)$ can be labeled as follows
\begin{center}
\includegraphics[width=0.75\textwidth]{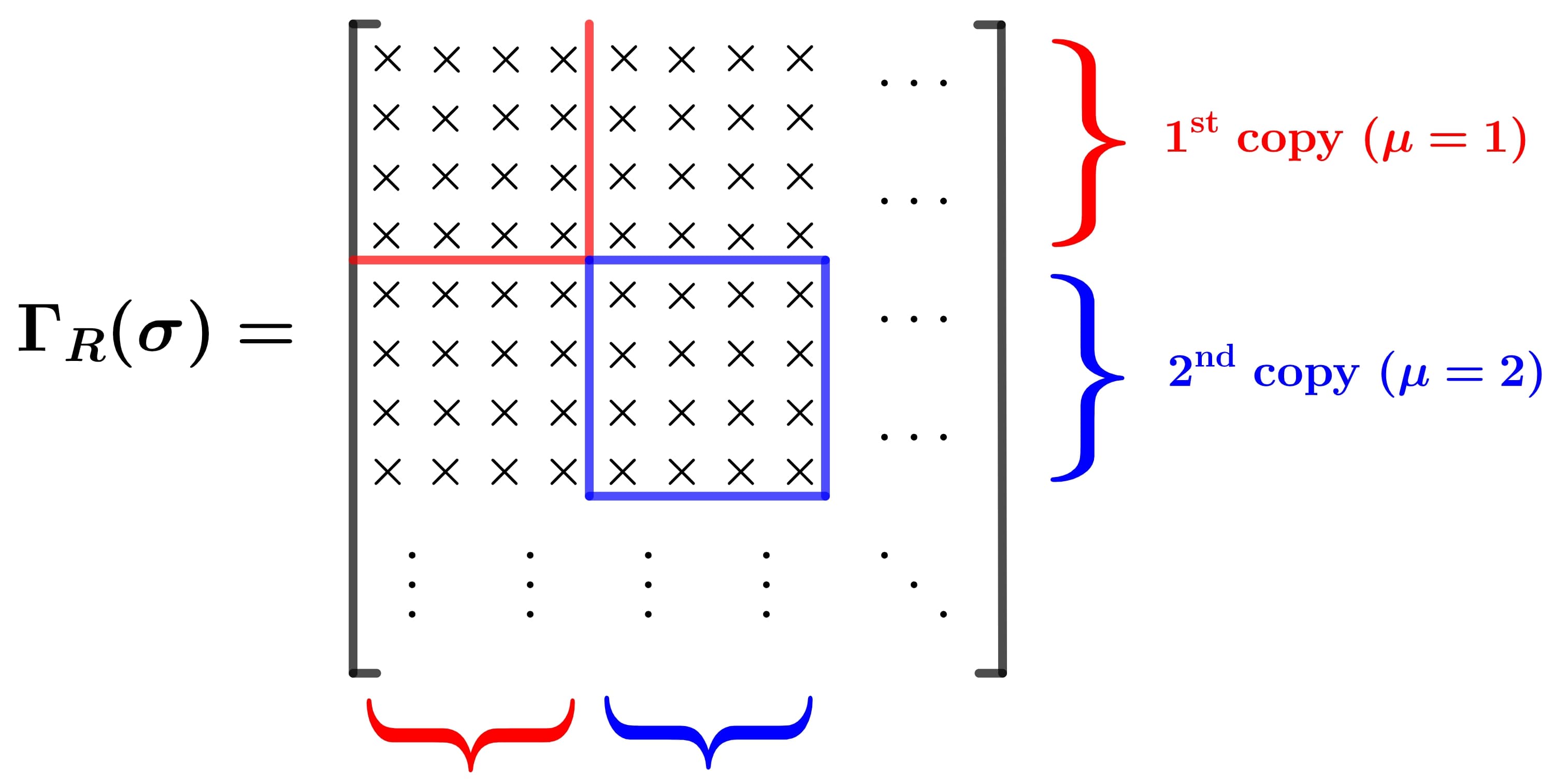}
\end{center}

\noindent
The restricted character $\chi_{{\tiny\yng(3,2,1)},({\tiny\yng(2,1)},{\tiny\yng(2,1)})11}$ is obtained by summing the elements circled in green below
\begin{center}
\includegraphics[width=0.75\textwidth]{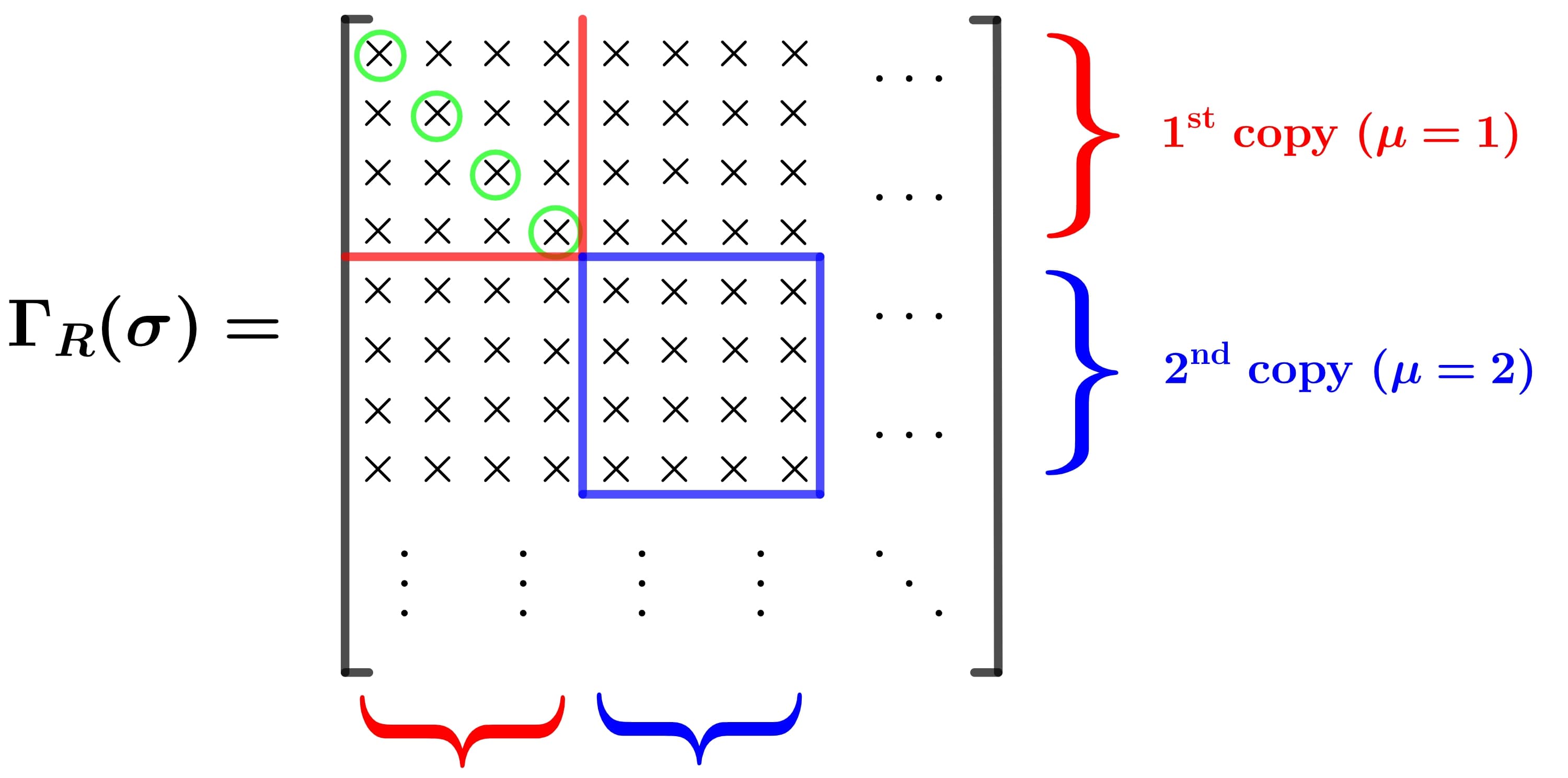}
\end{center}

\noindent
The restricted character $\chi_{{\tiny\yng(3,2,1)},({\tiny\yng(2,1)},{\tiny\yng(2,1)})22}$ is obtained by summing the elements circled in green below
\begin{center}
\includegraphics[width=0.75\textwidth]{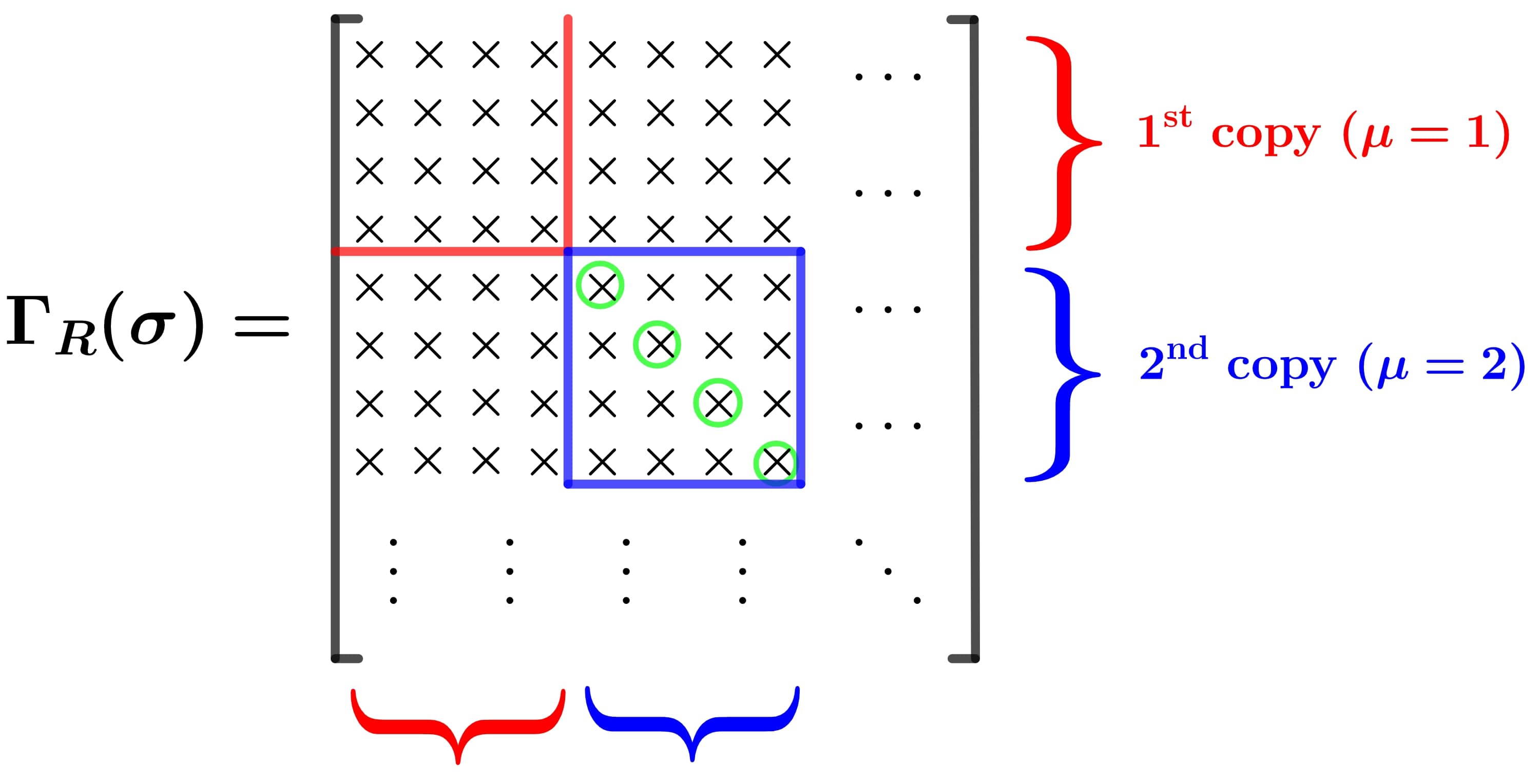}
\end{center}

\noindent
The restricted character $\chi_{{\tiny\yng(3,2,1)},({\tiny\yng(2,1)},{\tiny\yng(2,1)})12}$ is obtained by summing the elements circled in green below
\begin{center}
\includegraphics[width=0.75\textwidth]{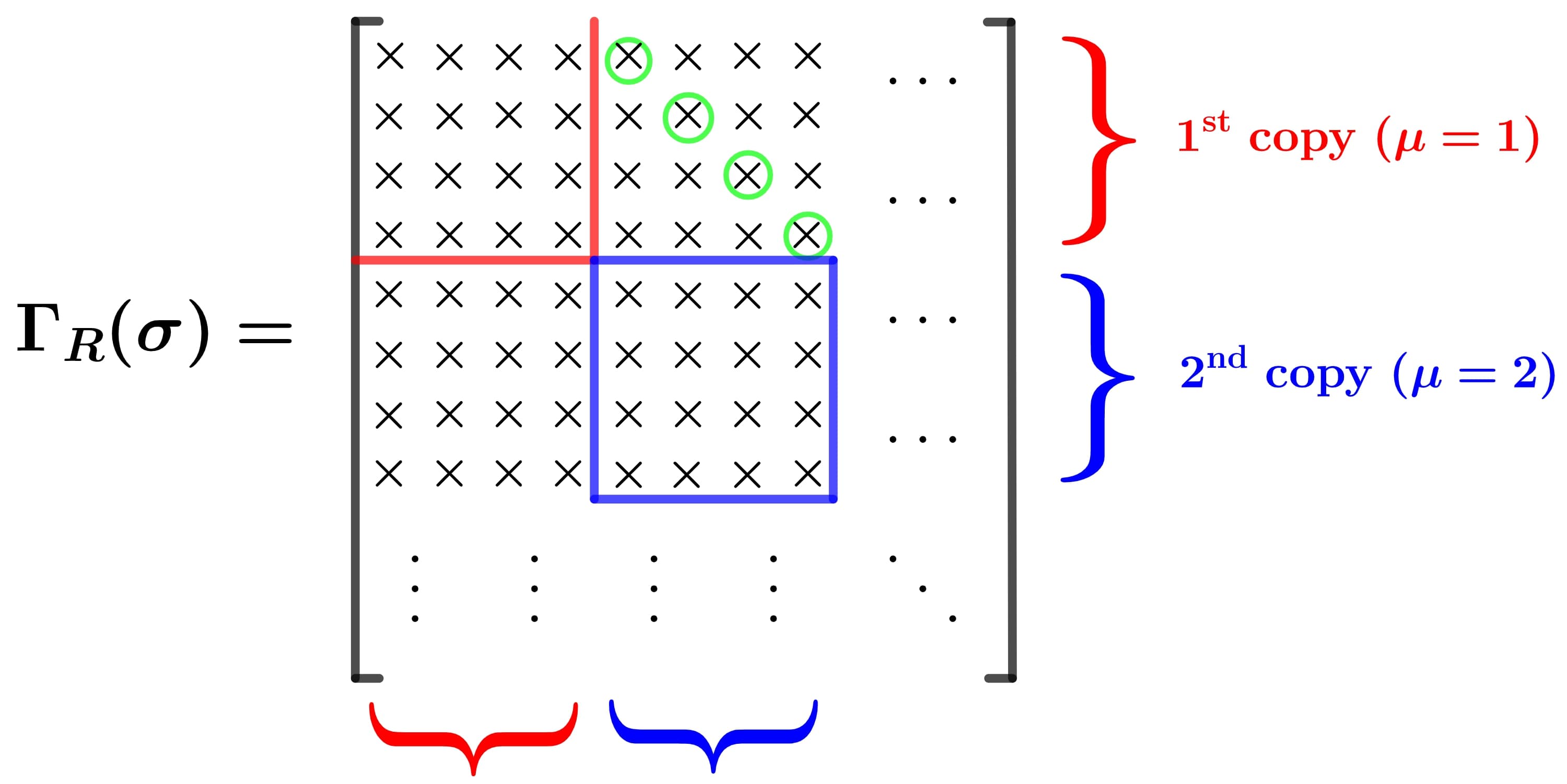}
\end{center}

\noindent
The restricted character $\chi_{{\tiny\yng(3,2,1)},({\tiny\yng(2,1)},{\tiny\yng(2,1)})21}$ is obtained by summing the elements circled in green below
\begin{center}
\includegraphics[width=0.75\textwidth]{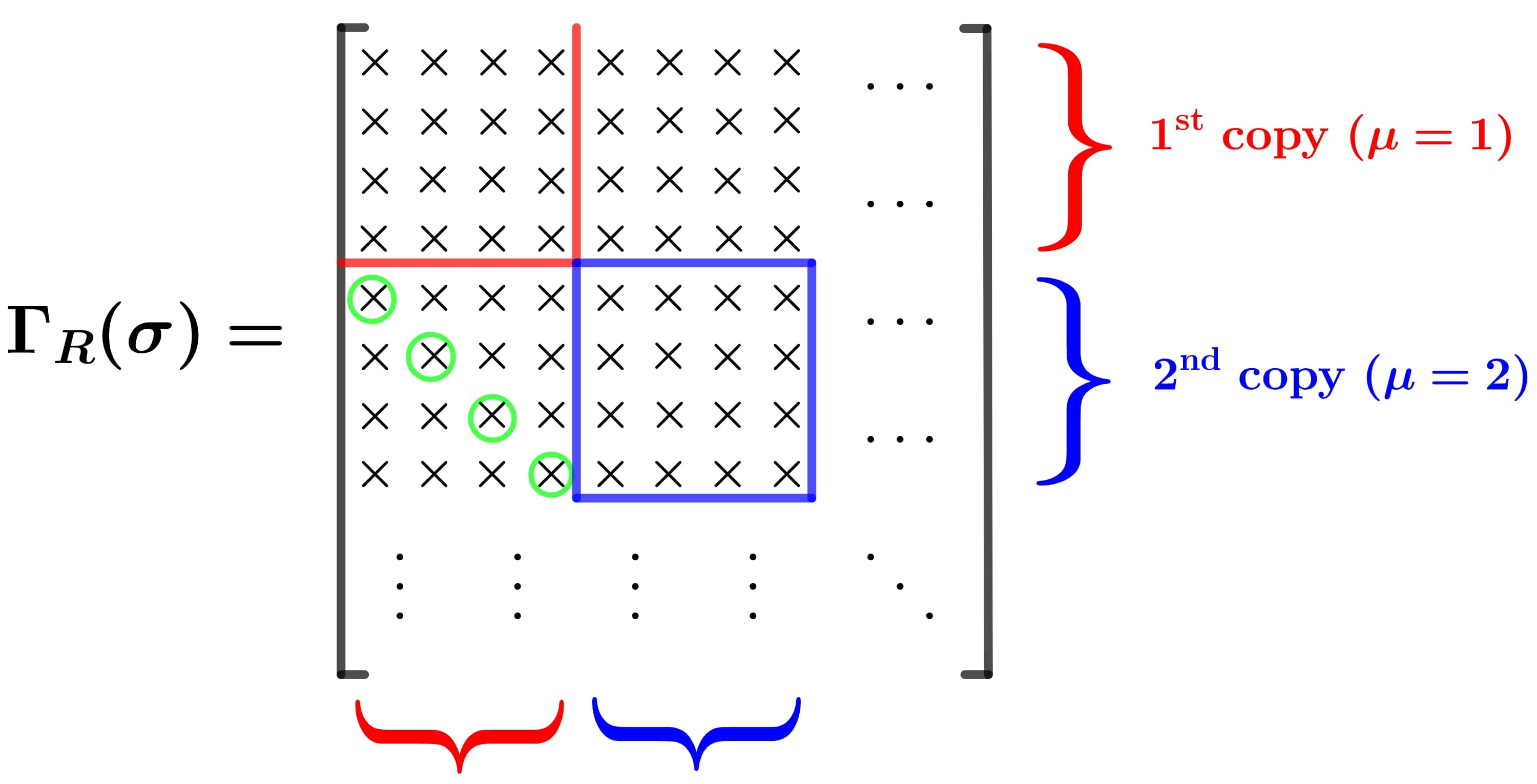}
\end{center}

Restricted characters obey some important orthogonality relations which generalize those of ordinary characters
\bea
\sum_{\rho\in S_{n+m}}\chi_{R,(r,s)\alpha\beta}(\rho^{-1})\chi_{T,(t,u)\gamma\delta}(\rho)&=&(n+m)!\delta_{RT}\delta_{rt}\delta_{su}\delta_{\gamma\beta}\delta_{\alpha\delta}
\eea
\bea
\sum_{R,(r,s)\alpha\beta}\chi_{R,(r,s)\alpha\beta}(\tau)\chi_{R,(r,s)\alpha\beta}(\sigma)&=&(n+n)!{{\rm dim}_r {\rm dim}_s\over {\rm dim}_R}\delta_{[\sigma]_r[\tau]_r}
\eea
where $\delta_{[\sigma]_r[\tau]_r}$ is 1 if $[\sigma]_r$ and $[\tau]_r$ belong to the same conjugacy class and it is zero otherwise. These orthogonality relations prove that the restricted characters are a complete set of functions on the restricted conjugacy classes and they ensure that we can again use them to perform a Fourier transform. This leads to the definition of the restricted Schur polynomials
\bea
\chi_{R,(r,s)\alpha\beta}(Z,Y)&=&{1\over n!m!}\sum_{\sigma\in S_{n+m}}\chi_{R,(r,s)\alpha\beta}(\sigma){\rm Tr}(\sigma Z^{\otimes n}Y^{\otimes m})
\eea

{\vskip 1.0cm}

\noindent
{\bf Summary of the key ideas:}
{\vskip 0.5cm}
\begin{itemize}
\item[1.] The restricted Schur polynomials $\chi_{R,(r,s)\alpha\beta}(Z,Y)$ labeled by a Young diagram with $n+m$ boxes ($R\vdash n+m$), a pair of Young diagrams one with $n$ boxes and one with $m$ boxes ($r\vdash n$, $s\vdash m$) as well as a pair of multiplicity labels $\alpha,\beta$ provide a complete set of holomorphic observables constructed using $n$ $Z$ fields and $m$ $Y$ fields.
\item[2.] These observables are a complicated linear combination of all possible trace structures. The fact that different trace structures mix is not a problem.
\item[3.] By using restricted Schur polynomials labelled by Young diagrams $R,r,s$ that have no more than $N$ rows, we don't need to worry about the trace relations.
\item[4.] The object
\bea 
P_{R,(r,s)\alpha\beta}&=&{1\over n!m!}\sum_{\sigma\in S_{n+m}}\chi_{R,(r,s)\alpha\beta}(\sigma)\sigma
\eea
is a projection operator (not correctly normalized!) acting in $V_N^{\otimes n+m}$. Using it we find that the restricted Schur polynomials diagonalize the free field two point function
\bea
\langle\chi_{R,(r,s)\alpha\beta}(Z,Y)\chi_{T,(t,u)\gamma\delta}(Z^\dagger,Y^\dagger)\rangle&=&\delta_{RT}\delta_{rt}\delta_{su}\delta_{\beta\gamma}\delta_{\delta,\alpha}{(n+m)!\over {\rm dim}_R}{{\rm dim}_r\over n!}{{\rm dim}_s\over m!}f_R \nonumber \\
\eea
All diagrams (not just the planar ones!) have been summed to obtain this result. 
\end{itemize}

{\vskip 0.5cm}

\noindent
{\bf Guide to further reading:}

The generalization of the group representation theory approach for single matrix models to that for multi-matrix models is highly non-trivial. The key series of paper which made this step possible are the works [\citen{Kimura:2007wy,Brown:2007xh,Brown:2008ij}]. The paper
[\citen{Brown:2007xh}] in particular stressed the notion of a Fourier transform and properly appreciated the role of the completeness of characters. Restricted Schur polynomials, as defined in this review, were first given in [\citen{Bhattacharyya:2008rb}]. The orthogonality relations for restricted characters were first proved in [\citen{Bhattacharyya:2008xy}]. Restricted Schur polynomials involving fermions were defined in [\citen{deMelloKoch:2012sie}]. It is possible to generalize restricted Schur polynomials beyond gauge group $U(N)$. For the generalization to gauge groups $SO(N)$ and $Sp(N)$ see [\citen{Kemp:2014apa,Kemp:2014eca}].

The definition of the restricted Schur polynomials given in [\citen{Bhattacharyya:2008rb}] was directly motivated by studies of open strings attached to giant gravitons given in [\citen{Balasubramanian:2004nb}]. This paper properly appreciated the role of using Young diagram labels, associated to subgroups of the symmetric group, for the open string excitations but did not define restricted characters or their orthogonality. Motivated by [\citen{Balasubramanian:2004nb}] restricted Schur polynomials for semi-classical open string excitations\footnote{By this we simply mean that each open string is described by a word using the letters of the theory, and this word is inserted into the Schur polynomial.} were defined in [\citen{deMelloKoch:2007rqf,deMelloKoch:2007nbd,Bekker:2007ea}]. The paper [\citen{deMelloKoch:2007rqf}] introduced the notion of a restricted character. To properly arrive at the definition given in [\citen{Bhattacharyya:2008rb}] it was necessary to combine the insights of [\citen{Balasubramanian:2004nb}] with those of [\citen{Kimura:2007wy,Brown:2007xh}].

The interpretation of restricted Schur polynomials as the dual gauge theory description of giant gravitons together with open string excitations has been established\cite{Balasubramanian:2002sa,Berenstein:2005vf,Berenstein:2005fa,Berenstein:2006qk,deMelloKoch:2007rqf,deMelloKoch:2007nbd,Bekker:2007ea} in a number of detailed studies. The semi-classical open string worldsheet dynamics is visible from the gauge theory description. In particular, the boundary conditions expected of open strings attached to giant gravitons in the gravitational description have been established using the gauge theory operators.  An interesting feature of the gauge theory description is the fact that fields can be exchanged between the giant graviton branes and the open string. Since these fields carry ${\cal R}$-charge, this corresponds to an exchange of momentum between the open string and the giant graviton i.e. to a force exerted between them. 

Finally, there have been numerous other applications to problems in gauge/gravity duality, impressive progress in counting arguments, as well interesting and important mathematical progress [\citen{Kimura:2008ac, Kimura:2009ur,Pasukonis:2010rv,Kimura:2012hp,Pasukonis:2013ts,Kimura:2013fqa,BenGeloun:2013lim,Mattioli:2014yva,Mattioli:2016eyp,Ramgoolam:2016ciq,Kimura:2011df,Kimura:2016bzo,Lin:2017dnz,BenGeloun:2017vwn,Lin:2017vfn,Lewis-Brown:2018dje,Kemp:2019log,Lewis-Brown:2020nmg,BenGeloun:2020yau,Lin:2021qso,Barnes:2022qli,Kemp:2023kma}] which we have not reviewed here. The reader is encouraged to explore these avenues herself.

\section{From restricted Schur polynomials to Gauss graph operators}\label{ExcitedGiant}

We will now turn to the question of a physical interpretation for the restricted Schur operators. As an example of what we want, consider the analogous question of the physical interpretation of single trace operators. We know that single trace operators with their bare dimension held fixed as we take $N\to\infty$ correspond to states of a closed string. The space of single trace operators reproduces the Fock space of a single closed string.

For Schur polynomials [\citen{Corley:2001zk}] have suggested the following interpretation for $\chi_R$: If the Young diagram has $p=O(1)$ columns and $O(N)$ rows making it tall and skinny
\begin{center}
\includegraphics[width=0.5\textwidth]{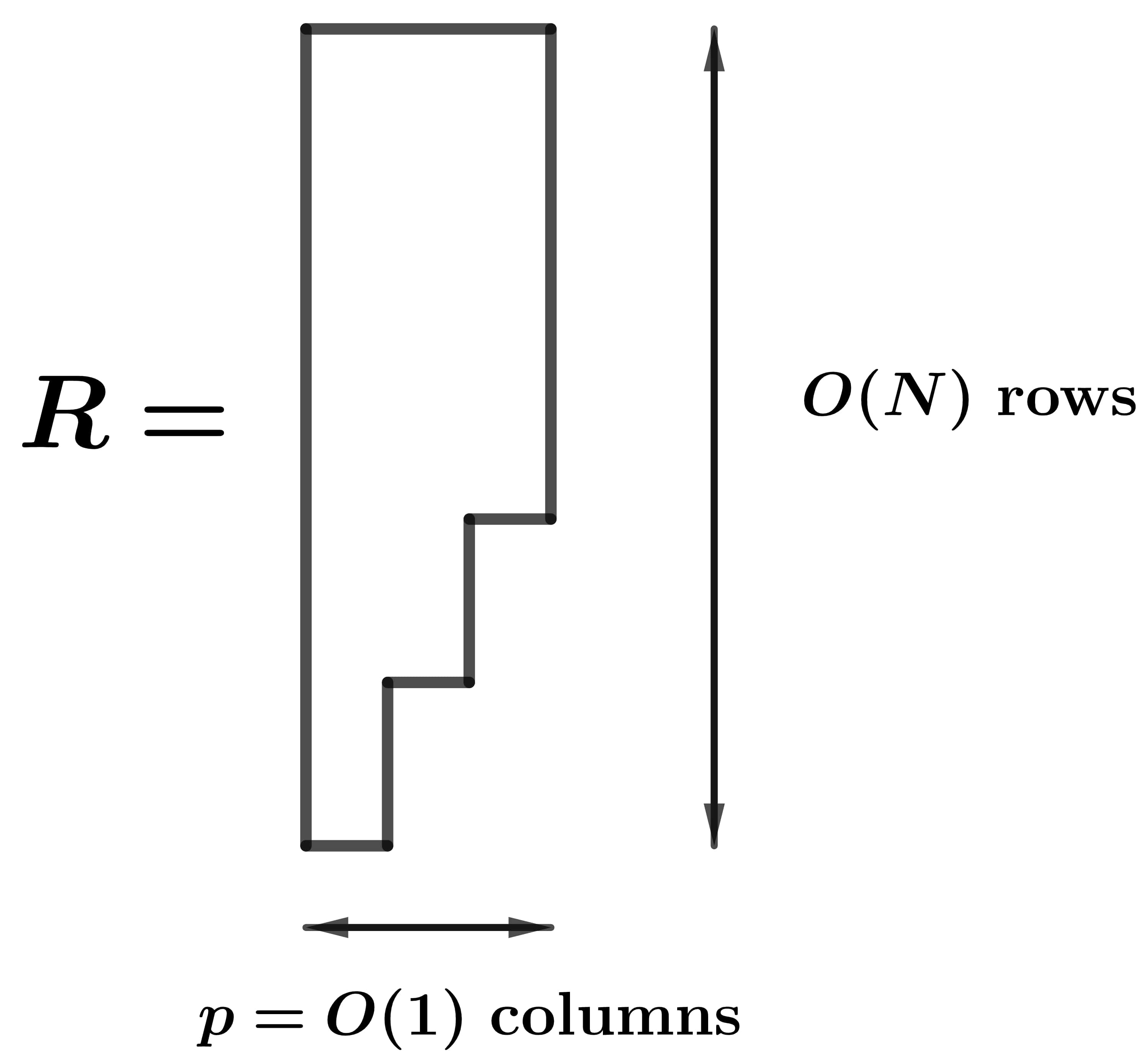}
\end{center}

\noindent
then $\chi_R$ is dual to a system of $p$ giant graviton branes - giant gravitons that wrap an S$^3$ contained in S$^5$ of the AdS$_5\times$S$^5$ spacetime. On the other hand, if $R$ is a Young diagram with $p=O(1)$ rows and $O(N)$ columns making it short and broad 
\begin{center}
\includegraphics[width=0.5\textwidth]{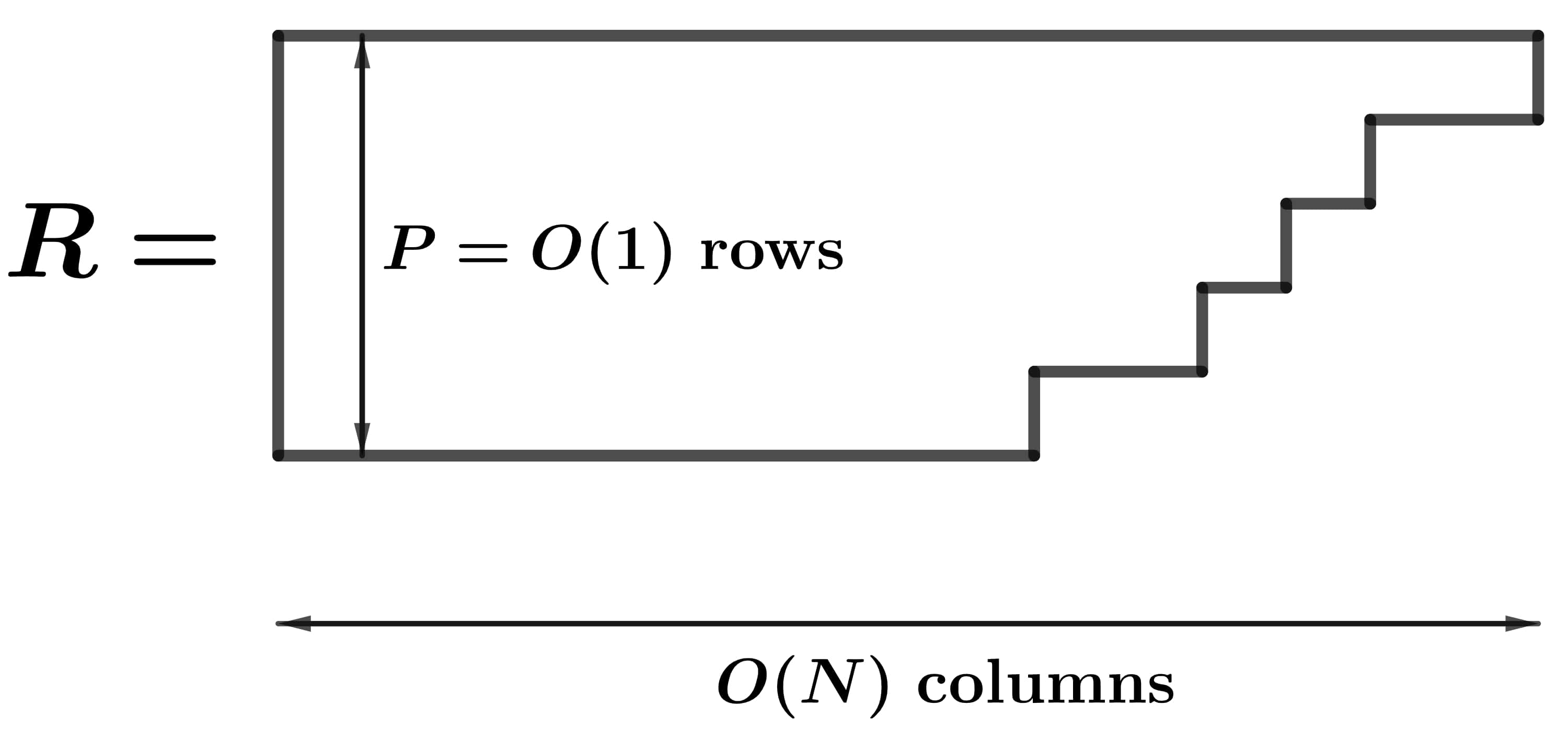}
\end{center}

\noindent
then $\chi_R$ is dual to a system of $p$ dual giant graviton branes - giant gravitons that wrap an S$^3$ contained in AdS$_5$ of the AdS$_5\times$S$^5$ spacetime. There is by now compelling evidence for this identification:
\begin{itemize}
\item This reproduces the correct cut off on angular momentum for giant graviton branes\cite{McGreevy:2000cw}. 
\item This agrees with the LLM description\cite{Lin:2004nb} of this ${1\over 2}$-BPS state. Both can be related to free fermion descriptions\cite{Corley:2001zk,Berenstein:2004kk}.
\item Using semi-classical approximations, the absorption/emission amplitudes for closed strings is reproduced\cite{Bissi:2011dc,Caputa:2012yj,Lin:2012ey}.
\end{itemize}

In the planar limit the dilatation operator was identified with the Hamiltonian of an integrable spin system\cite{Beisert:2010jr}. Perhaps it is interesting to consider the action of the dilatation operator on Schur polynomials?

Since the Schur polynomials all correspond to ${1\over 2}$-BPS states, they do not develop anomalous dimensions. Thus the action of the dilatation operator is too simple to be interesting.

Restricted Schur polynomials are dual to both BPS and non-BPS operators, so that the action of the dilatation operator on the restricted Schur polynomials is a rich problem with a lot of structure. In what follows we will study the action of the one loop dilatation operator on restricted Schur polynomials.

Consider a Schur polynomial labelled by the following Young diagram $R$
\begin{center}
\includegraphics[width=0.65\textwidth]{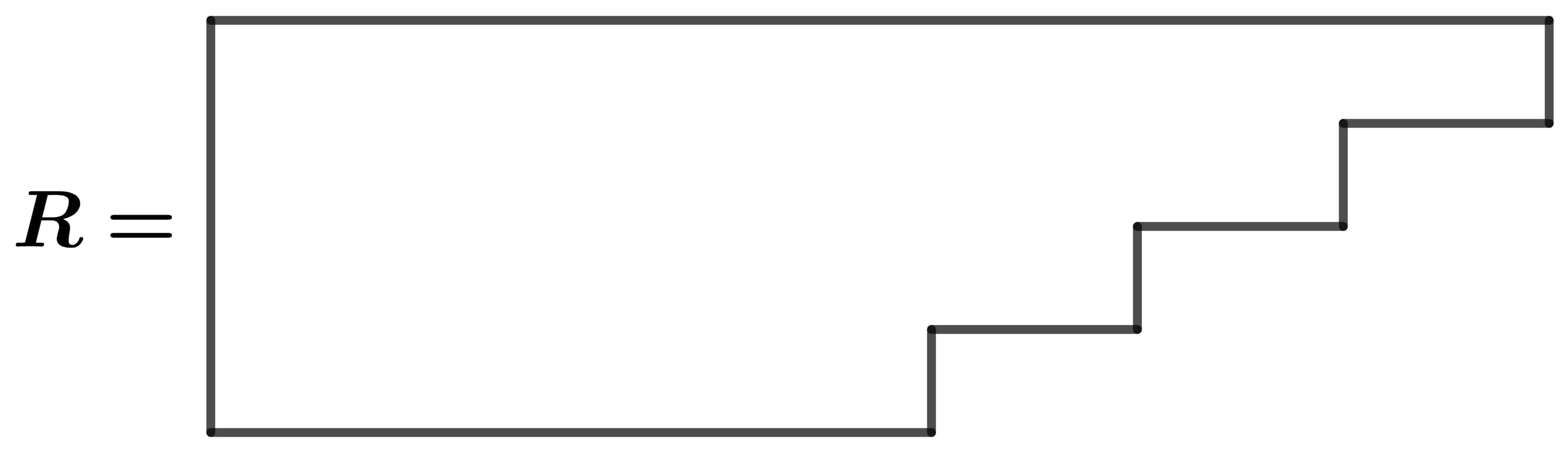}
\end{center}

\noindent
Each row corresponds to a dual giant graviton. The number of boxes in each row sets the momentum of the given giant graviton, and this in turn sets the size of the giant\cite{McGreevy:2000cw}. Long rows correspond to giants that have a large angular momentum and consequently, have a large size. A cartoon of the operator labelled by the above Young diagram $R$ is
\begin{center}
\includegraphics[width=0.5\textwidth]{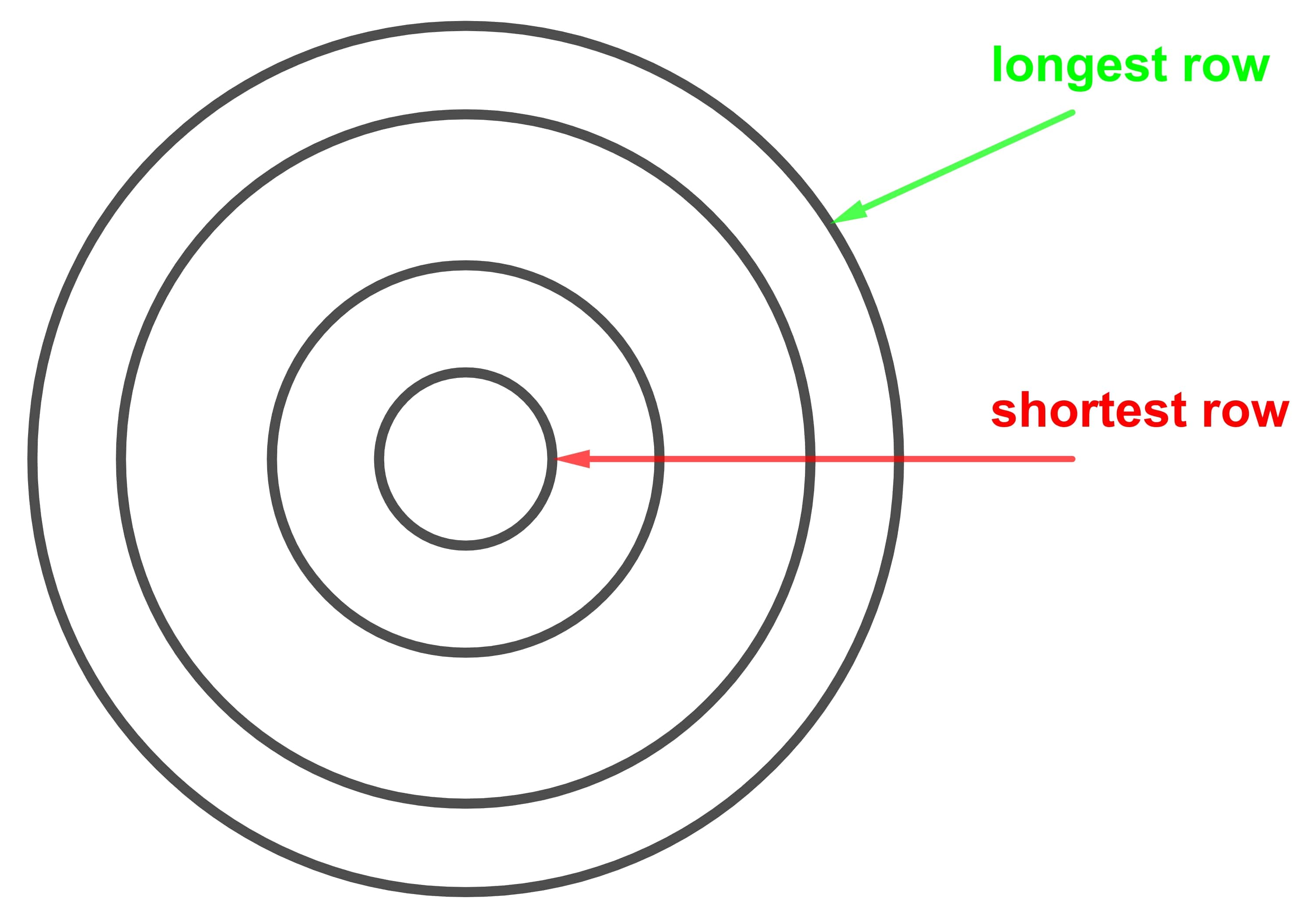}
\end{center}

\noindent
Now let's consider exciting this giant graviton system. In string theory this would correspond to attaching some open strings to the system of giant gravitons.
\begin{center}
\includegraphics[width=0.25\textwidth]{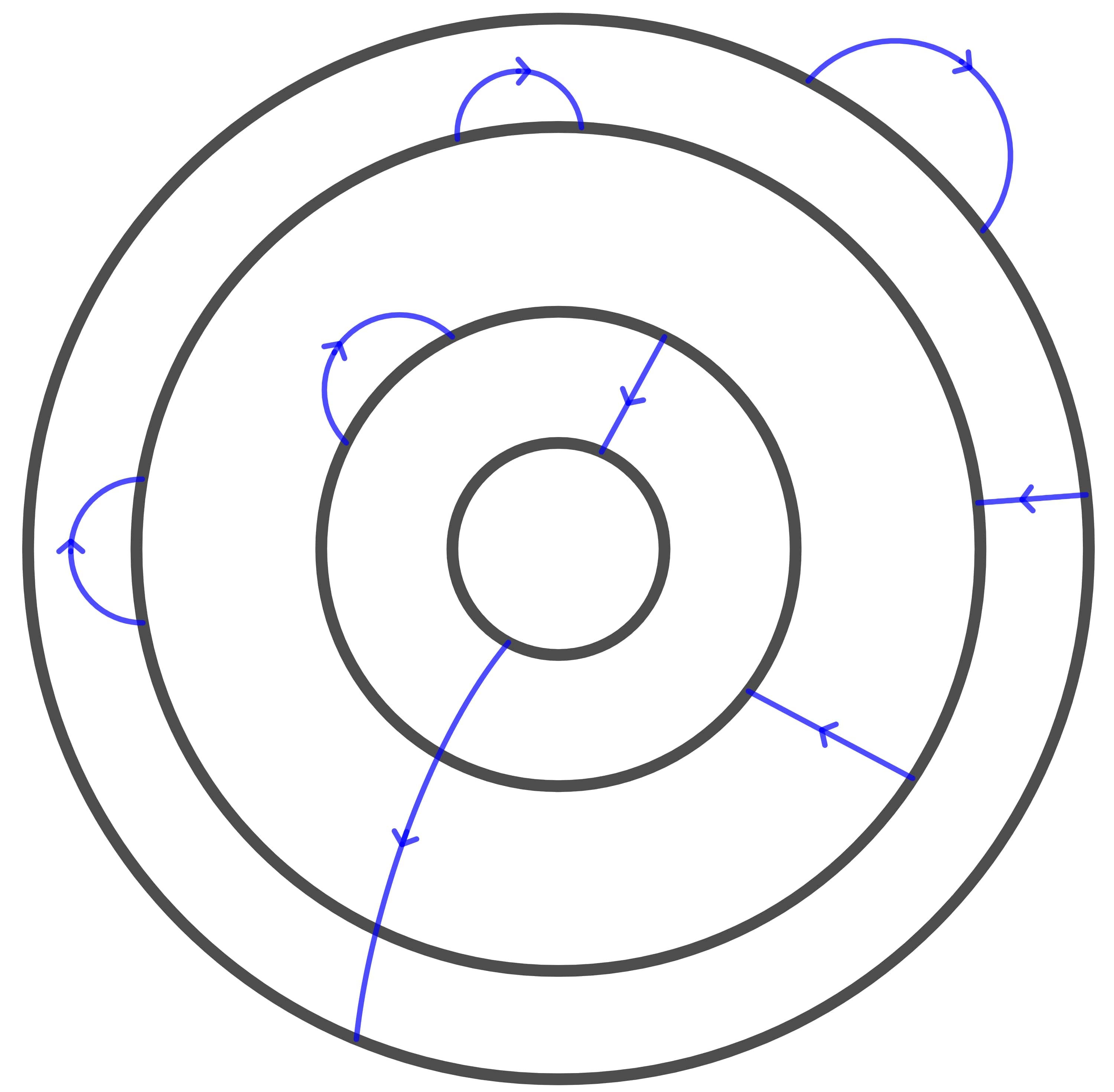}
\end{center}

\noindent
There is an interesting constraint on the state space of the allowed open string excitations. The world volume $\Sigma$ of the giant graviton is compact - topologically it is an S$^3$. It is described by an open string theory which at low energies is well approximated by an ordinary gauge theory. Gauge theories on compact spaces must always have a vanishing charge since by the Gauss Law we have
\bea
Q&=&\int_{\Sigma} \rho dV\,\,=\,\,\int_{\Sigma} \vec{\nabla}\cdot\vec{E} dV\,\,-\,\,\oint_{\partial\Sigma}\vec{E}\cdot d\vec{S}\,\,=\,\,0 
\eea
where in the last line we used the fact that the worldvolume of the giant has no boundary $\partial\Sigma=\partial {\rm S}^3=0$. This constrains the allowed open string configurations because the open string end points are charged
\begin{center}
\includegraphics[width=0.35\textwidth]{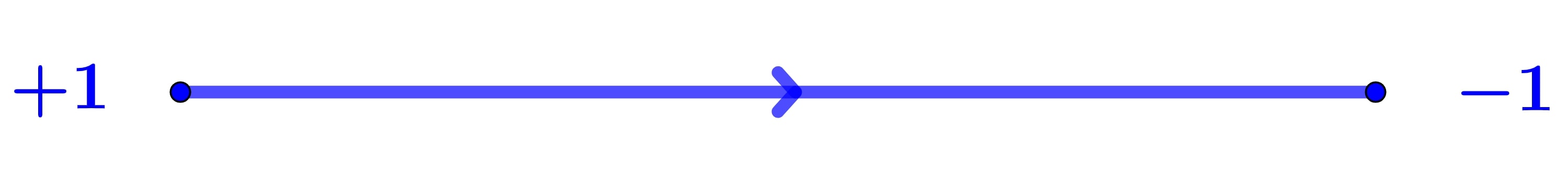}
\end{center}

\noindent
To get vanishing total charge, the number of open strings ending on a giant must equal the number of open strings starting from the giant. Thus, the states
\begin{center}
\includegraphics[width=0.55\textwidth]{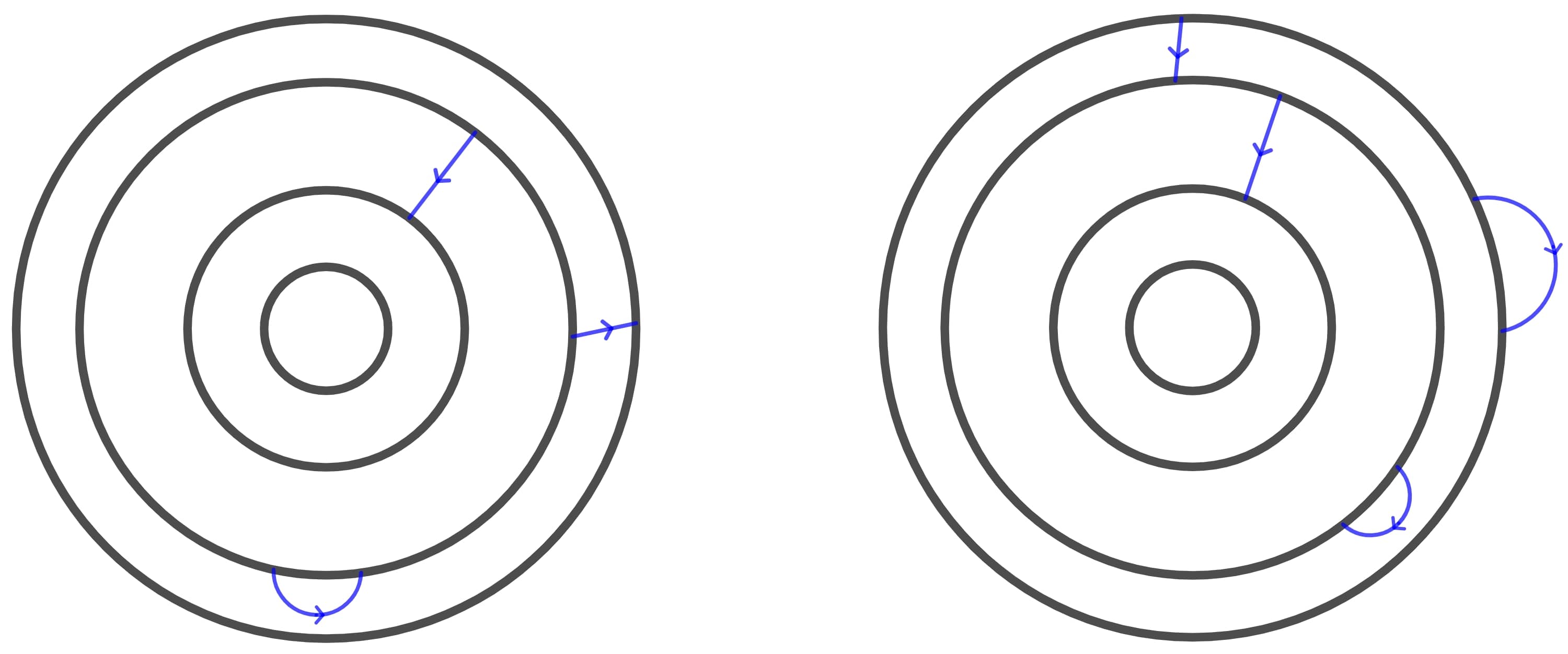}
\end{center}

\noindent
are forbidden, but the states
\begin{center}
\includegraphics[width=0.55\textwidth]{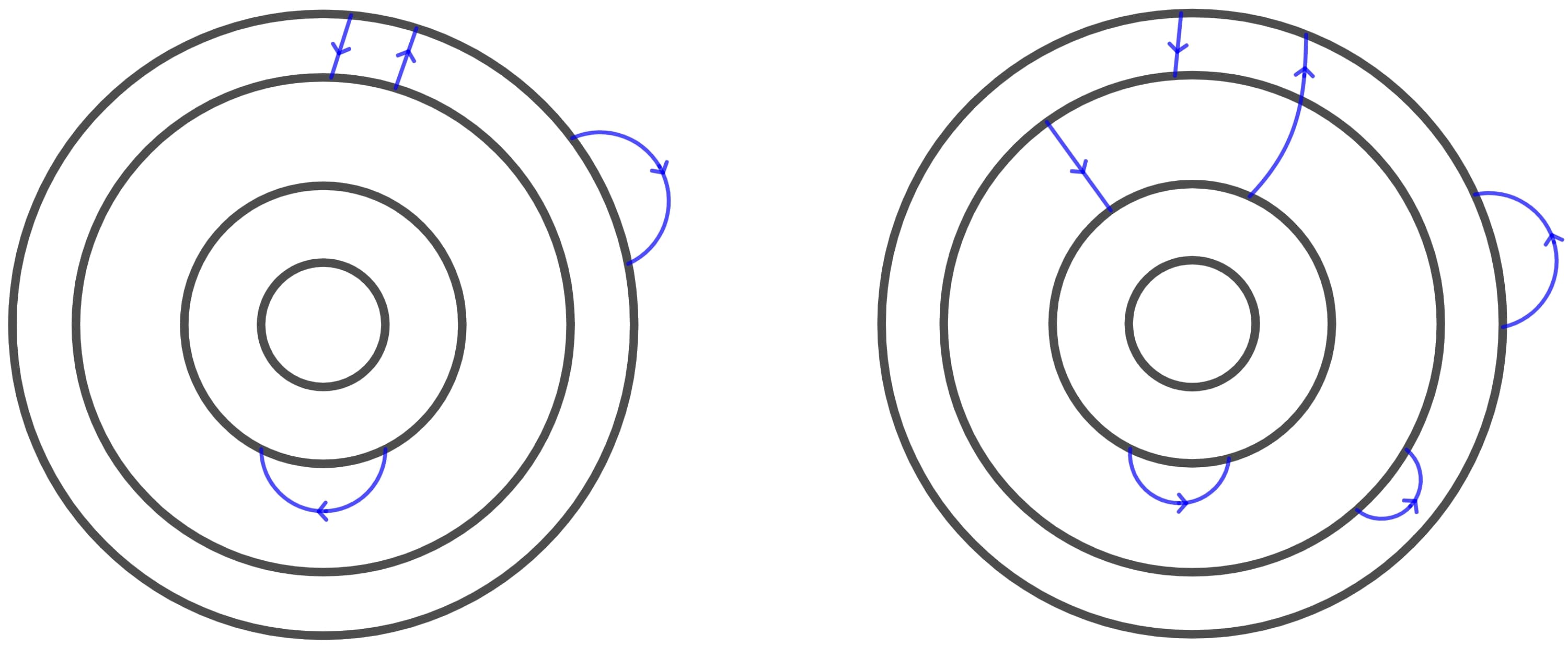}
\end{center}

\noindent
are allowed. A more convenient way to draw these states is to collapse each giant graviton brane into a node 
\begin{center}
\includegraphics[width=0.55\textwidth]{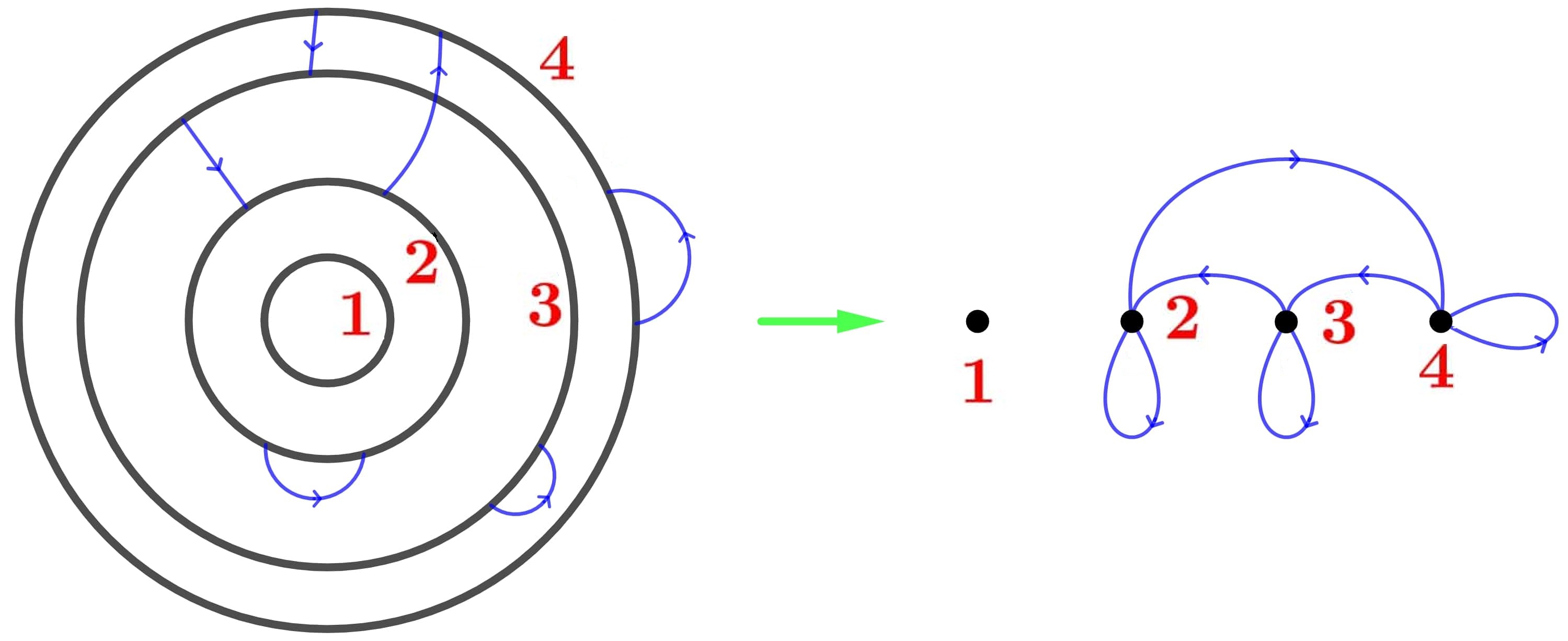}
\end{center}

\noindent
The red numbering is included to make it transparent how the diagram was simplified. We don't usually number the nodes. The second graph shown above is an example of a Gauss graph. The eigenstates of the one loop dilatation operator are labelled by Gauss graphs and this provides the physical interpretation of the restricted Schur polynomials. We will now explain this connection.
 
Consider restricted Schur polynomials constructed using $n$ $Z$ fields and $m$ $Y$ fields, with $n\gg m$. These will be a small deformation of a ${1\over 2}$-BPS operator so that we can still use the physical interpretation of the Schur polynomials.
\begin{center}
\includegraphics[width=0.55\textwidth]{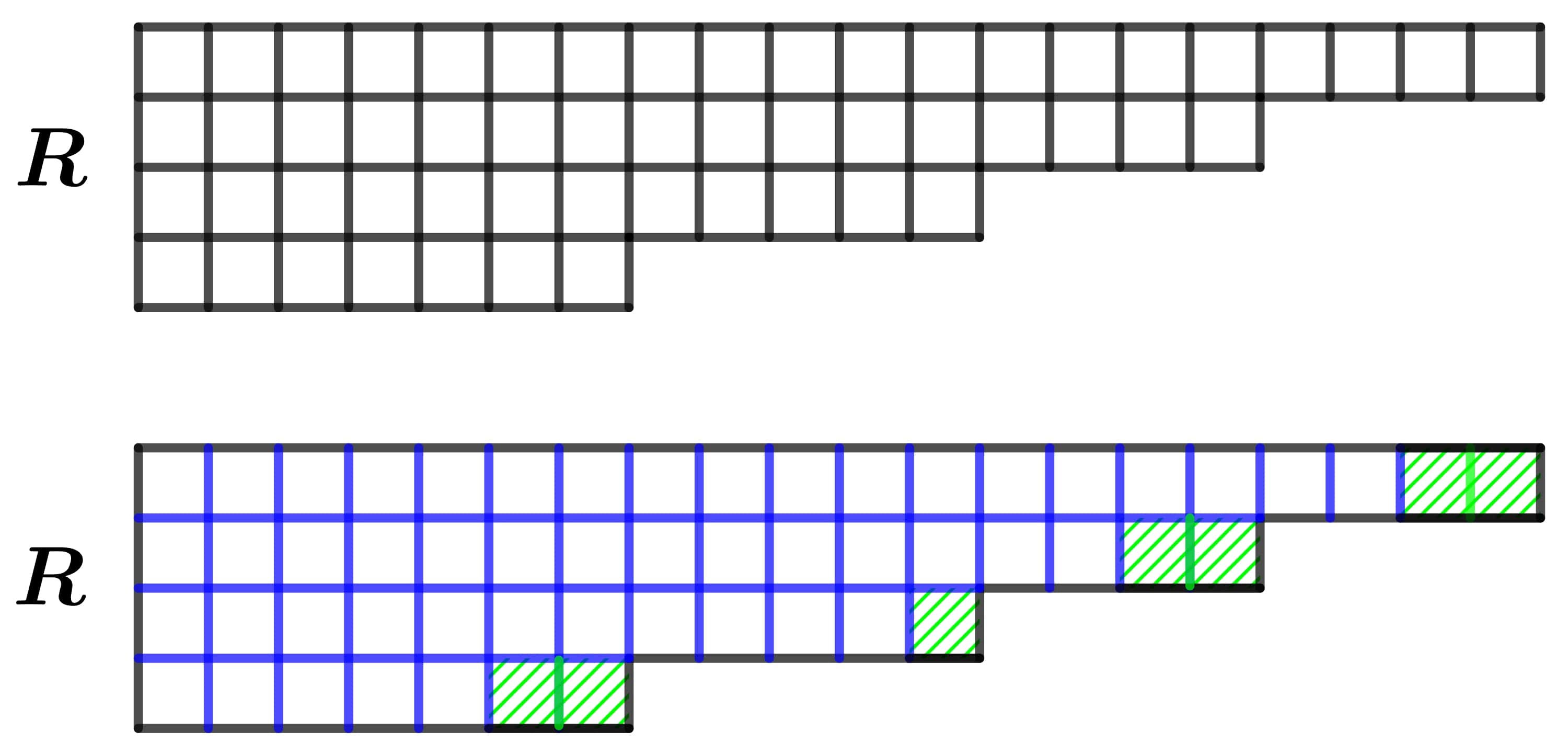}
\end{center}

The restricted Schur polynomial is given by $\chi_{R,(r,s)\alpha\beta}(Z,Y)$. The Young diagram shown in blue above is the Young diagram $r$ that organizes the $Z$ fields. We read the system of giant gravitons from $r$. There is a giant graviton for every row in $r$. The green boxes correspond to $Y$ fields. Each green box will be an edge in the Gauss graph. We assemble the green boxes to produce the Young diagram $s$ label of the restricted Schur polynomial. As usual, $\alpha$ and $\beta$ are multiplicity labels. We will now argue that the labels $s,\alpha,\beta$ can be traded for the Gauss graph. To do this we need a better understanding of the multiplicity labels $\alpha$ and $\beta$. A Young diagram $r$ with $p$ rows describes a system of $p$ giant graviton branes. Consequenlty we expect that the world volume description of this system involves a $U(p)$ gauge theory. To describe this $U(p)$ theory we introduce some vectors
\bea
\vec{e}_1&=&\left[\begin{array}{c} 1\\ 0\\ 0\\ \vdots \\ 0\end{array}\right]\quad
\vec{e}_2\,\,=\,\,\left[\begin{array}{c} 0\\ 1\\ 0\\ \vdots \\ 0\end{array}\right]\qquad\cdots
\qquad \vec{e}_p\,\,=\,\,\left[\begin{array}{c} 0\\ 0\\ 0\\ \vdots \\ 1\end{array}\right]
\eea
The generators of $U(p)$ are linear combinations of the $p^2$ matrices
\bea
(E_{ij})_{ab}&=&\delta_{ia}\delta_{jb}
\eea
In particular, $E_{11}$, $E_{22}$, $\cdots$ $E_{pp}$ are generators of the $U(1)^p$ subgroup of $U(p)$. The state in representation $s$ constructed for a restricted Schur polynomial with $m_1$ green boxes in the first row, $m_2$ boxes in the second row,..., $m_p$ green boxes in the $p^{\rm th}$ row is a state with $U(1)^p$ charges given by 
\bea
\vec{m}&=&(m_1,m_2,\cdots,m_p)
\eea
The multiplicity labels $\alpha$ and $\beta$ each run over the states of the $U(p)$ representation $s$ that have $U(1)^p$ charge $\vec{m}$. The number of such states has been studied in mathematics and it is known as the Kostka number and it is denoted $K_{s,\vec{m}}$.

It is helpful to consider an example. Consider the restricted Schur polynomial with 
\begin{center}
\includegraphics[width=0.55\textwidth]{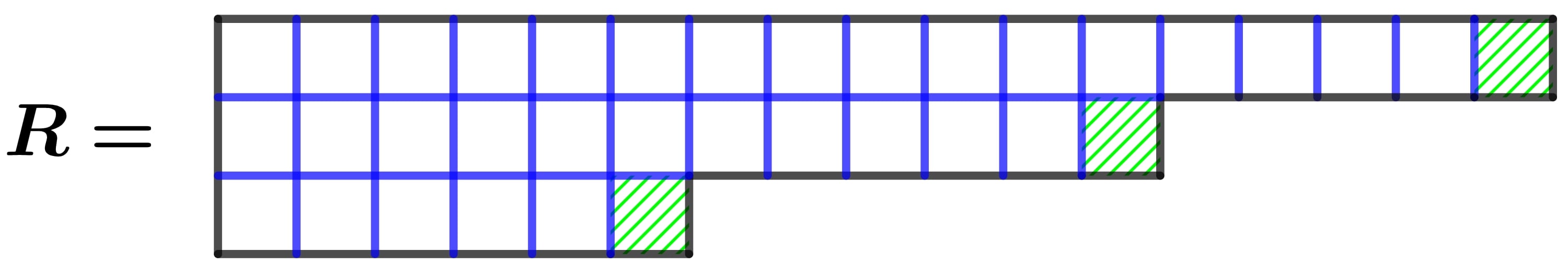}
\end{center}

\noindent
where $r$ has three rows so we are studying a problem with $U(3)$ symmetry. The space spanned by the two labels $s,\alpha$ is spanned by the six states
\bea
&& \vec{e}_1\otimes\vec{e}_2\otimes\vec{e}_3\quad
\vec{e}_1\otimes\vec{e}_3\otimes\vec{e}_2\quad
\vec{e}_2\otimes\vec{e}_1\otimes\vec{e}_3 \cr
&& \vec{e}_2\otimes\vec{e}_3\otimes\vec{e}_1\quad
\vec{e}_3\otimes\vec{e}_2\otimes\vec{e}_1\quad
\vec{e}_3\otimes\vec{e}_1\otimes\vec{e}_2
\eea
Now,
\bea
{\tiny\yng(1)}\otimes{\tiny\yng(1)}\otimes{\tiny\yng(1)}&=&{\tiny\yng(1)}\otimes({\tiny\yng(2)}\oplus{\tiny\yng(1,1)})\cr\cr
&=&{\tiny\yng(3)}\oplus{\tiny\yng(2,1)}\oplus{\tiny\yng(2,1)}\oplus{\tiny\yng(1,1,1)}\cr\cr
&=&{\tiny\yng(3)}\oplus 2\,{\tiny\yng(2,1)}\oplus{\tiny\yng(1,1,1)}
\eea
Notice that the representation ${\tiny\yng(2,1)}$ appears with multiplicity 2. Consider the $U(3)$ semi-standard tableaux for each of the irreducible representations and indicate the $U(1)^3$ charge $\vec{m}$ for each:
\bea
{\color{red}\young(1,2,3)\quad\vec{m}=(1,1,1)}\nonumber
\eea
\bea
\young(11,2)\quad\vec{m}=(2,1,0)\qquad\young(11,3)\quad\vec{m}=(2,0,1)\qquad\young(12,2)\quad\vec{m}=(1,2,0)\qquad{\color{red}\young(12,3)\quad\vec{m}=(1,1,1)}\nonumber
\eea
\bea
{\color{red}\young(13,2)\quad\vec{m}=(1,1,1)}\qquad\young(13,3)\quad\vec{m}=(1,0,2)\qquad\young(22,3)\quad\vec{m}=(0,2,1)\qquad\young(23,3)\quad\vec{m}=(0,1,2)\nonumber
\eea
\bea
\young(111)\quad\vec{m}=(3,0,0)\qquad\young(112)\quad\vec{m}=(2,1,0)\quad\young(113)\quad\vec{m}=(2,0,1)\cr\cr
\young(122)\quad\vec{m}=(1,2,0)\qquad{\color{red}\young(123)\quad\vec{m}=(1,1,1)}\qquad\young(133)\quad\vec{m}=(1,0,2)\cr\cr
\young(222)\quad\vec{m}=(0,3,0)\qquad\young(223)\quad\vec{m}=(0,2,1)\qquad\young(233)\quad\vec{m}=(0,1,2)\cr\cr
\young(333)\quad\vec{m}=(0,0,3)\nonumber
\eea
The entries coloured in red are those that have the correct $U(1)^3$ charge for our problem where there is one green box on each row. Thus, we have the following Kostka numbers
\bea
K_{{\tiny \yng(3)},(1,1,1)}=1\qquad K_{{\tiny \yng(2,1)},(1,1,1)}=2\qquad K_{{\tiny \yng(1,1,1)},(1,1,1)}=1
\eea
We now know that there are a total of 6 restricted Schur polynomials, given by
\bea
&&1\qquad{\rm of}\qquad\chi_{R,(r,{\tiny\yng(3)})}\qquad{\rm no\,\,mulitplicity\,\,labels}\cr\cr
&&1\qquad{\rm of}\qquad\chi_{R,(r,{\tiny\yng(1,1,1)})}\qquad{\rm no\,\,mulitplicity\,\,labels}\cr\cr
&&4\qquad{\rm of}\qquad\chi_{R,(r,{\tiny\yng(2,1)})\alpha\beta}\qquad\alpha,\beta=1,2
\eea

There is a second way to compute the Kostka numbers. Introduce the group
\bea
H&=&S_{m_1}\times S_{m_2}\times\cdots\times S_{m_p}
\eea
which is a subgroup of $S_m$. If we restrict $S_m$ to subgroup $H$, then $K_{s,\vec{m}}$ is equal to the multiplicity of the trivial representation of $H$, which appears when we have restricted $S_m$ representation $s$. This was proved by Kostka and we will simply use this result. It implies that
\bea
{1\over |H|}\sum_{\gamma\in H}\Gamma_s(\gamma)_{ab}&=&\sum_{\mu}B_{\mu a}B_{\mu b}
\eea
$B_{\mu a}$ is called a branching coefficient. $\mu$ is a restricted Schur polynomial multiplicity label. To understand this last equation, note that the left hand side is a projection operator onto irreducible representations that are each one dimensional. The vectors $v(\mu)_a$ with components $a$ are basis vectors for each representation. The trivial representation of $H$ is the representation given by
\bea
\gamma&=&1\qquad \forall\gamma\in H
\eea
Since it is a 1 dimensional representation, the restricted characters in this representation are very simple: they are given by
\bea
\chi^s_{\mu\alpha}(\sigma)&\equiv& \sum_{a,b=1}^{{\rm dim}_s}B_{\mu a}\Gamma_s(\sigma)_{ab}B_{\alpha b}
\eea
Since we have projected to the trivial representation of $H$ we have
\bea
\chi^s_{\mu\alpha}(\sigma)&=&\chi^s_{\mu\alpha}(\gamma\sigma)\,\,=\,\,\chi^s_{\mu\alpha}(\sigma\gamma)
\eea
for all $\gamma\in H$. Thus, $\chi^s_{\mu\alpha}(\sigma)$ is a function on the double coset
\bea
H\setminus S_m/H
\eea
In fact, $\chi^s_{\mu\alpha}(\sigma)$ are a complete set of functions on this double coset, so that we can perform a Fourier transform
\bea
O_{R,r,\sigma}(Z,Y)&=&\sum_{s,\mu,\alpha}\chi_{R,(r,s)\mu\alpha}(Z,Y)\chi^s_{\mu\alpha}(\sigma)
\eea
which has exchanged the following sets of labels
\bea
\sigma&\longleftrightarrow&s,\mu,\alpha
\eea
So, all the data related to the $Y$s (which we claim are related to the edges of the Gauss graphs) is now replaced by an element of the double coset $H\setminus S_m/H$. So the natural question now is to ask how we should interpret this double coset. We will take a small detour to answer this question. Imagine we want to draw all graphs with $p$ nodes and $m$ edges. Further, there are $m_i$ edges starting from and ending on node $i$. Draw the nodes and the starting/ending points as follows
\begin{center}
\includegraphics[width=0.55\textwidth]{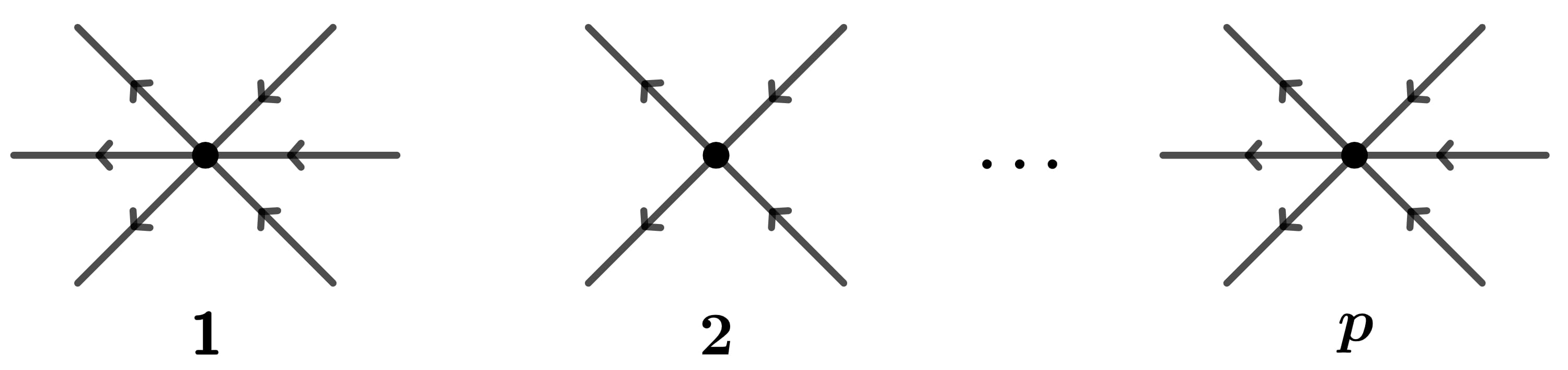}
\end{center}

\noindent
Label the start points and the endpoints, each with their own unique integer 1,2,...,$m$ as follows
\begin{center}
\includegraphics[width=0.55\textwidth]{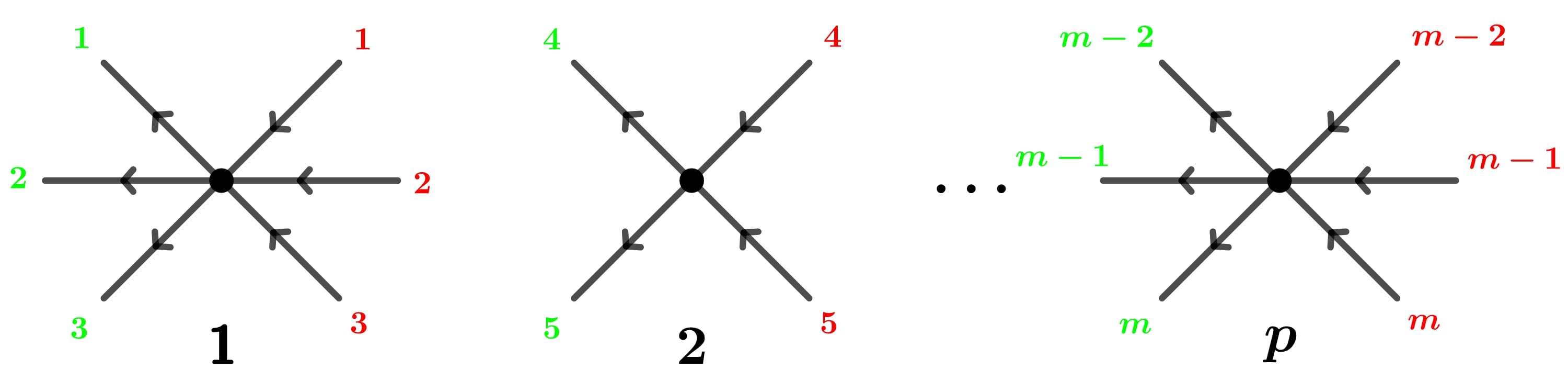}
\end{center}

\noindent
To get all possible graphs we now need to consider all possible associations of the endpoints with the starting points. Each such association defines a permutation 
\begin{center}
\includegraphics[width=0.4\textwidth]{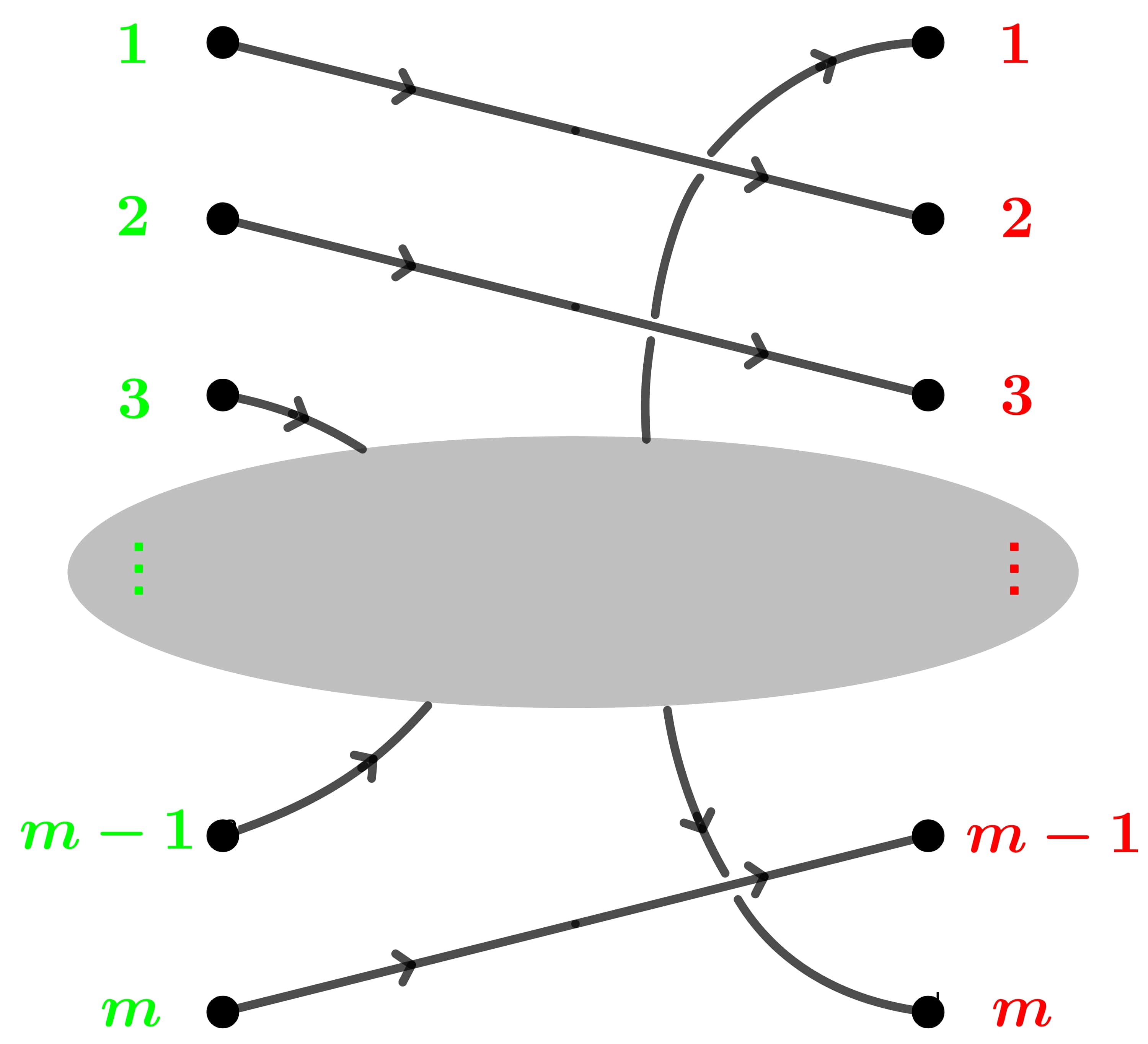}
\end{center}

\noindent
Not all permutations correspond to distinct graphs. Permutations that differ only by swapping start points that start at the same node give the same graph. Similarly, permutations that differ only by swapping end points that end at the same node give the same graph. Thus, two permutations $\sigma_1$ and $\sigma_2$ related as
\bea
\sigma_1&=&\gamma_1\sigma_2\gamma_2\qquad \gamma_1,\gamma_2\in H
\eea
give the same graph. Thus, graphs with $p$ nodes and $m$ edges together with the constraint that the number of edges starting on node $i$ equals the number of edges ending on node $i$ equals $m_i$ are given by elements of the double coset
\bea
H\setminus S_m/H\qquad H&=&S_{m_1}\times S_{m_2}\times\cdots\times S_{m_p}
\eea
Thus our Gauss graph operators are labelled by graphs. A very natural interpretation is to identify the nodes of the graph with giant graviton branes and to identify the directed edges with the open string excitations.

We can illustrate the above arguments with an example. Above we argued that there are 6 restricted Schur polynomials of the form
\begin{center}
\includegraphics[width=0.6\textwidth]{diagram30.jpg}
\end{center}

\noindent
Recall that they are given by
\bea
&&1\qquad{\rm of}\qquad\chi_{R,(r,{\tiny\yng(3)})}\qquad{\rm no\,\,mulitplicity\,\,labels}\cr\cr
&&1\qquad{\rm of}\qquad\chi_{R,(r,{\tiny\yng(1,1,1)})}\qquad{\rm no\,\,mulitplicity\,\,labels}\cr\cr
&&4\qquad{\rm of}\qquad\chi_{R,(r,{\tiny\yng(2,1)})\alpha\beta}\qquad\alpha,\beta=1,2
\eea
Fourier transforming to the Gauss graph basis these becomes $O_{R,r,\sigma}$. Thus, there should be 6 possible Gauss graphs labeled by the permutations $\sigma$. This is indeed the case and the 6 graphs are given by
\begin{center}
\includegraphics[width=0.6\textwidth]{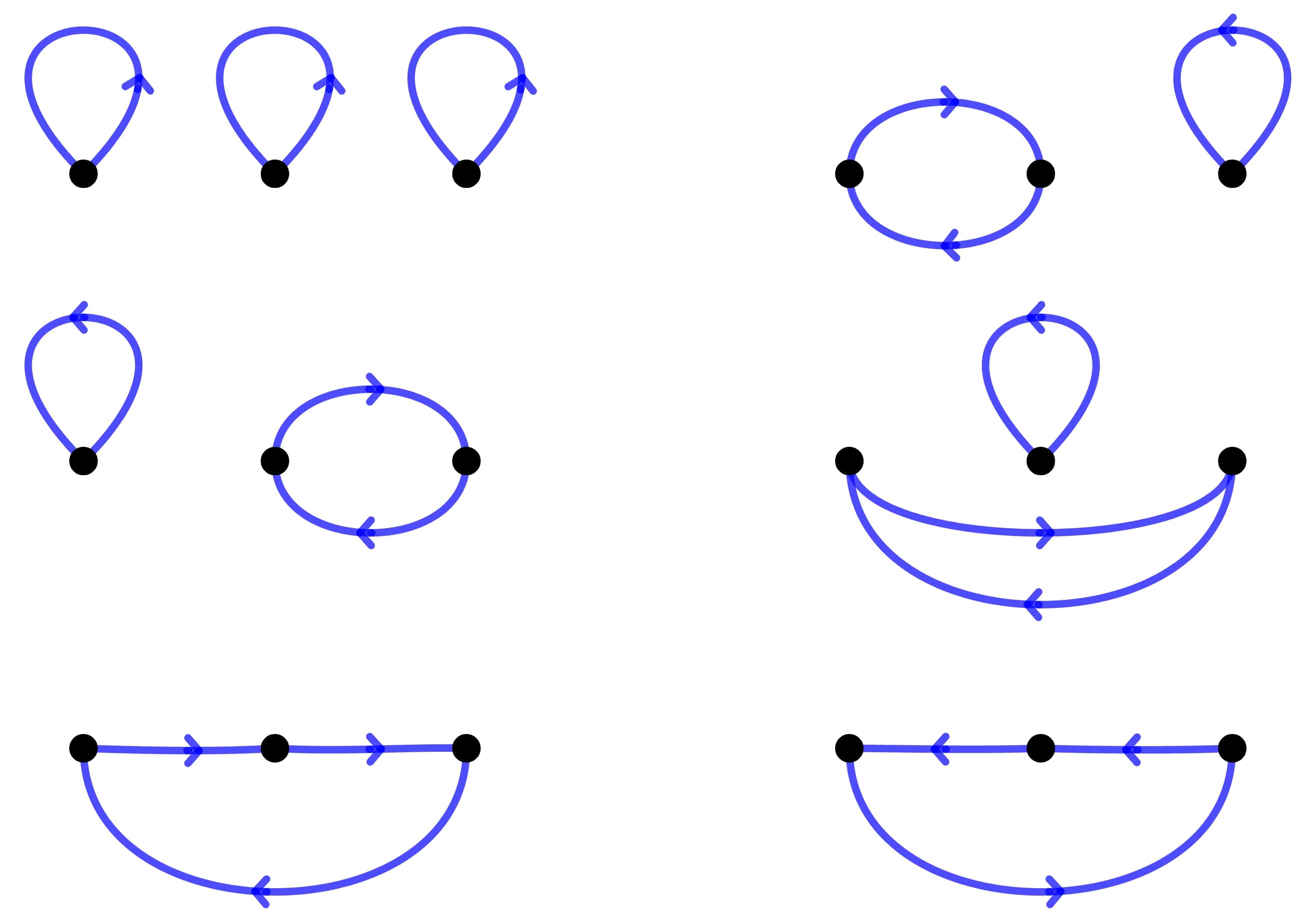}
\end{center}

The dilatation operator can be evaluated in the restricted Schur polynomial basis\cite{DeComarmond:2010ie,deMelloKoch:2011wah}. The dilatation operator in the SU(2) sector is given by
\bea
D&=&-g_{YM}^2 {\rm Tr}\left([Z,Y][\partial_Z,\partial_Y]\right)
\eea
Acting on restricted Schur polynomials, that have a normalized two point function
\bea
O_{R,(r,s)\alpha\beta}(Z,Y)&=&\sqrt{{\rm hooks}_r{\rm hooks}_s\over {\rm factors}_R{\rm hooks}_R}\chi_{R,(r,s)\alpha\beta}(Z,Y)
\eea
the dilatation operator acts as
\bea
DO_{R,(r,s)\alpha\beta}&=&\sum_{T,(t,u)\gamma\delta}N_{R,(r,s)\alpha\beta\,\,T,(t,u)\gamma\delta} O_{T,(t,u)\gamma\delta}
\eea
where
\bea
N_{R,(r,s)\alpha\beta\,\,T,(t,u)\gamma\delta}&=&- g_{YM}^2\sum_{R'}{c_{RR'} d_T n m\over d_{R'} d_t d_u (n+m)}
\sqrt{f_T \, {\rm hooks}_T\, {\rm hooks}_r \, {\rm hooks}_s \over f_R \, {\rm hooks}_R\, {\rm hooks}_t\, {\rm hooks}_u}\times\cr\cr
&&\times{\rm Tr}\Big(\Big[ \Gamma_R((1,m+1)),P_{R\to (r,s)\alpha\beta}\Big]I_{R'\, T'}\Big[\Gamma_T((1,m+1)),P_{T\to (t,u)\gamma\delta}\Big]I_{T'\, R'}\Big)\cr
&&
\eea
Some comments are in order
\begin{itemize}
\item Operators $O_{R,(r,s)\alpha\beta}$ and $O_{T,(t,u)\gamma\delta}$ only mix if $R$ and $T$ agree after a single box is dropped from each. Denote the Young diagram obtained by dropping a box from $R$, ($T$) as $R'$, ($T'$) respectively. We must have $R'=T'$ for a non-zero matrix element $N_{R,(r,s)\alpha\beta\,\,T,(t,u)\gamma\delta}$. The intertwiner $I_{R'T'}$ is a map from the carrier space of $T'$ to that of $R'$ so that, for example $\Gamma_{R'}(\sigma)I_{R'T'}=I_{R'T'}\Gamma_{T'}(\sigma)$, $\sigma\in S_{n+m-1}$. Similarly, the intertwiner $I_{T'R'}$ is a map from the carrier space of $R'$ to that of $T'$.
\item $c_{RR'}$ is the factor of the box that must be dropped from $R$ to obtain $R'$.
\item The (not normalized) projector $P_{R\to (r,s)\alpha\beta}$ is given by
\bea
P_{R\to (r,s)\alpha\beta}&=&{1\over n!m!}\sum_{\sigma\in S_{n+m}}\chi_{R(r,s)\alpha\beta}(\sigma)\Gamma_R(\sigma)
\eea
\item The above result for the matrix elements is exact in the sense that it includes the complete $N$ dependence - no approximation has been made.
\end{itemize}

Now consider the action of the dilatation operator on the Gauss graph operators. This computation is performed using an approximation known as the distant corners approximation. To state what the distant corners approximation is, consider the Young diagram shown below
\begin{center}
\includegraphics[width=0.6\textwidth]{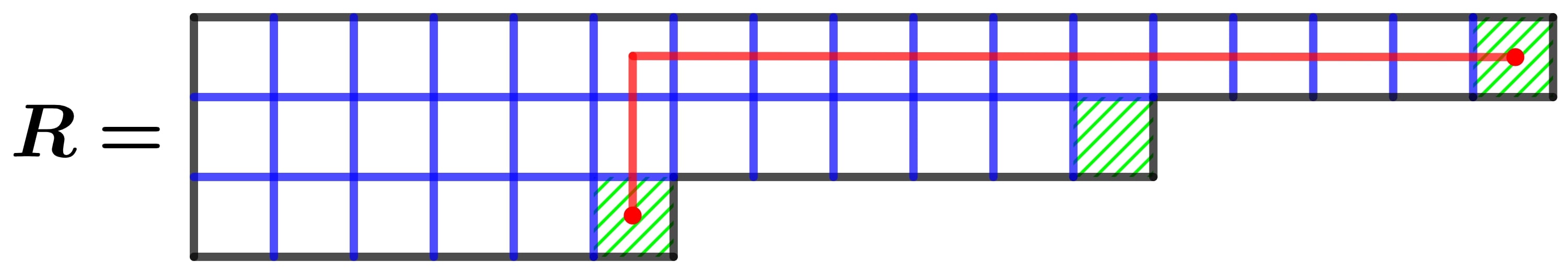}
\end{center}

\noindent
This approximation amounts to assuming that the red elbow is a length of order $N$. The point of making this approximation is that it leads to a significant simplification in the action of the symmetric group. For this discussion we will make use of a specific representation known as Young's orthogonal representation. States are labelled by Young-Yamanouchi patterns. A Young-Yamanouchi pattern is a Young diagram with $m$ boxes and with all boxes labelled with a unique integer taken from the set $\{1,2,\cdots,m\}$. If the boxes are dropped in the sequence specified by their integer labelling, a valid Young diagram is obtained at each step. Here are two Young diagrams with the boxes dropped according to their labelling
\bea
\young(43,21)\,\,\to\,\,\young(43,2)&\to&\young(43)\,\,\to\,\,\young(4)\cr\cr
\young(42,31)\,\,\to\,\,\young(42,3)&\to&\young(4,3)\,\,\to\,\,\young(4)
\eea
Each Young-Yamanouchi pattern labels a state and the number of Young-Yamanouchi patterns is equal to the dimension of the representation. Giving a matrix acting in this vector space is the same as giving a rule for how the matrices act on the Young-Yamanouchi pattern. Young's orthogonal representation gives a rule for the adjacent two cycles, that is, two cycles of the form $(i\,i+1)$. Given the matrices for these adjacent two cycles, we can determine the matrices representing the rest of group by taking products. Here is an example of Young's rule
\bea
\Gamma_{\tiny\yng(6,3)}\Big((12)\Big)\,\young(986531,742)&=&{1\over 4}\,\young(986531,742)
+\sqrt{1-\left({1\over 4}\right)^2}\,\young(986532,741)\cr
&&
\eea
The ``4" in this rule is the length of the hook stretching between boxes 1 and 2. One way of defining this length is in terms of the factors of the boxes. Box 1 has factor $f_1=N+5$ and box 2 has factor $N+1$. The hook length is the difference between these two factors $f_1-f_2=4$. In the limit of a large Young diagram, taken so that that this elbow length is infinite the ``${1\over 4}$" terms in the above formula vanish. In this case the action of the symmetric group is particularly simple: permutations swap the labels of boxes that belong to different rows. Here is an example:
\begin{center}
\includegraphics[width=0.7\textwidth]{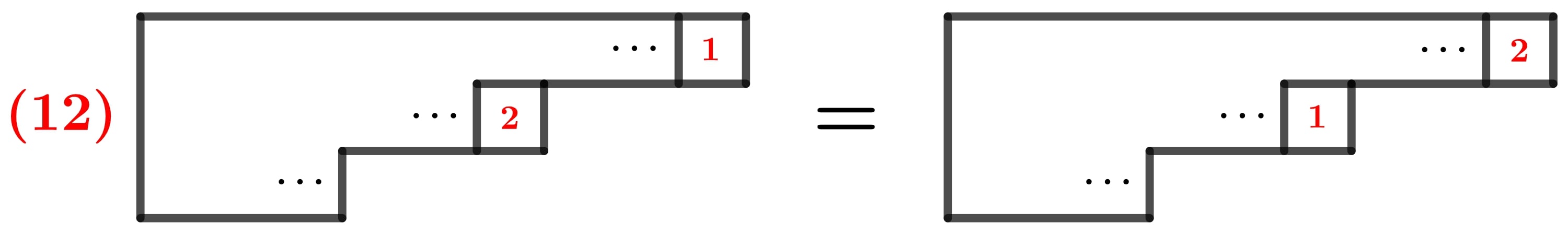}
\end{center}

\noindent
States that have the labelled boxes in the same row are invariant under the permutation. In the distant corners approximation, the action of the dilatation operator on the Gauss graph operators is
\bea
DO_{R,r,\sigma}(Z,Y)&=&\sum_{i<j=1}^p n_{ij}(\sigma)\Delta_{ij}O_{R,r}(\sigma)
\eea
The action of operator $\Delta_{ij}$ is given by three terms
\bea
  \Delta_{ij}&=&\Delta_{ij}^{+}+\Delta_{ij}^{0}+\Delta_{ij}^{-}
\eea
Denote the row lengths of $r$ by $r_i$. The Young diagram $r_{ij}^+$ is obtained by removing a box from row $j$ and adding it to row $i$. The Young diagram $r_{ij}^-$ is obtained by removing a box from row $i$ and adding it to row $j$. In terms of these Young diagrams we have
\bea
\Delta_{ij}^0O_{R,r,\sigma}(Z,Y)&=&-(2N+r_i+r_j)O_{R,r,\sigma}(Z,Y)
  \label{0term}
\eea
\bea
\Delta_{ij}^+O_{R,r,\sigma}(Z,Y)&=&\sqrt{(N+r_i)(N+r_j)}O_{R^+_{ij},r^+_{ij},\sigma}(Z,Y)
  \label{pterm}
\eea
\bea
\Delta_{ij}^{-}O_{R,r,\sigma}(Z,Y)&=&\sqrt{(N+r_i)(N+r_j)}O_{R^-_{ij},r^-_{ij},\sigma}(Z,Y)
  \label{mterm}
\eea
Notice that $\Delta_{ij}$ acts only on the $R,r$ labels, so that the dilatation operator does not mix operators with different Gauss graph labels. Finally, $n_{ij}(\sigma)$ counts the number of edges running between nodes $i$ and $j$. For example
\begin{center}
\includegraphics[width=0.6\textwidth]{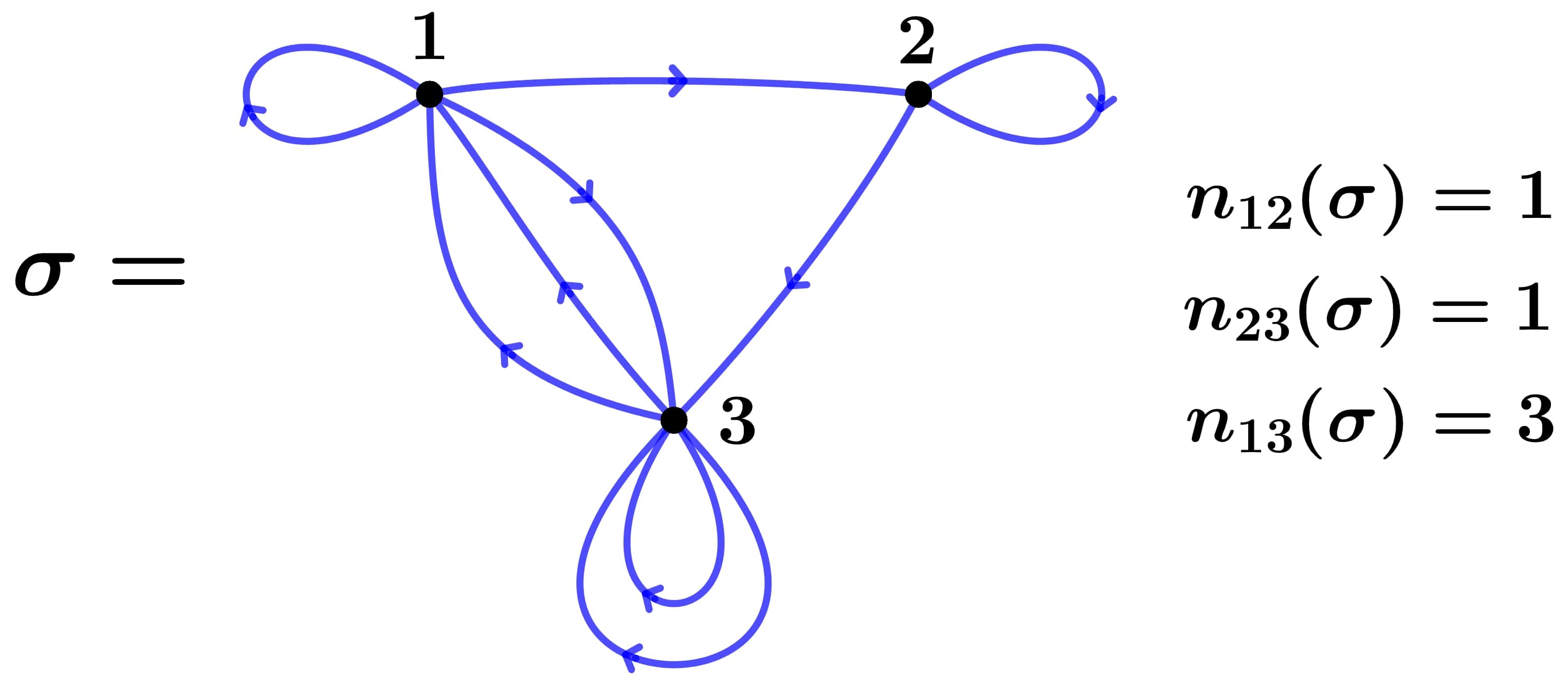}
\end{center}

\noindent
With this action it is easy to identify the BPS operators: they are the operators labelled by Gauss graphs that have no edges stretched between any nodes. Here is an example of a BPS operator constructed using Young diagrams $R,r$ that have three rows and a total of 6 $Y$ fields
\begin{center}
\includegraphics[width=0.6\textwidth]{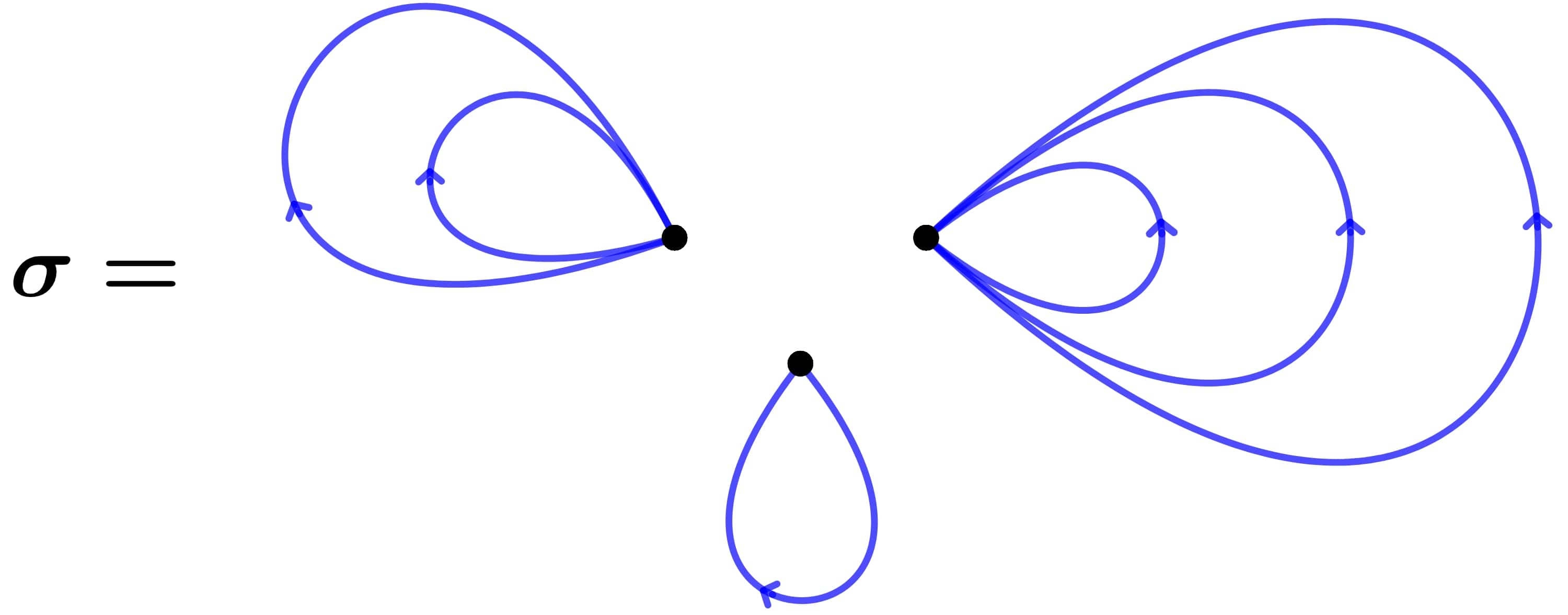}
\end{center}

\noindent
Another very robust result obtained using this dilatation operator is the fact that the energy gap (or equivalently, the size of the anomalous dimension) between BPS and non-BPS operators is given by
\bea
E_{\rm gap}\sim{\lambda\over N}
\eea
where $\lambda$ is the 't Hooft coupling.

{\vskip 0.5cm}

\noindent
{\bf Guide to further reading:}

The action of the dilatation operator acting on restricted Schur polynomials has been studied in a number of papers\cite{DeComarmond:2010ie,deMelloKoch:2011wah,deMelloKoch:2010zrl,Carlson:2011hy,deMelloKoch:2012ck,deMelloKoch:2011ci}. The form of the one loop dilatation operator was determined in [\citen{DeComarmond:2010ie,deMelloKoch:2011wah}]. Using these explicit results, the papers [\citen{deMelloKoch:2010zrl,DeComarmond:2010ie}] performed a numerical diagonalization of the dilatation operator. The result was rather surprising, exhibiting anomalous dimensions that were evenly spaced with the spacing given by an integer time $g_{YM}^2$. In an attempt to explain this simple result, the paper [\citen{Carlson:2011hy}] introduced the displaced corners approximation, for Young diagrams with either two rows or two columns. This was extended to arbitrary Young diagrams in [\citen{deMelloKoch:2011wah}] and, again with the help of numerical analysis, the Gauss graph description was discovered. The paper [\citen{deMelloKoch:2012ck}] achieved an analytic description of the Gauss graph basis. This analytic description was heavily influenced by the group theoretic description of graphs\cite{deMelloKoch:2011uq,deMelloKoch:2012tb}. As we have explained above, the action of the dilatation operator on the Gauss graph operators $O_{R,r}(\sigma)$ still has a non-trivial action on the $R,r$ labels. This action is diagonalized in [\citen{deMelloKoch:2011ci}]; see also [\citen{Lin:2014yaa}] for an interesting discussion of the problem. 

For the action of the dilatation operator on the basis constructed in [\citen{Brown:2007xh,Brown:2008ij}] see [\citen{Brown:2008rs}]. For further interesting discussions related to the material of this section see [\citen{Berenstein:2013md,Kimura:2013fqa,Berenstein:2013eya,deMelloKoch:2013eww,deMelloKoch:2013not,Berenstein:2014pma,Diaz:2014ixa,Diaz:2015tda,Jiang:2019xdz,deCarvalho:2020pdp,Suzuki:2021sma,Su:2022ntc}].

\section{Counting Operators}\label{PakMan}

Our operators $O_{R,r, \sigma}$ have 3 labels $R$, $r$ and $\sigma$. Let's consider the case that our operator is constructed from a total of $O(N^2)$ fields so that $R$ has $O(N^2)$ boxes. We will also assume that there are order $N^2$ $Z$ fields and order $N^2$ $Y$ fields. An over estimate of the number of $R,r$ labels is given by counting the partitions of $n+m$ and $n$ respectively. The number of partitions of $n+m$ grows as
\bea
&\sim& {1\over 4(n+m)\sqrt{3}}\exp \left(\pi\sqrt{2(n+m)\over 3}\right)\,\,\sim\,\,{c_1\over N^2}e^{c_2 N}
\eea
where $c_1$ and $c_2$ are numbers that are order 1. The number of Gauss graphs is given by the sum over the squares of Littlewood-Richardson numbers. The biggest Littlewood-Richardson number for 3 partitions $R,r,s$ has been estimated in [\citen{PakPanova}]. Applying that result we find ($c_3$ and $c_4$ are both order $N^0=1$ constants)
\bea
f^{R}_{rs}\sim c_3e^{c_4 N^2}
\eea
so that the number of Gauss graphs grows as $e^{N^2}$.

There is a well defined probability distribution on the space of partitions (which is the space of Young diagrams). This distribution is singular (i.e. a $\delta$ function measure) and it is concentrated\cite{limitcurve} on Young diagrams that have the so called VKLS shape\cite{VershikKerov,LoganShepp}. The largest Littlewood-Richardson number is obtained when $R,r$ and $s$ all have the VKLS shape.\footnote{To understand this with finite number of boxes, note that we discussed the multiplicity labels for $S_3 \times S_3$ subgroup of $S_6$ in section 5. When $R,r$ and $s$ are only close to the VKLS shape, we got the Littlewood-Richardson number. All others decompositions didn't have it.}

The fact that there is a typical Young diagram and that the growth in the number of Gauss graphs is fast enough to account for the entropy of a black hole suggests that restricted Schur polynomials or perhaps Gauss graph operators, with the typical Young diagram labels, may provide operators dual to microstates of a black hole. The labels would be something like what we have shown below.
\begin{center}
\includegraphics[width=0.6\textwidth]{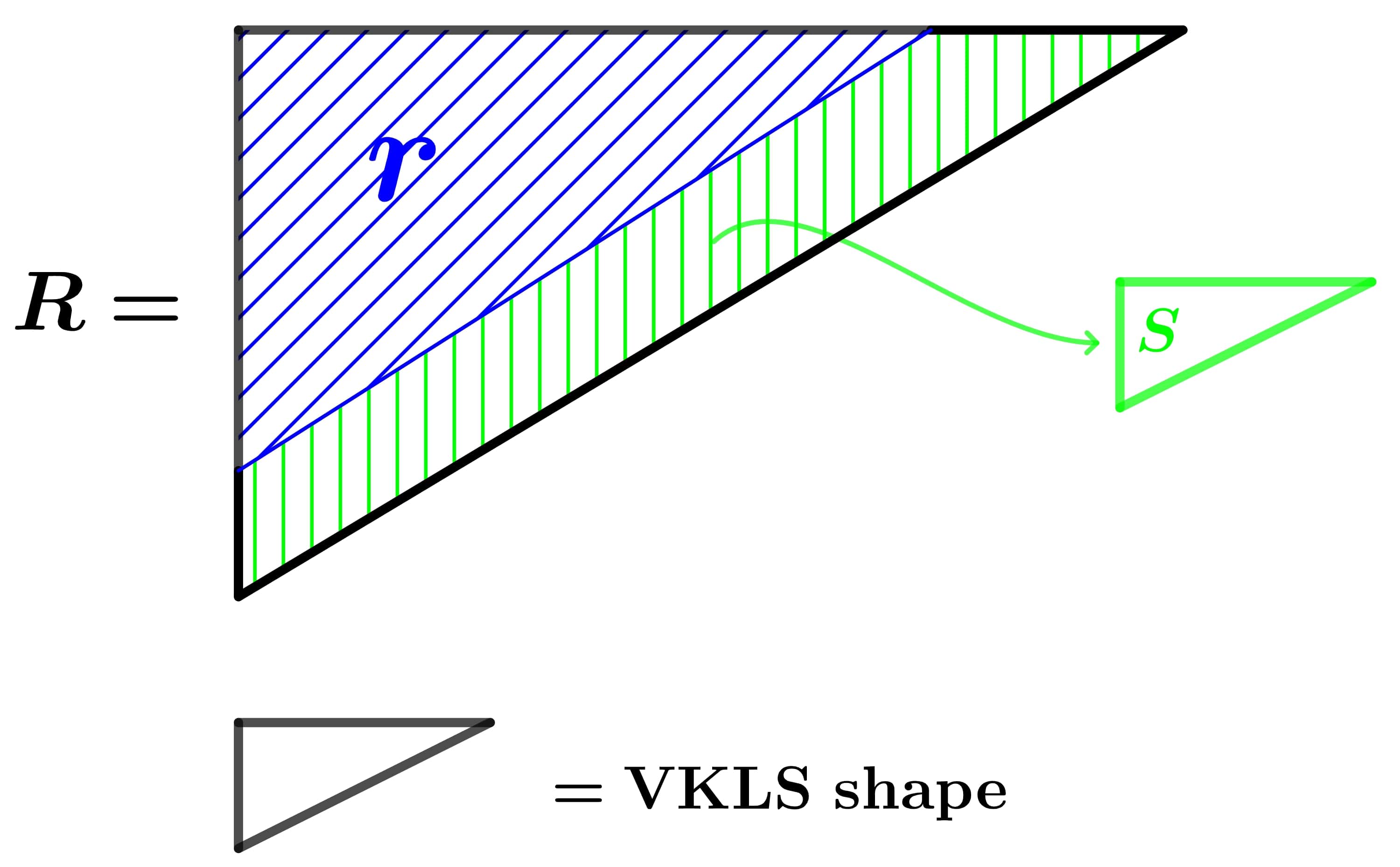}
\end{center}

Notice that the green boxes are not separated as they were in the distant corners limit. This has a number of important consequences for the construction of the Gauss graph operators. Start from a Young diagram that does exhibit the distant corners scenario
\begin{center}
\includegraphics[width=0.6\textwidth]{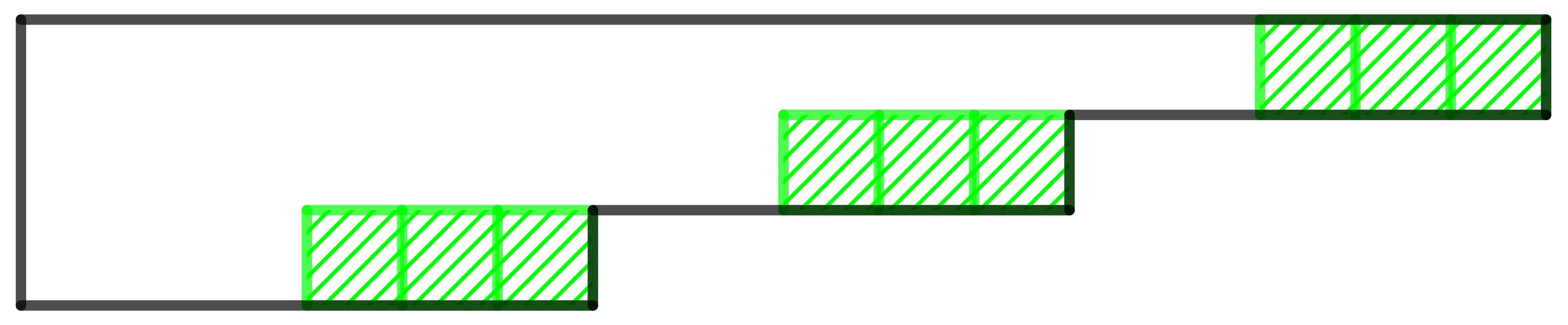}
\end{center}

\noindent
The green boxes on their own define a skew Young diagram. Young's orthogonal representation can still be used to construct a representation, from the skew Young diagram. The key difference is that, while the representation constructed from a Young diagram corresponds to an irreducible representation, the representation constructed from a skew Young diagram is in general reducible. For the case we consider here, the skew Young diagram is given by
\begin{center}
\includegraphics[width=0.3\textwidth]{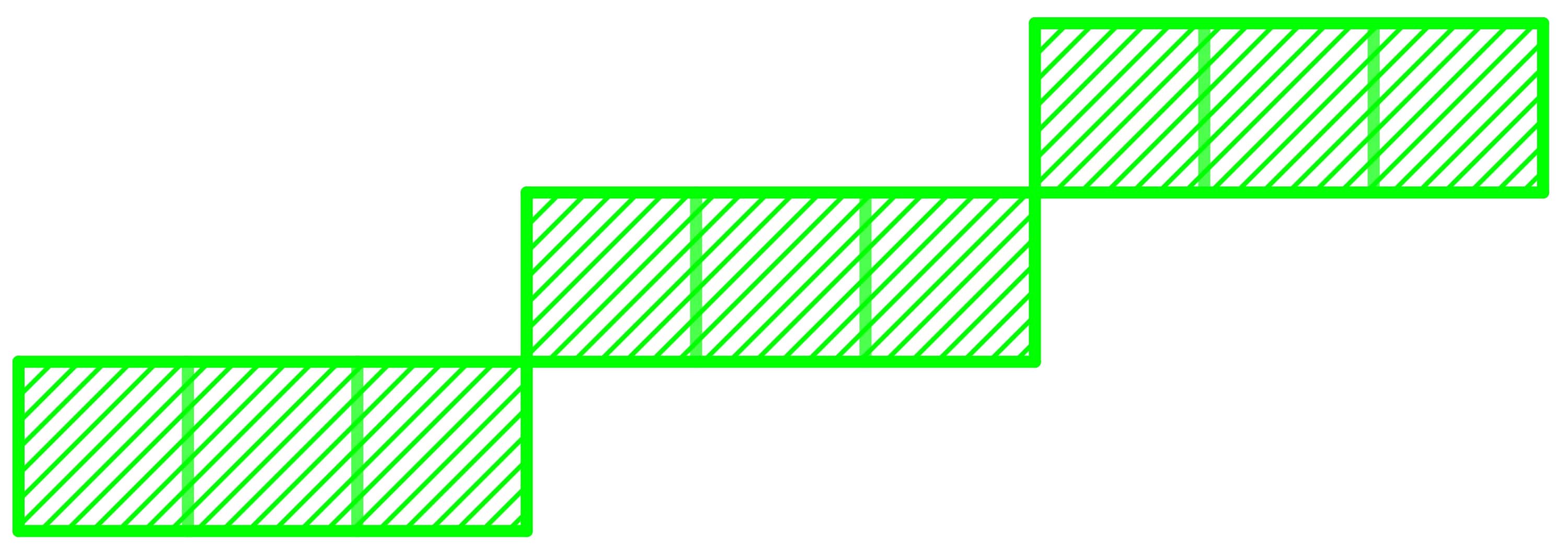}
\end{center}

\noindent
This skew Young diagram is known as a strip. The green boxes in the different rows don't overlap. If we decompose the reducible representation associated to the strip, the multiplicities are simply given by the Kostka numbers. This is what is used in the Gauss graph construction we derived above and it gives us the physics of the Coulomb branch of the gauge theory.

A small generalization is to consider the following Young diagram
\begin{center}
\includegraphics[width=0.45\textwidth]{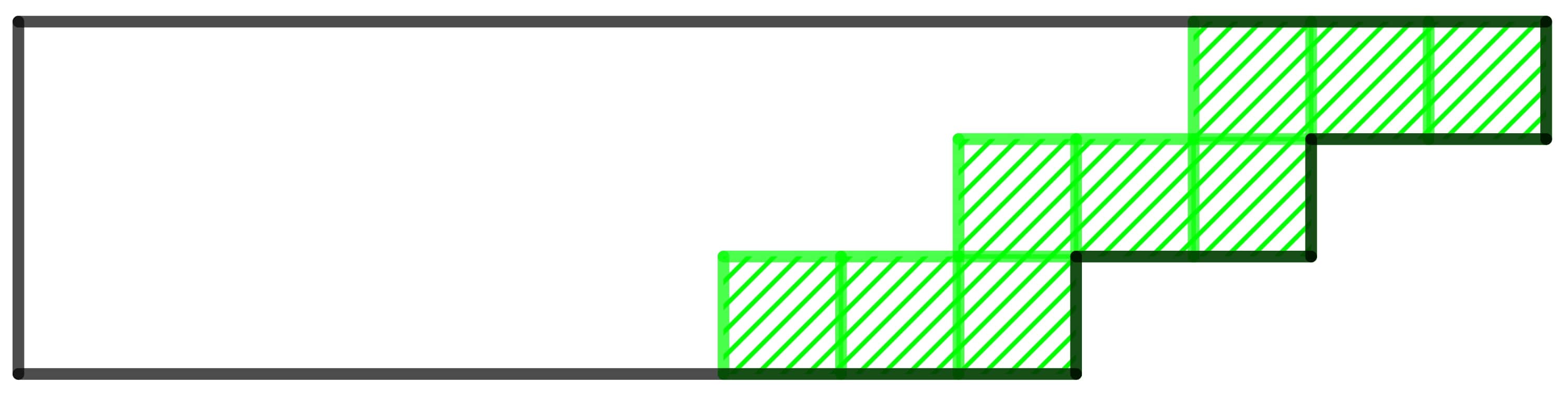}
\end{center}

\noindent
Now there is overlap between the green boxes in different rows. However, the overlap always involves only a single box. This case has been studied recently by mathematicians [\citen{smallkostka}]. The resulting skew Young diagram, shown below, is known as a ribbon
\begin{center}
\includegraphics[width=0.25\textwidth]{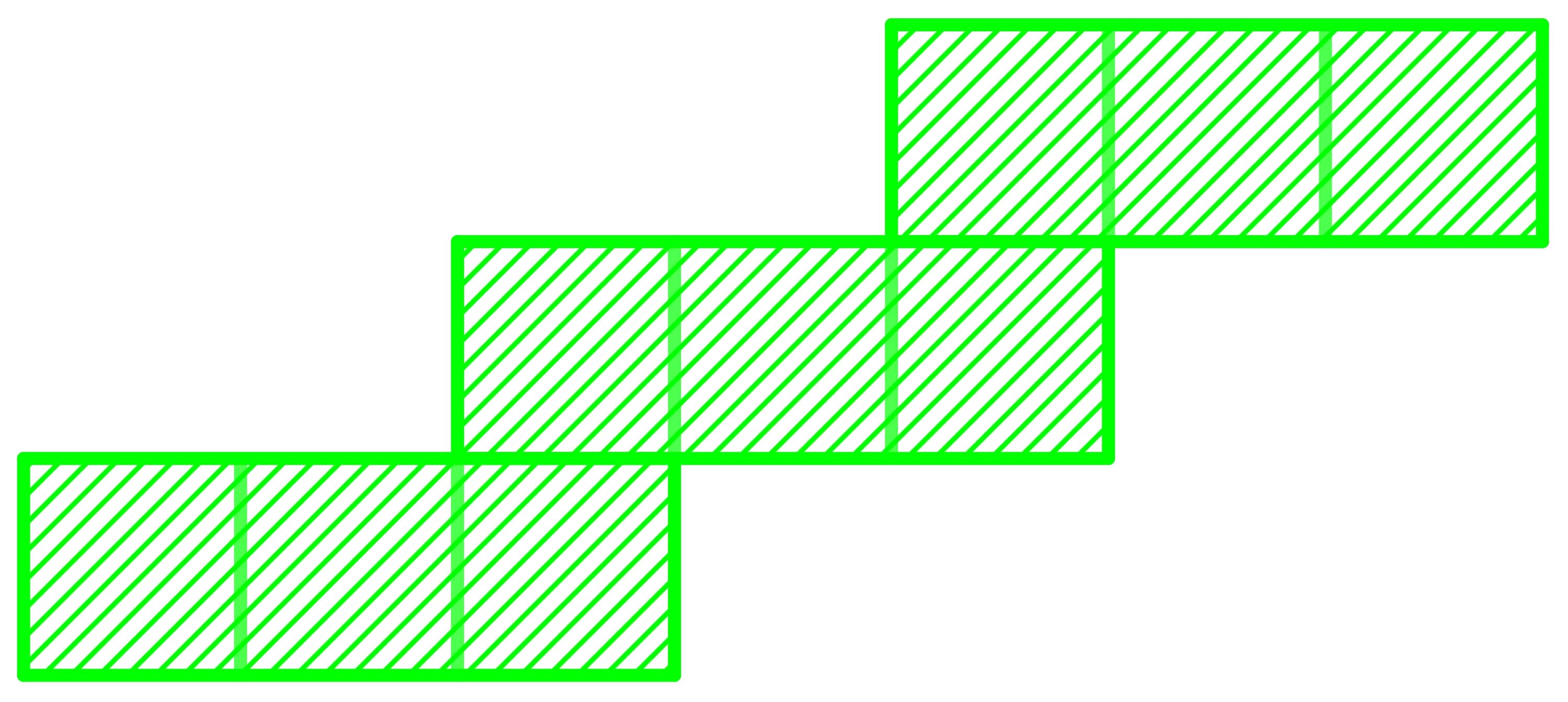}
\end{center}

\noindent
The multiplicities related to the ribbon are known as the small Kostka numbers.

The case that is relevant for black hole microstates is a generalization of the ribbon to the case shown below, where we allow arbitrary overlaps between the green boxes in different rows.
\begin{center}
\includegraphics[width=0.5\textwidth]{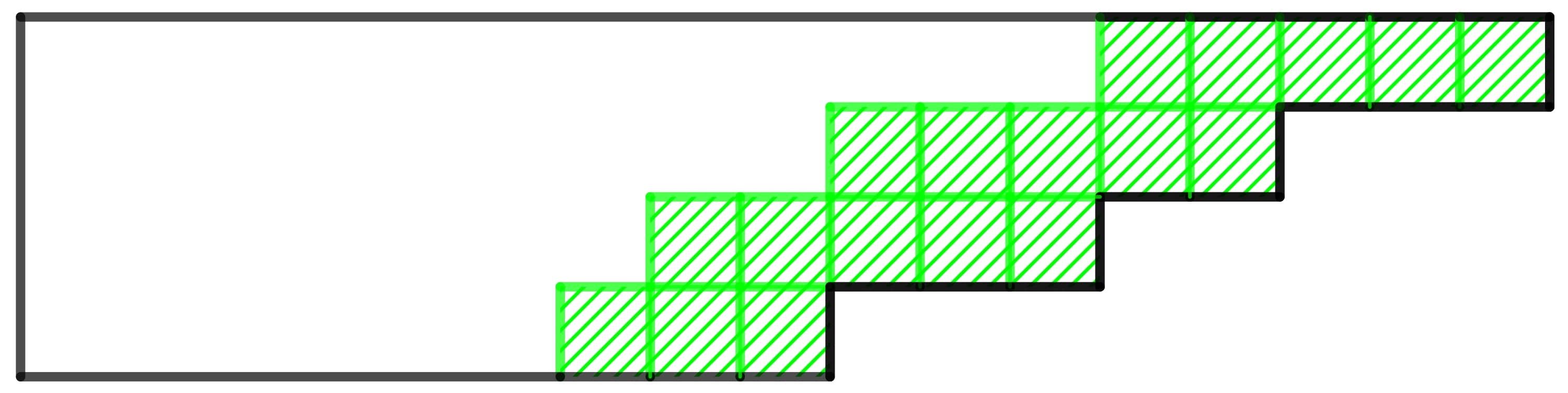}
\end{center}
The multiplicity problem for this case has not been solved. We expect that solving this problem will shed light on the construction of the operators dual to black hole microstates.

The natural setting in which we can consider the construction of black hole microstates, is for the ${1\over 16}$-BPS black hole. To construct operators dual to microstates for this black hole it is necessary to construct composite operators using field strengths $f$, fermions $\psi_{j}$ $(j=1,2,3)$ and scalars $\phi^i$ ($i=1,2,3$). A simple toy model, that we expect captures many of the essential features of the full problem, is provided by a two-matrix model which uses two of the scalars. Consider the problem of state counting and for simplicity consider a free theory. A basis for the local operators would be provided by the restricted Schur polynomials, and as we have explained, these are labelled by Young diagrams. A typical state will belong to those configurations with the highest entropy and for these the Young diagram labels will take the VKLS shape. The maximal Littlewood-Richardson number, which fixes the range of the multiplicity labels of the two-matrix restricted Schur polynomial, is \cite{PakPanova}
\bea
f^{R}_{rs} = 2^{\frac{l}{2}}
\eea
where $l$ is the total number of boxes $l\,=\,n+m$. For the free theory $l$ can be identified with the energy $E$. The total number of restricted Schur polynomials grows as the square of this number. Taking the log of this number leads to the following entropy
\bea
S = \log (f^{R}_{rs})^2 \simeq E \log{2}
\eea
where $R,r,s$ will each take on the VKLS shape. This estimate gives an entropy of the expected size and this result may have a direct application to small AdS black holes. For large AdS black holes this is an over estimate because the finite $N$ trace relations have not been taken into account. To account for the trace relations, the Young diagram labels should all be truncated to ensure that they have no more than $N$ rows. The corresponding limit curve problem under this condition has been studied by Logan and Shepp \cite{LoganShepp} and it leads to a ``truncated VKLS'' curve. The transition from the VKLS curve to the tVKLS curve is illustrated below 
\begin{center}
\includegraphics[width=0.6\textwidth]{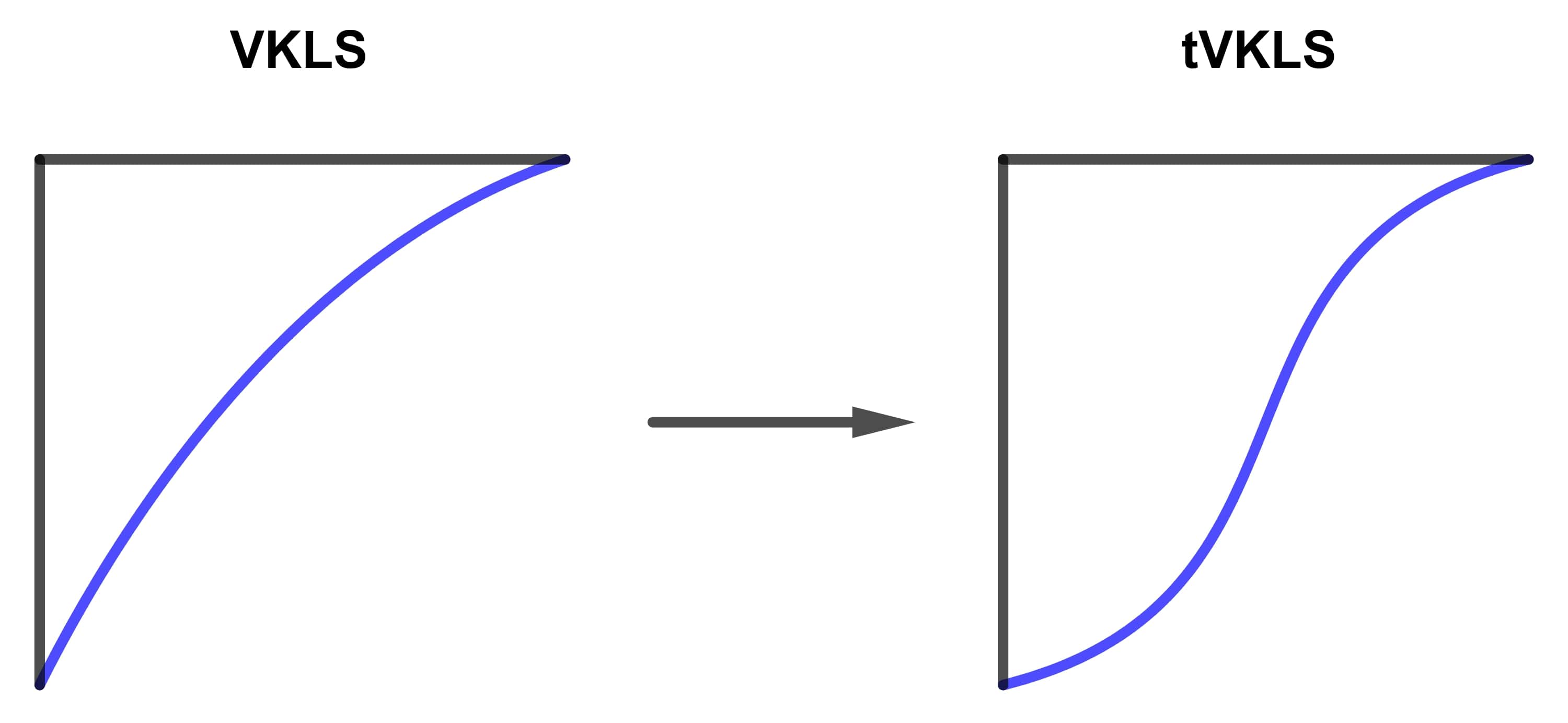}
\end{center}
The maximum value of the Littlewood-Richardson number when the Young diagram labels take on the truncated VKLS shape was also considered in [\citen{PakPanova}]. The maximal value can be written as
\bea
\left(f^{R}_{rs}\right)_{N} = (l+1)^{\frac{1}{2} N^2}
\eea
With this value the entropy becomes  
\bea
S= N^2 \log{E} 
\eea
which matches the expected value of the entropy for a large AdS black hole. It is an interesting open problem, currently under investigation\cite{dkkl}, to provide a physicist's standard of proof for the above entropy value.

{\vskip 0.5cm}

\noindent
{\bf Guide to further reading:}

The physics of very heavy operators (with a bare dimension of order $N^2$) has been explored in detail using gauge theory/gravity duality. For nice pioneering papers which tried to recover the broad features of black hole thermodynamics from the ${1\over 2}$-BPS sector of ${\cal N}=4$ super Yang-Mills theory, see [\citen{Balasubramanian:2005kk,Balasubramanian:2005mg}]. For interesting attempts to reconstruct spacetime see [\citen{Berenstein:2005aa,Berenstein:2008eg,deMelloKoch:2008hen,Berenstein:2010dg}]. Recent studies of the thermodynamics of matrix models, using the techniques of this review are [\citen{Berenstein:2018hpl,Berenstein:2023srv,OConnor:2023ptf,OConnor:2023mss,OConnor:2024udv}].

\section*{Acknowledgments}
This review is based on lectures delivered at Seoul National University in June 2024. We are very grateful to Seok Kim for hospitality at Seoul National University and for extremely useful and encouraging discussions. We would also like to thank the participants at the lectures for creating a stimulating scientific atmosphere through their questions and discussions. RdMK is supported by a start up research fund of Huzhou University, a Zhejiang Province talent award and by a Changjiang Scholar award. MK is supported by the National Research Foundation of Korea (NRF) Grants funded by the Korea government~(MSIT), NRF2023K2A9A1A01098740, NRF2023R1A2C200536011,  NRF-2022R1I1A1A01069032 and NRF-2020R1A6A1A03047877 (Center for Quantum Space Time). ALM is supported by a grant from ``The World Academy of Sciences (TWAS)," grant reference number: No. 20-120 RG/MATHS/AF/AC\_I – FR3240314136.

\end{document}